\documentclass[a4paper,usenatbib]{mnras}
\usepackage{graphicx}
\usepackage[fleqn]{amsmath}
\usepackage[T1]{fontenc}
\usepackage{ae,aecompl}
\usepackage{amssymb,amsfonts,hyperref,color}
\title[Massive lopsided transition discs II]{Gas and dust
  hydrodynamical simulations of massive lopsided transition discs --
  II.  Dust concentration}

\author[Baruteau and Zhu]{Cl{\'e}ment Baruteau$^1$\thanks{E-mails:
    clement.baruteau@irap.omp.eu (C.B.); zhzhu@astro.princeton.edu
    (Z.Z.)}
  and Zhaohuan Zhu$^{2,3}$\\
  $^1$CNRS / Institut de Recherche en Astrophysique et Plan{\'e}tologie, 14 avenue Edouard Belin, 31400 Toulouse, France\\
  $^2$Department of Astrophysical Sciences, 4 Ivy Lane, Peyton Hall,
  Princeton University, Princeton, NJ 08544, USA\\
  $^3$Hubble Fellow }

\date{Accepted 2015 October 27.  Received 2015 October 27; in original form 2015 August 4}

\begin{document}
\label{firstpage}
\pagerange{\pageref{firstpage}--\pageref{lastpage}} \pubyear{2015}
\maketitle

\begin{abstract}
  We investigate the dynamics of large dust grains in massive lopsided
  transition discs via 2D hydrodynamical simulations including both
  gas and dust.  Our simulations adopt a ring-like gas density profile
  that becomes unstable against the Rossby-wave instability and forms
  a large crescent-shaped vortex. When gas self-gravity is discarded,
  but the indirect force from the displacement of the star by the
  vortex is included, we confirm that dust grains with stopping times
  of order the orbital time, which should be typically a few
  centimetres in size, are trapped ahead of the vortex in the
  azimuthal direction, while the smallest and largest grains
  concentrate towards the vortex centre. We obtain maximum shift
  angles of about 25 degrees.  Gas self-gravity accentuates the
  concentration differences between small and large grains. At low to
  moderate disc masses, the larger the grains, the farther they are
  trapped ahead of the vortex. Shift angles up to 90 degrees are
  reached for 10 cm-sized grains, and we show that such large offsets
  can produce a double-peaked continuum emission observable at mm/cm
  wavelengths. This behaviour comes about because the large grains
  undergo horseshoe U-turns relative to the vortex due to the vortex's
  gravity. At large disc masses, since the vortex's pattern frequency
  becomes increasingly slower than Keplerian, small grains concentrate
  slightly beyond the vortex and large grains form generally
  non-axisymmetric ring-like structures around the vortex's radial
  location. Gas self-gravity therefore imparts distinct trapping
  locations for small and large dust grains which may be probed by
  current and future observations, and which suggest that the formation
  of planetesimals in vortices might be more difficult than previously
  thought.
\end{abstract}


\section{Introduction}
\label{sec:intro}
Transition discs are protoplanetary discs around young (1-10 Myr)
stars with little or no emission in the near- and mid-IR but strong
emission at longer wavelengths \citep[see the review
by][]{Espaillat14}. Modeling of the Spectral Energy Distribution (SED)
of transition discs implies that their inner parts form dust cavities
largely devoid of small dust grains, followed by cold dust populating
an outer disc. Interferometric imaging shows that the cavities also
have a deficit in (sub-)mm continuum emission and are mostly devoid of
large dust grains. Cavities typically span a few tens of Astronomical
Units (AU) like in LkCa 15 \citep[$\sim$50 AU;][]{Andrews11} and AB
Aur \citep[$\gtrsim$70 AU;][]{Pietu05}, with minimum and maximum sizes
so far in TW Hya \citep[$\sim$4 AU;][]{Hughes07} and HD 142527
\citep[$\sim$140 AU;][]{Casassus13,Avenhaus14}. A subset of transition
discs shows significant near-IR excesses. These so-called
pre-transition discs feature an optically thick inner disc of
typically $\sim$1 AU between the star and the cavity, and some of them
have accretion rates exceeding a few $\times 10^{-8} M_{\odot}\,{\rm
  yr}^{-1}$ \citep[][and references therein]{Espaillat14}. There is
growing observational evidence that gas is present in the dust
cavities of (pre-)transition discs, though at a reduced level compared
to the outer disc \citep{Carmona14,Bruderer14}.  The decrease in gas
surface density inside the cavity varies from 1 to 4 orders of
magnitude \citep{vanderMarel2015}.

Several origins for the diversity of cavity sizes and accretion rates
have been explored. One is grain growth and the consequent decrease in
the dust emissivity: cavities would be only apparent. In the inner
disc, however, grain growth is shown to be limited by destructive
collisions to mm/cm particles, which should be detected in the
sub-mm. It therefore appears difficult for fragmentation-limited grain
growth to simultaneously account for the near-IR dip in the SED of
transition discs and their deficit in sub-mm continuum emission
\citep{Birnstiel2012}. Alternatively, large dust and gas cavities
could be cleared via photoevaporative winds. A broad range of
accretion rates is expected depending on the level of high-energy
stellar radiation (extreme UV and X-Ray), consistent with the
diversity of accretion rates observed in transition discs \citep[see
the review by][]{AlexanderPP6}. However, current models of
photo-evaporation cannot account for discs with both large cavities
($\gtrsim 10$ AU) and substantial accretion rates \citep[$\gtrsim
10^{-9} M_{\odot}\,{\rm yr}^{-1}$;][]{Owen11}, nor discs with moderate
drops in the gas surface density within the cavity.

A third mechanism for sculpting the cavities of transition discs is
dynamical clearing by one or several companions.  These can be stellar
companions. For example, \cite{Casassus15b} describe how a highly
mutually inclined stellar companion to HD 142527 can generate the fast
radial flows inside the cavity of its transition disc
\citep{Rosenfeld14}. These fast radial flows could explain how disc
holes could be simultaneously transparent and still contain material
that accretes at rates $\gtrsim 10^{-9} M_{\odot}\,{\rm yr}^{-1}$
\citep{Rosenfeld14}. Companions can also be planets. The salient point
is that the size of the gap opened by a massive planet differs in the
gas and in the dust. Gaps in the gas are narrow. For example, for a
Jupiter-mass planet orbiting a Solar-mass star, the half-width of the
gap does not exceed $\sim$40$\%$ of the star-planet separation, even
in discs with very low turbulent activity \citep{crida06}. Gaps in the
dust depend on the size of the dust particles or, more generally, on
the coupling between dust and gas (quantified by the Stokes number,
see Section~\ref{sec:model_pc}).  Dust grains up to few tens of
microns are tied to the gas, so the width of their gap is basically
the same as the gas \citep{pm04,Fouchet07,Zhu12}.  Dust in the mm/cm
range, however, is not tied to the gas and drifts generally inward as
a result of gas friction. Radial drift can, however, stall at
locations known as dust traps where the gas pressure is maximum
\citep[e.g.,][]{Pinilla12,JohansenPP6}. Simulations of disc-planet
interactions show that the outer edge of a planet gap is generally a
robust pressure maximum while the inner edge is not: large dust grains
initially beyond the planet's orbit can be trapped efficiently at the
gap's outer edge while the large dust grains initially inside the
planet's orbit migrate inward \citep[e.g.,][]{Zhu12,Zhu14}. This
implies that a single massive planet could simultaneously carve a
large cavity in mm-grains and a narrow gap in $\mu$m-grains, with
therefore no or limited impact on the accretion rate. Large cavities
in $\mu$m-grains and strong reductions in the accretion rate are not
natural outcomes of disc-planet interactions, though they can be
obtained by invoking several massive planets in the disc
\citep{Zhu11}.

Recent high-resolution interferometric imaging (especially with ALMA)
has highlighted that the mm emission rings of transition discs are
often non-axisymmetric. Several of them feature large-scale lopsided
structures with distinct crescent or horseshoe shapes. These include
HD 142527 \citep{Casassus13,Fukagawa14}, Oph IRS 48
\citep{vanderMarel13}, AB Aur \citep{Tang12}, SR21
\citep{Andrews11all,Perez14}, LkH$\alpha$ 330 \citep{Isella13} and HD
135344B \citep{Brown09,Perez14}.  Intensity variations along the
lopsided rings are by factors of $\sim$2-3 for SR21, LkH$\alpha$, AB
Aur and HD 135344B, $\sim$30 for HD 142527 and $>$100 for Oph IRS
48. In contrast, the gas emission is generally more symmetric along
the lopsided rings \citep{vanDishoeck2015}. These findings are so far
best explained by the presence of a long-lived vortex in the gas that
would efficiently trap mm-grains.

Probably the most promising vortex-forming instability in the context
of lopsided transition discs is the Rossby-Wave Instability
\citep[RWI,][]{lovelace99}. It is essentially the form of the
Kelvin-Helmholtz instability in a differentially rotating disc, and a
thin-disc version of the Papaloizou-Pringle instability
\citep{PP84}. It is a linear instability associated with a radial
minimum in the gas potential vorticity which, in practice, can be
triggered where the pressure has a radial maximum. This may occur (i)
near the disc's centrifugal radius due to mass infall from the
protostellar cloud \citep{Bae15}, (ii) at the outer edge of a
gap-opening planet \citep[e.g.,][]{Lyra09,LinMK12}, (iii) at the edges
between magnetically active and inactive (dead) regions
\citep[e.g.,][]{VT06,Regaly12,Faure14,Lyra15,Flock15}, or (iv) at the
edge of a disc undergoing photo-evaporation, though to our knowledge
this latter scenario has not been examined yet. The planet gap
scenario has received the most attention, but detailed modeling is
still needed to reproduce observed distributions of gas and dust, like
in the Oph IRS 48 disc \citep{ZhuStone14}. The dead/active regions
scenario needs more exploration, in particular to determine the
efficiency of non-ideal magneto-hydrodynamic effects at suppressing
the magneto-rotational instability in the outer regions of transition
discs \citep{Lesur14,Bai15,Gressel15}.

Another route to producing lopsided dust rings in transition discs has
been recently examined by \cite{MC15}. The authors proposed a model
where a massive gas disc exhibits a mode with azimuthal wavenumber
$m=1$ and pattern frequency comparable to the disc's Keplerian
frequency. Mutual gravitational interactions will cause the star and
disc to rotate about their centre-of-mass, which, if the disc is
massive enough, can sustain an $m=1$ perturbation where the gas
describes horseshoe-shaped streamlines. Interestingly, \cite{MC15}
found that dust grains with Stokes numbers around unity, that is with
aerodynamic stopping times around the orbital period, are not trapped
at the horseshoe centre: they are shifted azimuthally by up to 45
degrees ahead of the horseshoe centre. These should be mm/cm-grains
depending on the gas surface density at the vortex's radial location,
thus a significant shift of the grains location relative to the
horseshoe centre should be detectable via high-resolution
interferometry in the (sub-)mm.

Nonetheless, \cite{MC15} left un-addressed several important issues.
They envisioned an $m=1$ mode based on the mutual interaction between
the lopsided disc and the star, but the trigger was not specified. We
propose that it could be the RWI. Also, gas self-gravity was discarded
although its acceleration can be as important as, if not larger than,
the indirect acceleration imparted by the star.  This we show in an
accompanying paper (Zhu \& Baruteau, submitted, hereafter Paper I),
where we present results of hydrodynamic simulations with and without
gas self-gravity, with lopsided gas distributions mediated by the
RWI. While Paper I deals with gas only, the present paper presents
results of simulations with both gas and dust to investigate the
concentration of dust grains when varying the mass of the gas disc,
with a focus on large disc masses for which gas self-gravity should
matter most. The model set-up is described in
Section~\ref{sec:setup}. Results of simulations are presented in
Section~\ref{sec:results}, first without gas self-gravity in
Section~\ref{sec:wo}, then with self-gravity in
Section~\ref{sec:w}. These results are used to generate synthetic
images of dust continuum observations, which are presented in
Section~\ref{sec:synthetic}. Concluding remarks follow in
Section~\ref{sec:conclusion}.

\section{Model set-up}
\label{sec:setup}
We carried out 2D hydrodynamical simulations using a modified version
of the grid-based code FARGO \citep{fargo1} called
\href{http://fargo.in2p3.fr/-FARGO-ADSG-}{FARGO-ADSG} \citep{BM08a,
  BM08b}, which includes optional energy equation and gas
self-gravity. The gas model is detailed in Paper I and briefly
recalled in Section~\ref{sec:model_gas}. For this work, a particle
integrator has been implemented in FARGO-ADSG for the dust. The dust
model is described in Section~\ref{sec:model_pc}. Conversion of our
results of simulations from code units to physical units follows in
Section~\ref{sec:codeunits}.

\subsection{Gas}
\label{sec:model_gas}
The gas model and the initial conditions of the simulations are chosen
to generate a long-lived elongated vortex through the
RWI. Specifically, the initial surface density of the gas has a
(modified) Gaussian radial profile of the form
\begin{multline}
  \Sigma(r,\varphi) = \Sigma_0 \left[10^{-2}
    + \exp\left\{-\frac{(r-r_0)^2}{2\sigma^2}\right\}\right]\nonumber\\
  \times\left[
    1+10^{-3}\times\cos\varphi\times\sin\left(\pi\frac{r-r_{\rm
          in}}{r_{\rm out}-r_{\rm in}}\right) \right],
          \label{eq:Sigma0}
\end{multline}
where $\Sigma_0$ is a free parameter that varies the disc mass, $r_0$
is a reference radius that defines the code's unit of length, $\sigma
= 2H(r_0)$ with $H=c_{\rm s} / \Omega_{\rm K}$ the disc's pressure
scale height, $c_{\rm s}$ the sound speed and $\Omega_{\rm K}$ the
Keplerian angular frequency. We take $H$ equal to $0.1r$ and constant
in time (the disc is assumed to be isothermal). The second bracket in
the above equation corresponds to a small perturbation of azimuthal
wavenumber $m=1$ applied smoothly between the grid's inner edge
($r_{\rm in} = 0.2r_0$) and outer edge ($r_{\rm out} = 2r_0$).

The gas disc that we simulate is therefore a narrow ring of gas rather
than an extended outer disc beyond a cavity, as is observed in
transition discs. This is mainly for numerical convenience, as it
allows us to reduce the radial extent of the computational grid that
simulates the disc, and to focus on the dynamics of the gas and dust
near the vortex. In particular, the low densities at the disc edges
minimize edge effects in the direct and indirect contributions to the
disc's gravitational potential \citep{HPS92,Adams89}.

The simulations presented in this paper are summarised in
Table~\ref{tab:models}. In the simulations with gas self-gravity, the
disc gravitationally accelerates itself and the star, while in the
simulations without self-gravity, the disc only accelerates the
star. In all the simulations, the star remains at the origin of the
reference frame, and the indirect terms that account for the
acceleration of the star by the disc are therefore included in the
equations of motion, regardless of whether self-gravity is included or
not. Our median value $\Sigma_0 = 5\times 10^{-3}$ corresponds to an
initial disc-to-star mass ratio of 1.7\%. The Toomre parameter, $Q
\approx c_{\rm s}\Omega / \pi G \Sigma$, with $\Omega$ the disc's
angular frequency and $G$ the gravitational constant, is initially
equal to 13, 5 and 2.5 at $r=r_0$ in models g2, g5 and g10,
respectively.
\begin{table}
  \centering
  \caption{\label{tab:models}Simulation models}
  \smallskip
  \begin{minipage}{\hsize}
    \centering
    \begin{tabular}{l c r}
      Name & Gas self-gravity & $\Sigma_0$ [code units]\\
      \hline
      \hline
      g0p2 & Yes & $0.2\times10^{-3}$\\
      g0p5 & Yes & $0.5\times10^{-3}$\\
      g2n & No & $2\times10^{-3}$\\
      g2 & Yes & $2\times10^{-3}$\\
      g5n & No & $5\times10^{-3}$\\
      g5 & Yes & $5\times10^{-3}$\\
      g10n & No & $10\times 10^{-3}$\\
      g10 & Yes & $10\times10^{-3}$\\
      \hline
    \end{tabular}\par
  \end{minipage}
\end{table}

A small turbulent viscosity is included in the
simulations. Preliminary runs have shown that a constant alpha viscous
parameter $\alpha = 10^{-6}$ is a good compromise between a strong,
long-lived vortex \citep[which requires low viscosities, particularly
with gas self-gravity as it inhibits the RWI;][]{GN88} and convergence
with increasing grid resolution (which would favour large
viscosities).

The hydrodynamical equations are solved on a polar grid with $N_{\rm
  r} = 300$ cells logarithmically spaced between $r_{\rm in}$ and
$r_{\rm out}$, and $N_{\rm s} = 600$ sectors evenly spaced between
$\varphi=0$ and $2\pi$. This grid resolution is sufficient for our
purposes and helps maintain a tractable computational cost when
including many dust particles.  Although the logarithmic radial
spacing is only required for the Fast Fourier Transform algorithm in
the calculation of the gas self-gravity \citep{BM08b}, it is also used
without self-gravity to keep the same grid resolution in all
simulations. When self-gravity is included, a softening length
$\varepsilon = 0.3H(r)$ is used in the self-gravitating potential to
mimic the effect of a finite vertical thickness. Standard outflow
(zero-gradient) boundary conditions are used in the simulations.

\subsection{Dust}
\label{sec:model_pc}
Dust is modelled by Lagrangian particles that feel the star's gravity
(direct and indirect acceleration terms), gas drag, but that do not
interact together (no dust self-gravity, no collisions nor growth). An
important point is that when gas self-gravity is included, the
particles also feel the self-gravitating acceleration of the gas. It
implies that differences in the gas and dust velocities are only
imparted by the pressure force felt by the gas.  We discard possibly
important effects of dust feedback onto the gas
\citep[e.g.,][]{CZS15}.

The dust particles that we simulate are much smaller than the
molecular mean-free path in our disc models (see
Section~\ref{sec:codeunits}).  The drag force exerted by the gas
therefore corresponds to the so-called Epstein regime, and the
particles Stokes number (St), which is a dimensionless measure of the
particles stopping time (${\tau_{\rm s}} = {\rm St}\,\Omega^{-1}$), is
given by
\begin{equation}
{\rm St} = \frac{\pi}{2} \frac{s \rho_{\rm pc}}{f_{\rm D}\Sigma_{\rm gas}} 
\approx \frac{0.015}{f_{\rm D}} \times 
\left(\frac{s}{1{\rm mm}}\right)
\left(\frac{\rho_{\rm pc}}{1{\rm g\;cm}^{-3}}\right)
\left(\frac{10{\rm g\;cm}^{-2}}{\Sigma_{\rm gas}}\right)
\label{eq:St}
\end{equation}
with $s$ the physical radius of the particles, $\rho_{\rm pc}$ their
internal mass density (we choose $\rho_{\rm pc} = 1$ g cm$^{-3}$),
$\Sigma_{\rm gas}$ the gas surface density at the particles location,
and $f_{\rm D} = \sqrt{1 + 9\pi{\cal M}^2/128}$ is a dimensionless
coefficient that features the Mach number of the dust relative to the
gas, ${\cal M} = |{\bf v} - {\bf v_{\rm gas}}|/c_{\rm s}$
\citep[][${\bf v}$ and ${\bf v_{\rm gas}}$ denote the velocity vector
of the dust and gas, respectively]{SJP07}. In practice $f_{\rm D}
\approx 1$ in our simulations. We consider several particle sizes that
are fixed in the simulations. Thus, the particles Stokes number does
vary in the simulations according to the gas surface density at the
particles location. Stokes numbers typically range from
$\sim$$10^{-2}$ to $\sim$$40$, which allows us to probe different
trapping efficiencies via gas drag.

We use 10,000 particles to simulate dust grains of a given size. Their
initial distribution is uniform between $r=0.8r_0$ and $1.2r_0$,
unless otherwise stated. Their initial velocities are Keplerian,
except in simulations with gas self-gravity, for which the initial
azimuthal velocity of the particles includes the gas self-gravitating
radial acceleration \citep[much like in disc-planet simulations with
self-gravity, see][]{BM08b}.  The number of dust particles may seem
rather small but, as we will see in Section~\ref{sec:results}, dust
particles quickly concentrate at well-defined locations in the disc,
either at a single point inside or near the vortex, or along a ring
around the vortex's radial location. We find 10,000 particles to be
enough to describe the dust's azimuthal density distribution when it
forms ring-like structures around the vortex's radial location.

The equations of motion for the dust particles are solved with a
leapfrog integrator in polar coordinates. The integrator basically
corresponds to the 2D leapfrog integrator in cylindrical coordinates
implemented by \cite{Zhu14} in the code ATHENA. We carried out the
various tests described in the Appendix of \cite{Zhu14} to validate
our particle integrator. The particle integration scheme is
parallelized with MPI using FARGO's domain decomposition into rings.
The timestep used in the integration of the dust particles is
identical to that used in the gas equations, it is therefore
constrained by the Courant-Frederichs-Levy (CFL) condition specific to
FARGO. Test simulations using a fixed timestep, chosen to be shorter
than the minimum stopping time of the particles, showed negligible
changes to our results. In practice, the effective time step in our
simulations is shorter than the particles stopping time.

\subsection{Conversion from code to physical units}
\label{sec:codeunits}
As mentioned in Section~\ref{sec:model_gas}, the radial location $r_0$
where the initial gas surface density is maximum defines the code's
unit of length. The unit of mass is the mass of the central star, and
the unit of time is the Keplerian orbital period at $r=r_0$ divided by
$2\pi$. Whenever time is expressed in orbits, it refers to the orbital
period at $r=r_0$, which we denote by $T_{\rm orb}$.

To keep our results as general as possible, we do not specify physical
units of length and mass in the simulations, since the gas and dust
dynamics are set by dimensionless numbers like the Toomre parameter,
the disc-to-star mass ratio or the particles Stokes number. All our
results are thus shown in code units, but it is very easy to convert
them into physical units. Assuming the mass unit is the Sun's mass and
the length unit is 65 AU, our median value for $\Sigma_0$ of $5\times
10^{-3}$ corresponds to $\approx$10 g cm$^{-2}$ and $T_{\rm orb}\sim$
500 yrs.  For comparison, the gas surface density at the edge of the
$\sim$$70$ AU cavity in the J1604-2130 disc is estimated as $\sim$10 g
cm$^{-2}$, and that at the edge of the $\sim$$45$ AU cavity in the
LkCa 15 disc is $\sim$60 g cm$^{-2}$ \citep{vanderMarel2015}.
Conversion from Stokes number to dust size is via Eq.~(\ref{eq:St}),
using $f_{\rm D} = 1$. If, for example, dust grains have an internal
density of 1 g cm$^{-3}$, and that the gas surface density at their
location is 10 g cm$^{-2}$ (5$\times 10^{-3}$ in code units), then St
$=1$ particles correspond to $\approx$ 6 cm grains. Using the same
units, the largest particles that we have simulated are a few metres
in size, which remains much smaller than the minimum molecular mean
free path in our disc models ($\sim$ 300 metres in models g10 and
g10n) and justifies that the dust grains that we simulate are in the
Epstein regime.

\section{Results}
\label{sec:results}
This section describes our results of hydrodynamical simulations,
first for models without gas self-gravity in Section~\ref{sec:wo},
then for models with self-gravity in Section~\ref{sec:w}. Contrary to
Paper I where its effect is examined, the indirect term in the star's
acceleration on the gas and dust, due to the displacement of the star
by the vortex, is included in all the simulations presented in this
paper.

\subsection{Models without gas self-gravity}
\label{sec:wo}
\begin{figure}
\centering
\includegraphics[width=0.85\hsize]{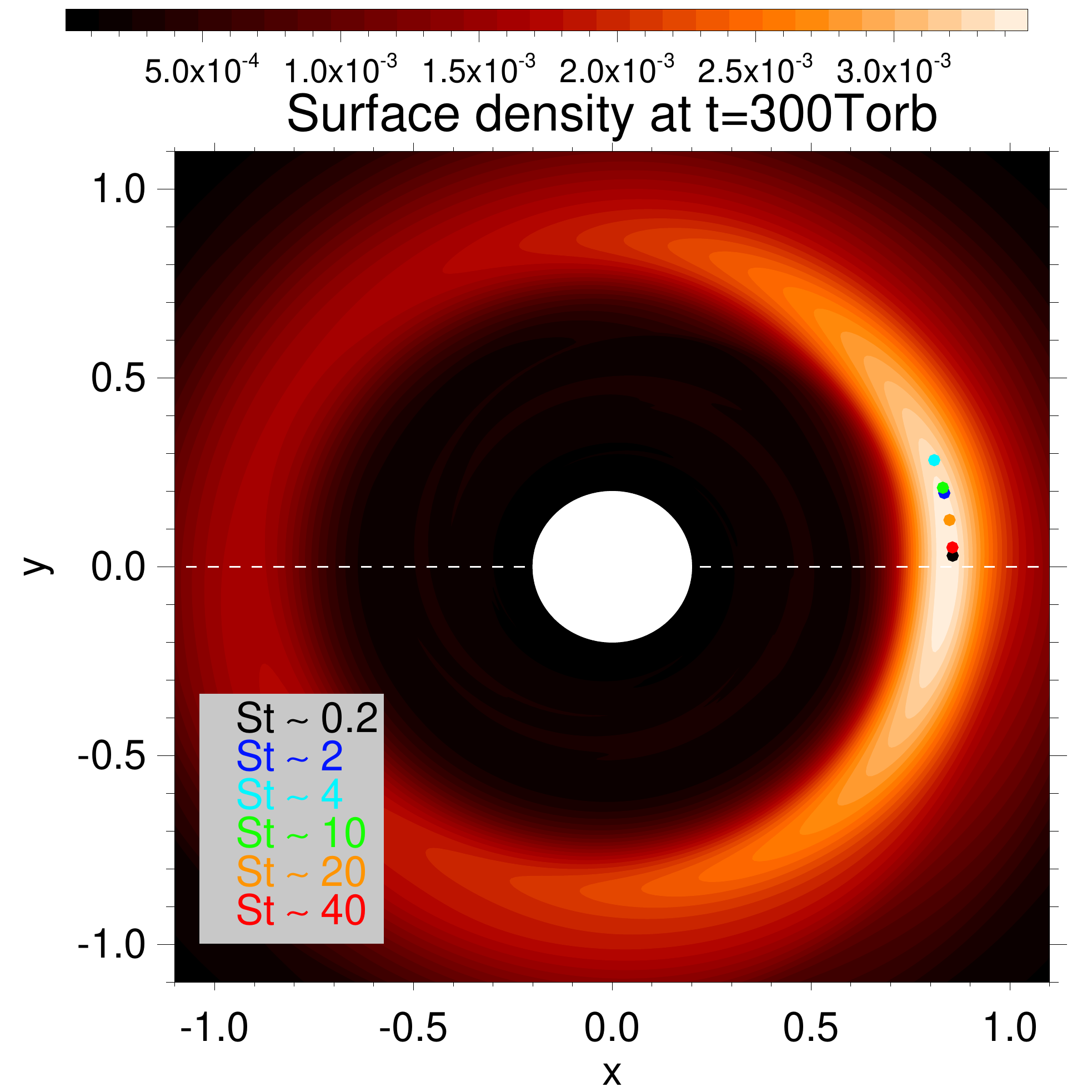}
\includegraphics[width=0.85\hsize]{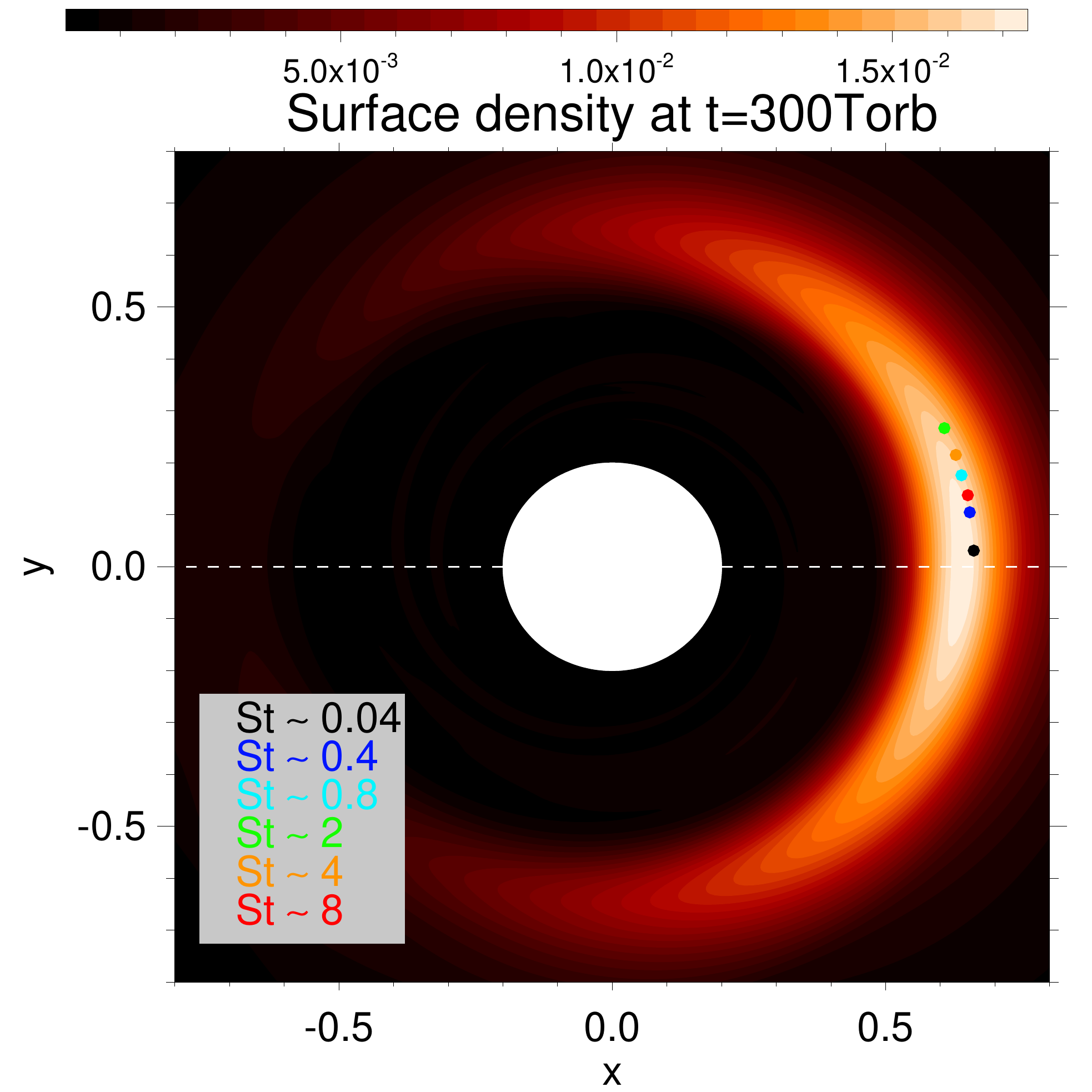}
\includegraphics[width=0.85\hsize]{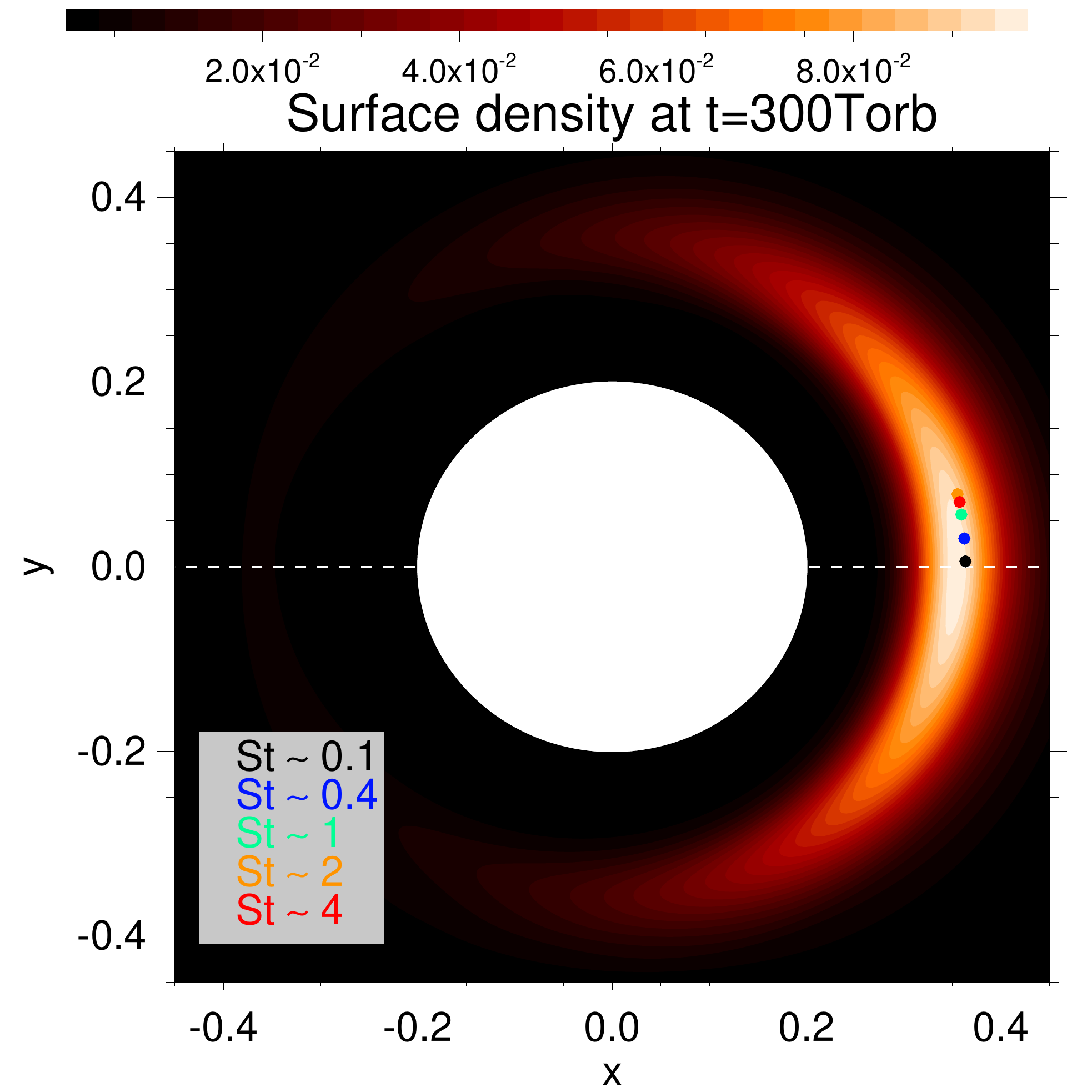}
\caption{\label{fig:g1n}Results of simulations without gas
  self-gravity for models g2n, g5n and g10n (from top to bottom in the
  panels). Contours of the gas surface density are shown at 300
  orbits. In all panels, the grey dashed line passes by the star and
  the location in the disc where the gas surface density in the vortex
  is maximum. The circles filled with different colours show the
  location of dust particles of various sizes. Their Stokes number at
  the time of the simulation is indicated in the lower-left corner of
  the panels.}
\end{figure}

In our models without self-gravity, the RWI quickly sets in and forms
a single, elongated vortex in less than 30 orbits that lives over the
duration of our simulations (500 orbits). Contours of the gas surface
density are displayed at 300 orbits in Figure~\ref{fig:g1n} for models
g2n, g5n and g10n from top to bottom in the panels. Cartesian
coordinates $\{x,y\}$ are used with the star at the origin. To ease
the comparison between the models, the maximum in the gas surface
density is set at $y=0$ in the panels. The vortex's radial location,
where the azimuthally-averaged surface density of the gas peaks, can
thus be read along the x-axis (it also corresponds to the radial
location where the gas pressure is maximum since the disc temperature
is uniform).

The first notable result, also shown in Paper I, is that the vortex
migrates inward fairly rapidly, at a rate that increases nearly
proportionally to the disc's initial surface density: in 300 orbits,
the vortex's radial location has decreased by about 15\%, 35\% and
65\% in models g2n, g5n and g10n, respectively. The vortex's migration
results from an asymmetric wave emission by the vortex \citep{PLP10}. 
Vortex migration is therefore due to pressure
  torques, not to gravitational torques, at least in the absence of
  gas self-gravity. The waves emitted by the vortex cannot be seen in
the panels of Figure~\ref{fig:g1n} because their density contrast
relative to the background disc is smaller than the vortex's, and
because the radial profile of the gas density drops rapidly away from
the vortex's radial location. The wave that propagates inward from the
vortex's radial location carries a negative flux of angular momentum
and thus exerts a positive torque on the vortex. Conversely, the wave
that propagates outward from the vortex's radial location carries a
positive flux of angular momentum and exerts a negative torque on the
vortex.  In the absence of gas self-gravity, geometrical effects and
the radial gradient of surface density favour the outer wave, and the
net exchange of angular momentum between the vortex and the disc leads
to inward migration of the vortex \citep{PLP10}. We also notice that
the vortex strengthens during its inward migration with both the
surface density at the vortex centre and the azimuthal contrast in
surface density along the vortex increasing over time. At 300 orbits,
the surface density contrasts along the vortex are by factors of
roughly 2.5, 12 and 25 in models g2n, g5n and g10n, respectively.

The location of the particles is marked by filled circles in
Figure~\ref{fig:g1n}, with different colours used for different
particle sizes. Recall that in the simulations the size of the
particles is fixed, but their Stokes number evolves with time
according to the gas surface density at the particles location (see
Eq.~\ref{eq:St}).  The Stokes numbers shown in the lower-left corner
in the panels are those at the particles location at 300 orbits. Since
initial Stokes numbers range from about 0.01 to 30, particles are
quickly trapped inside the vortex as a result of gas drag. They
eventually concentrate at specific locations inside the vortex where a
force balance is reached between gas drag, the centrifugal force and
the star's gravity (direct and indirect terms; force balance is meant
to be in the non-rotating frame centred onto the star). We find
qualitatively the same trend as in \cite{MC15}: while particles with
St $\ll 1$ and $\gg 1$ are trapped towards the vortex centre,
particles with St $\sim 1$ are shifted ahead of the vortex centre in
the azimuthal direction. The orientations and relative amplitudes of
the aforementioned forces implies that the force balance for St $\sim
1$ particles is reached ahead of the vortex and not behind it
\citep[see the lower-right panel in figure 3 of][]{MC15}. We quantify
the azimuthal offset of the particles with respect to the vortex by
measuring the azimuth of the particles relative to the location in the
vortex where the gas surface density peaks. We call it the shift angle
of a particle. At 300 orbits, the largest shift angles are about 20
degrees in model g2n for St $\sim 4$, 23 degrees in model g5n for St
$\sim 2$, and 13 degrees in model g10n for St $\sim 2$. Although quite
significant, our maximum shift angles are smaller than in \cite{MC15},
who found that St$=0.5$ particles could be shifted by as much as 45
degrees.  Comparison between our models, and with \cite{MC15}, shows
that shift angles do not depend solely on the particles Stokes number
but also on the velocity differences between gas and dust, which are
intimately related to the shape and strength of the vortex. We point
out that the smallest shift angles are obtained in model g10n, which
is the one that maximises the indirect term in the star acceleration
by having the most massive, asymmetric and closest vortex. We also
note a tiny shift of the particles beyond the vortex centre in the
radial direction.

\begin{figure}
\centering
\includegraphics[width=0.49\hsize]{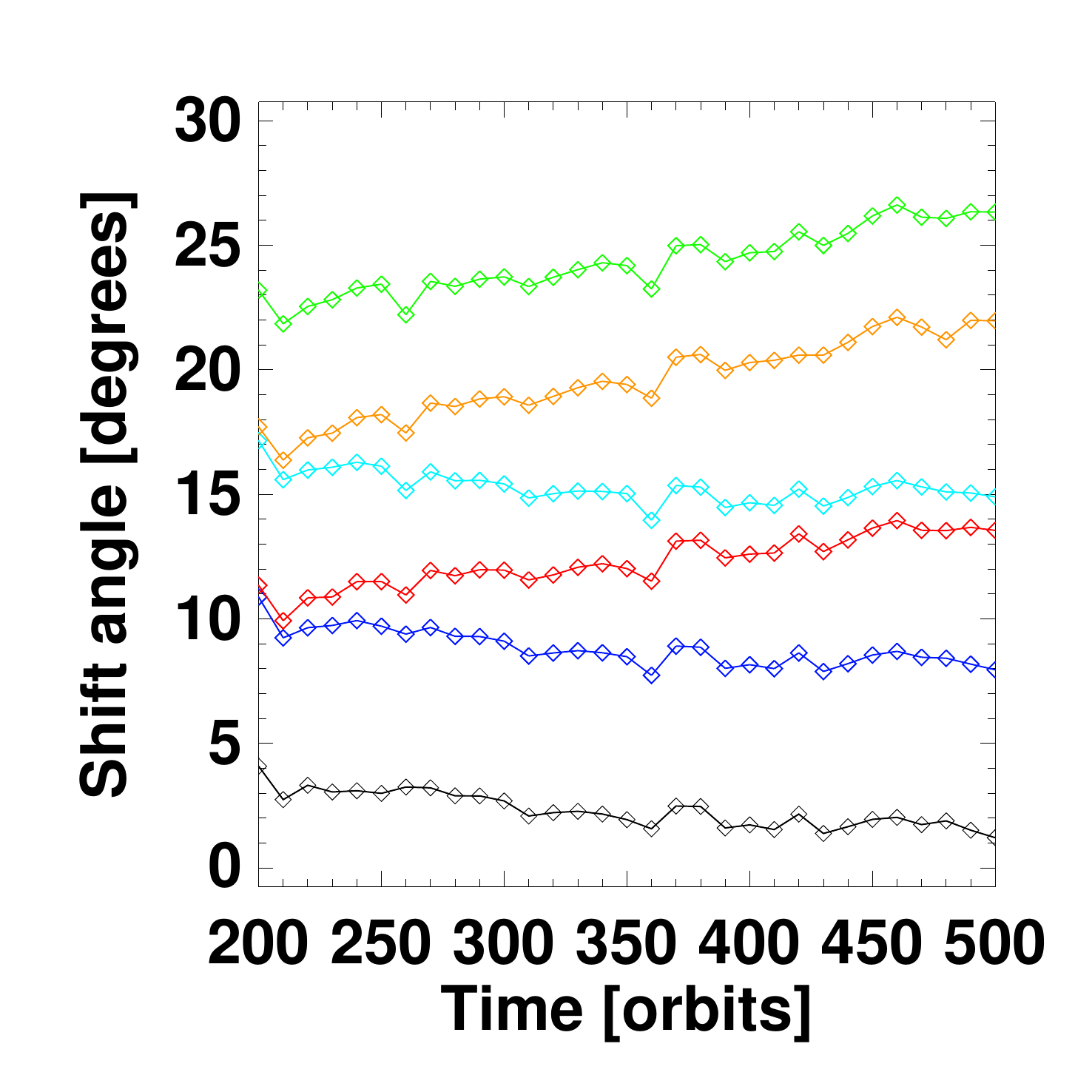}
\includegraphics[width=0.49\hsize]{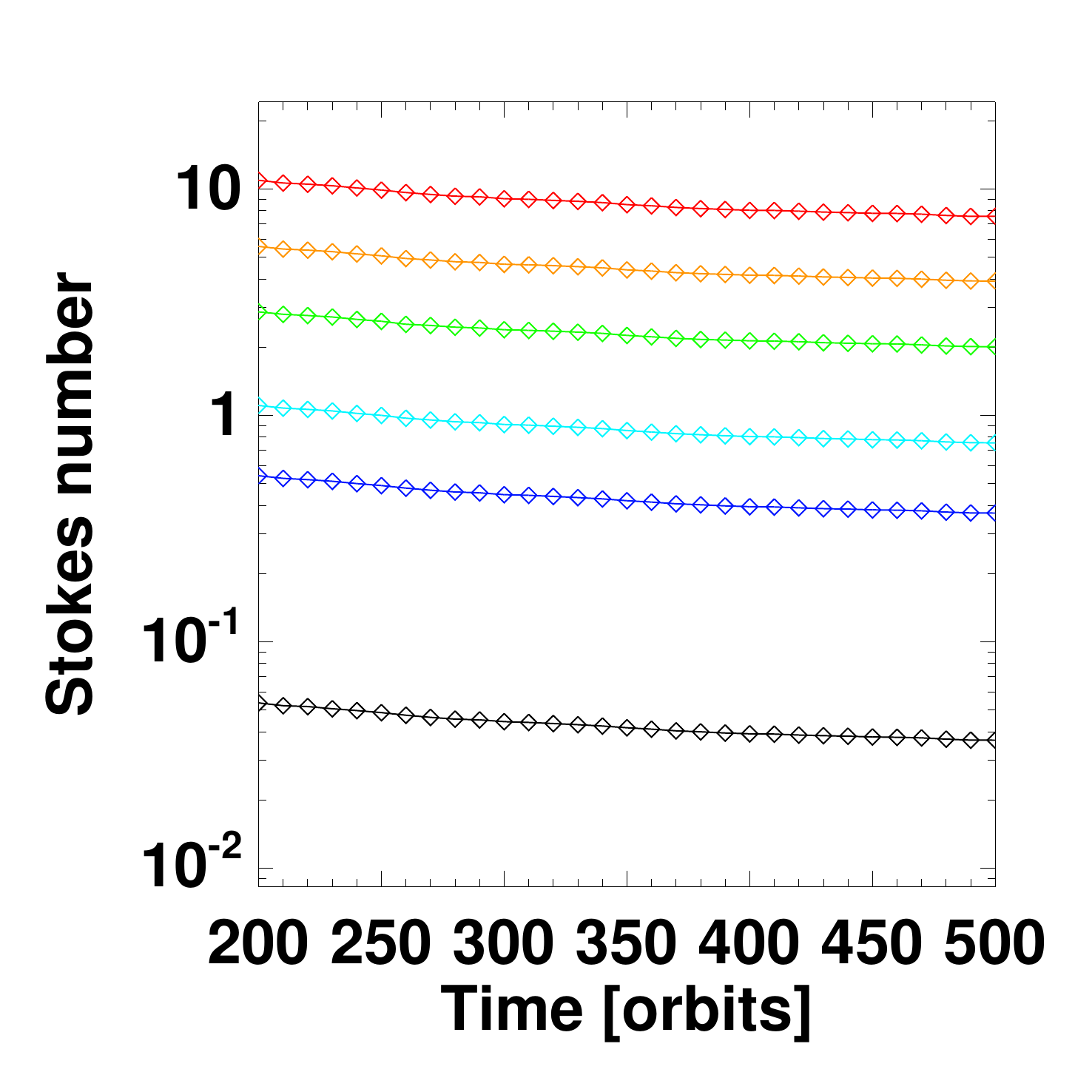}
\caption{\label{fig:g1n_shift_ts}Time evolution of the particles shift
  angle (left panel) and Stokes number (right panel) in model
  g5n. Shift angles are defined relative to the location in the vortex
  where the gas surface density is maximum. The colour of the symbols
  refers to the same particle size between both panels and is the same
  as in the middle panel of Figure~\ref{fig:g1n}.}
\end{figure}
\begin{figure*}
\centering
 \includegraphics[width=0.325\hsize]{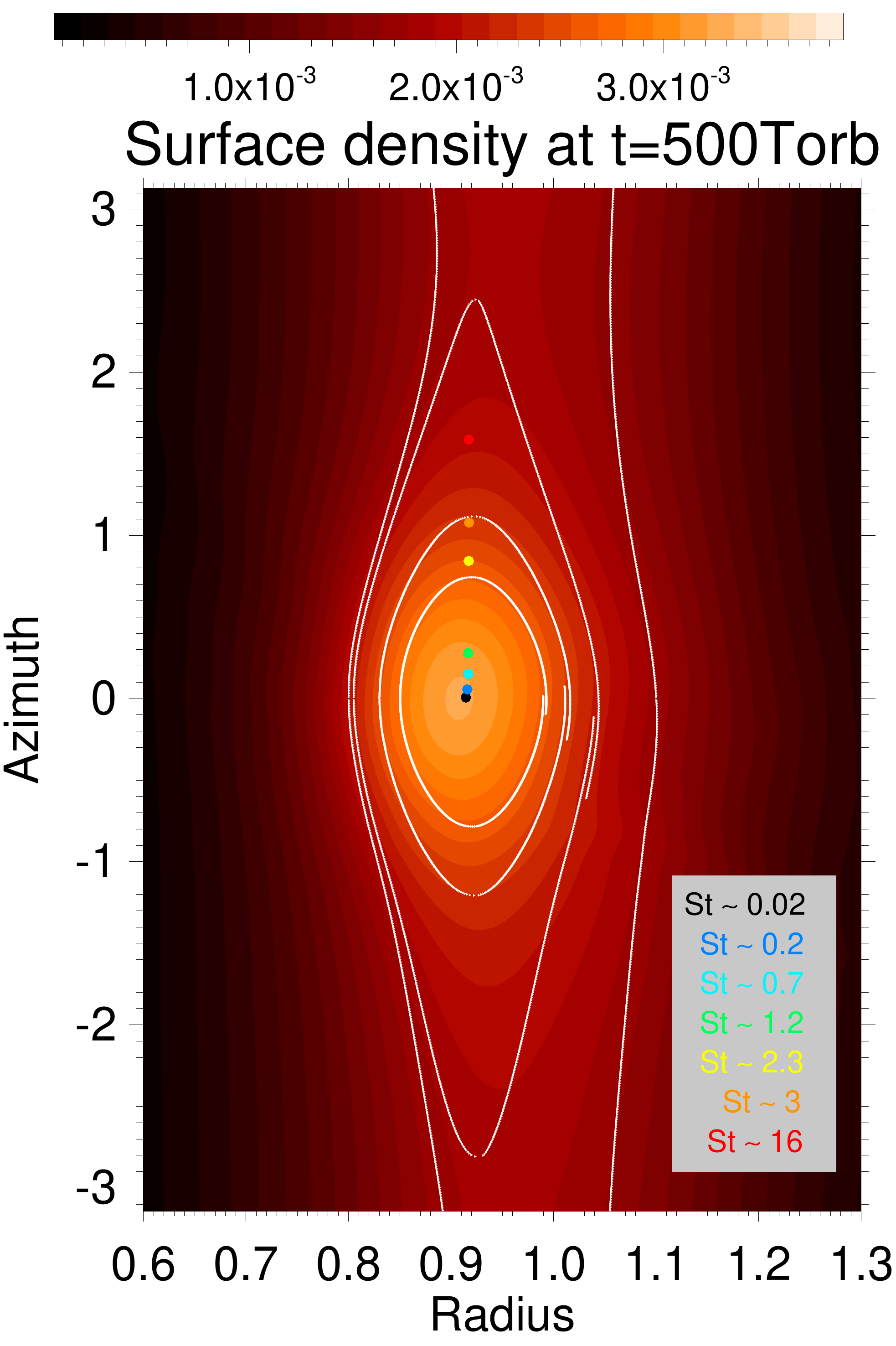}
 \includegraphics[width=0.325\hsize]{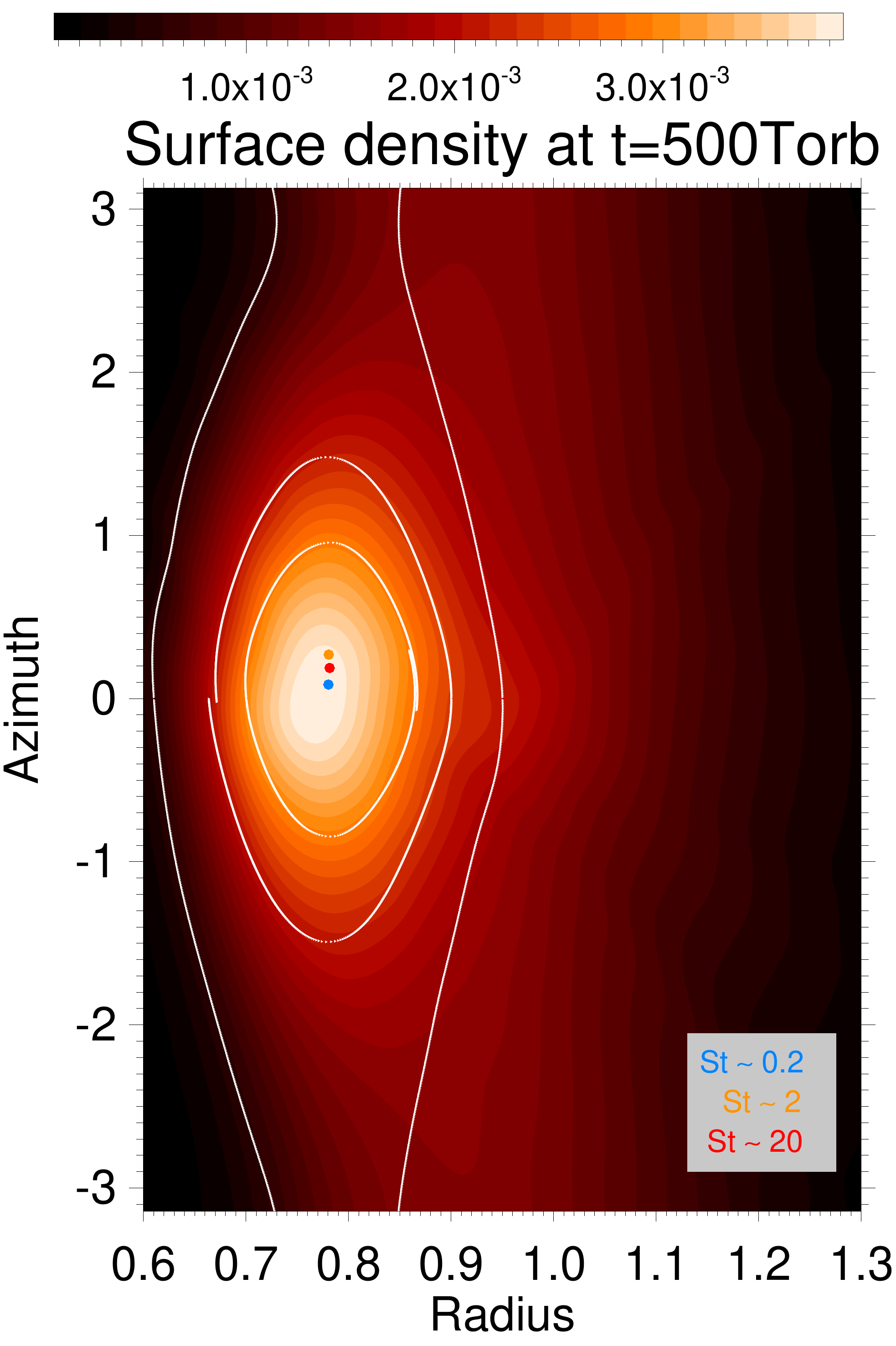}
 \includegraphics[width=0.325\hsize]{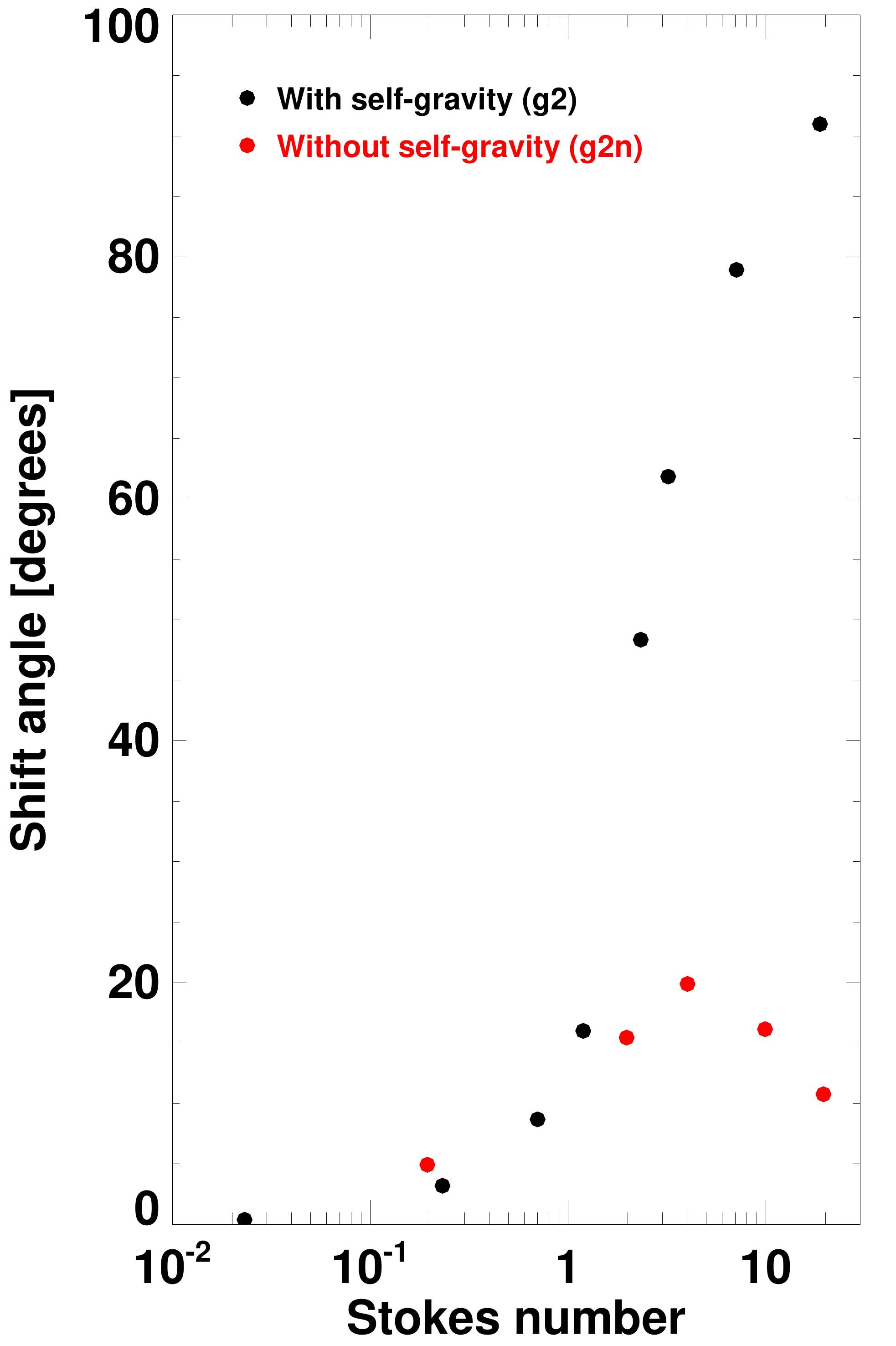}
 \caption{\label{fig:g13_rf2}Results of simulations with gas
   self-gravity (left panel, model g2) and without (middle panel,
   model g2n).  In both panels, contours of the gas surface density
   are shown in polar coordinates at 500 orbits. Streamlines in the
   frame rotating at the vortex's pattern frequency are overplotted by
   solid white curves. The filled circles with varying colours are the
   instantaneous location of dust particles of various sizes. Their
   Stokes number at 500 orbits is indicated in the lower-right corner
   of the panels. The right panel displays the particles shift angle
   at 500 orbits as a function of their instantaneous Stokes number
   for models g2 (black) and g2n (red).}
\end{figure*}
Since particle shift angles depend on the gas velocity and density
inside the vortex, they should vary as the vortex moves inward. This
is illustrated in Figure~\ref{fig:g1n_shift_ts}, which displays the
time evolution of the particles shift angle (left panel) and Stokes
number (right panel). We see that the Stokes numbers all decrease by
about 45\% between 200 and 500 orbits, which is due to the increase in
the gas surface density during vortex migration (the vortex migrates
from $r \approx 0.72$ to $r \approx 0.55$ between 200 and 500 orbits).
Variations in the particles shift angle are not as straightforward to
explain. We observe that shift angles tend to decrease with time for
particles with St $\lesssim1$, and to increase with time for St
$\gtrsim 2$. At the end of the simulation, St $\sim 2$ particles reach
a shift angle $\gtrsim 25$ degrees, which is the largest offset that
we have obtained over the duration of our simulations without
self-gravity. Tests of convergence with varying grid resolution are
provided in Section~\ref{sec:conv} of the Appendix.

As a brief summary of our results without self-gravity, we confirm the
qualitative trend found by \cite{MC15} that dust grains with Stokes
number near unity are shifted azimuthally ahead of the horseshoe
centre. The particles' shift is therefore not specific to their model,
where gas horseshoe streamlines result from a manually imposed
circular motion of the star about the system barycentre, and can be
obtained self-consistently in hydrodynamical simulations of discs
unstable against the RWI.

\subsection{Models with gas self-gravity}
\label{sec:w}
For models with self-gravity, the RWI takes longer to form a fully
developed vortex than in models without self-gravity. The amplitude of
the $m=1$ mode in the Fourier analysis of the gas surface density
peaks at about 40, 60 and 50 orbits for models g2, g5 and g10,
respectively (it is about twice as long as in the models without
self-gravity). Its stationary value is about twice as small with
self-gravity than without, illustrating that gas self-gravity hinders
the RWI \citep{GN88}.

For some of the models with self-gravity, the dust concentration is
sensitive to the initial distribution of the particles, in particular
whether dust is introduced before or after the vortex forms
(specifically, the ring-like structures formed by the dust grains in
model g5, which will be shown in Section~\ref{sec:g5}, tend to be
disrupted when particles are introduced before the vortex forms - they
rather take the form of arc rings). However, when dust particles are
introduced after the vortex reaches a quasi-steady state, their
concentration shows overall marginal dependence on the initial
distribution. For comparison purposes, all simulations with
self-gravity have been restarted at 250 orbits (long after the vortex
forms) by introducing the particles between $r=1.2$ and 1.4 with a
uniform distribution.

\begin{figure*}
\centering
\includegraphics[width=0.24\hsize]{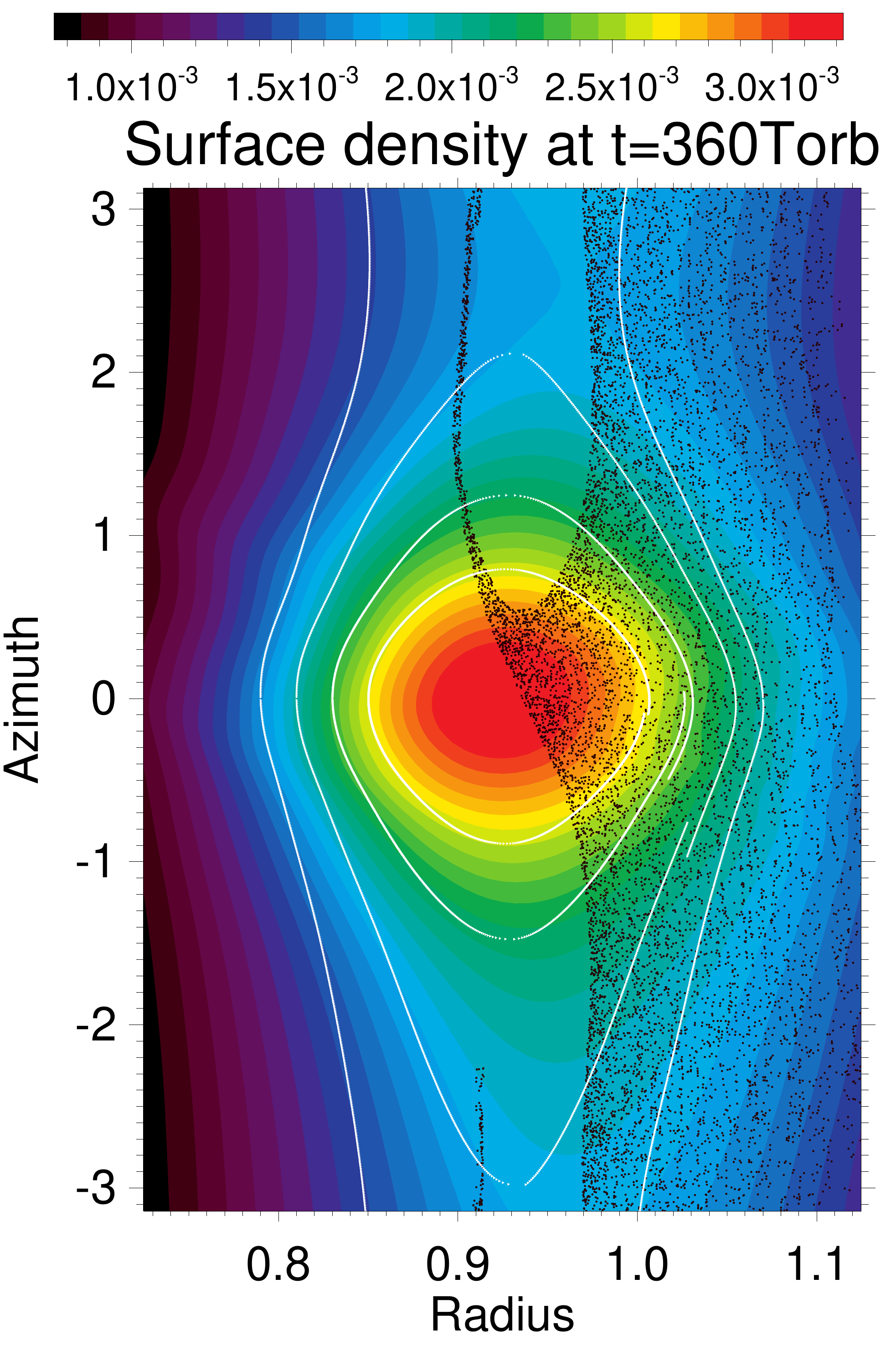}
\includegraphics[width=0.24\hsize]{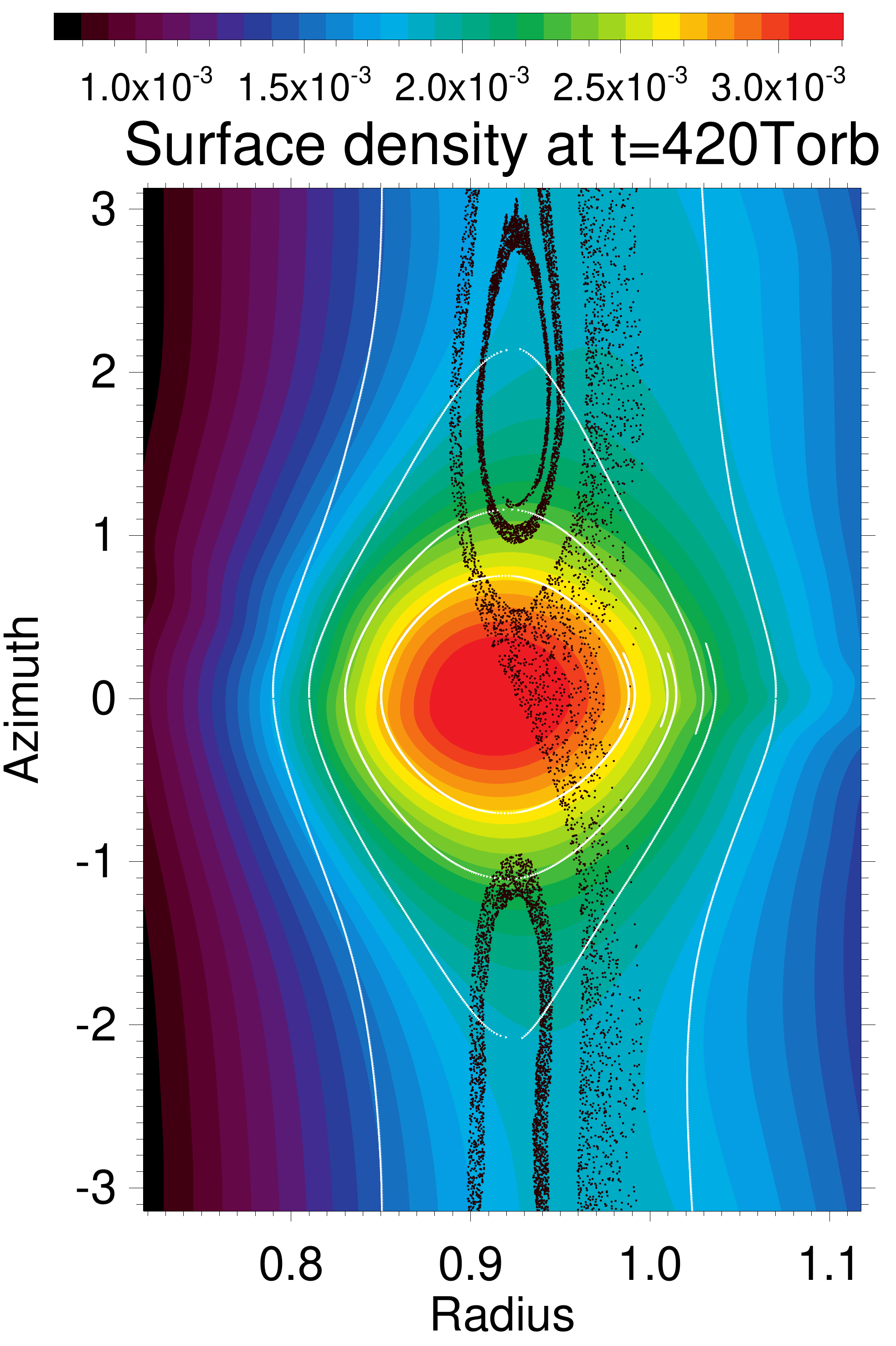}
\includegraphics[width=0.24\hsize]{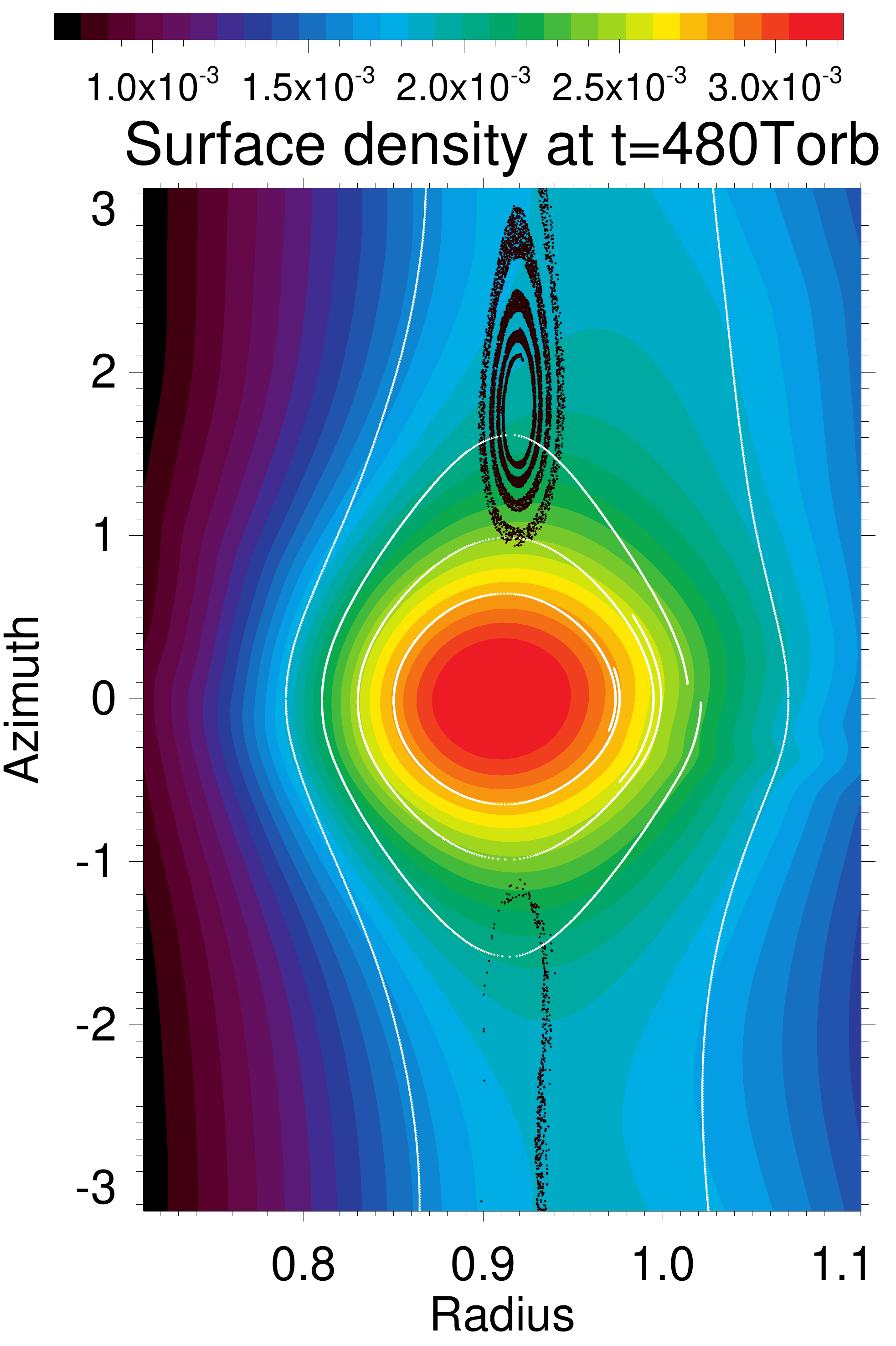}
\includegraphics[width=0.24\hsize]{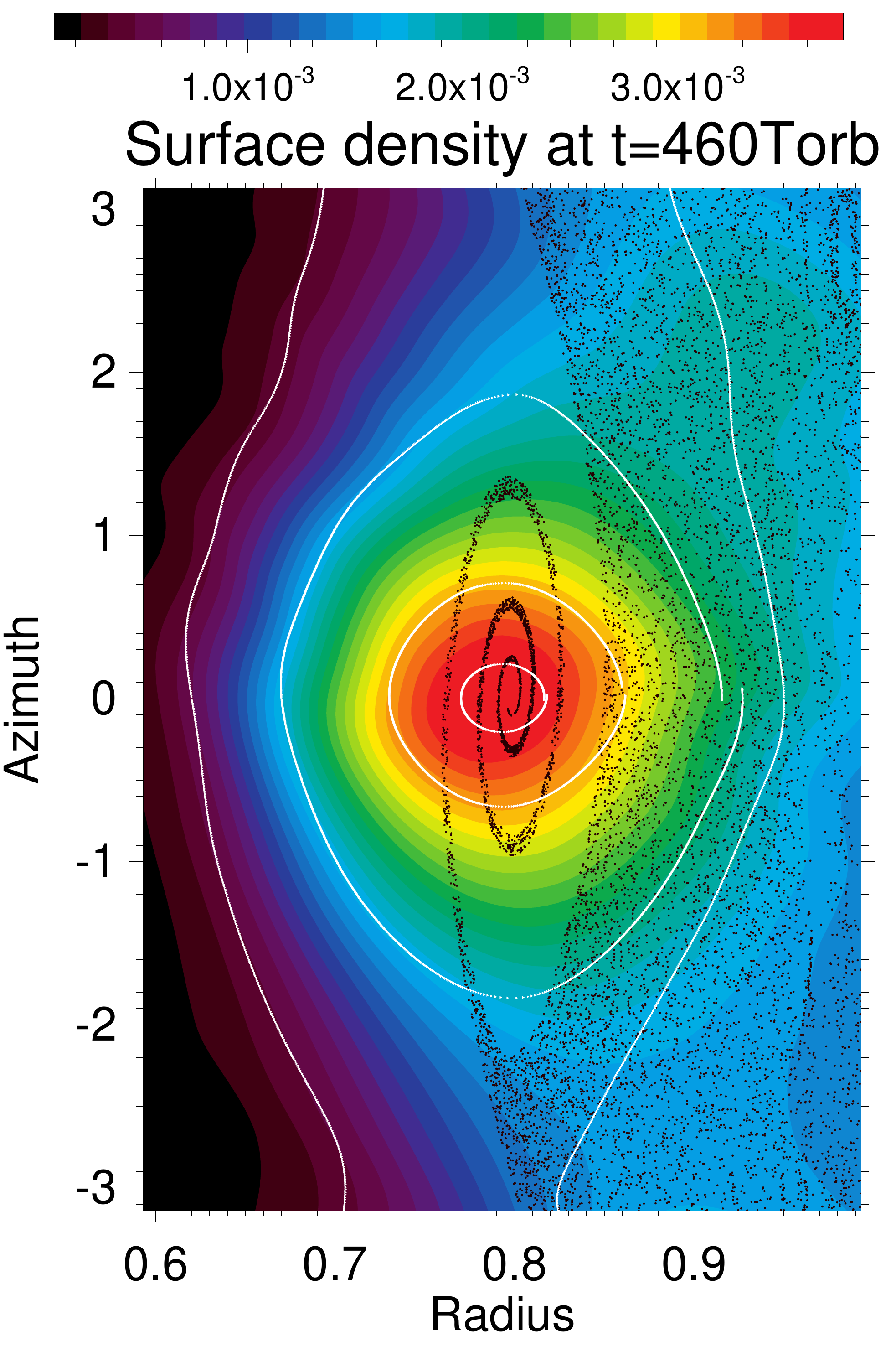}
\caption{\label{fig:ph}Results of simulations with large (typically
  metre-sized) particles with model g2 (three leftmost panels) and
  model g2n (right panel). Contours of the gas surface density are
  shown along with the particles location depicted by black dots, and
  gas streamlines overplotted by white curves. With self-gravity,
  large particles perform horseshoe U-turns relative to the vortex and
  drift much ahead of the vortex. Without self-gravity, large
  particles drift toward the vortex centre.}
\end{figure*}

\subsubsection{Results of model g2}
The left panel of Figure~\ref{fig:g13_rf2} displays contours of the
gas surface density at 500 orbits for model g2. Results are shown with
polar coordinates $\{r,\varphi\}$ and the maximum in the gas surface
density is set at $\varphi=0$. For comparison, the results of model
g2n (without self-gravity) are shown at the same time in the middle
panel of Figure~\ref{fig:g13_rf2}, using the same colour scale for the
density contours (recall that, without self-gravity, particle
concentration is found to be independent of the initial particle
distribution).  We first see that vortex inward migration is much less
rapid with self-gravity, and consequently the gas surface density in
the vortex is smaller with self-gravity at the same time.  Actually,
the vortex's shape and strength at 500 orbits in model g2 are,
overall, very similar to those of model g2n at $\sim$$130$ orbits (not
shown here). A qualitative comparison of the two models at these times
indicates that the density contrast in the vortex's spiral waves is
smaller with self-gravity, which is consistent with inward migration
being slower in that case. A detailed interpretation for why gas
self-gravity slows down the vortex's inward migration is beyond the
scope of this paper. Part of it could be due to self-gravity
inhibiting the RWI and thus weakening the vortex when the disc's
surface density near the vortex's location is large enough
\citep[][see also Paper I]{GN88}. It is also possible that gas
self-gravity shifts the location of the vortex's Lindblad resonances
(where the flow becomes supersonic relative to the vortex). By analogy
with the analytic study of \cite{PH05} for planet-disc interactions,
gas self-gravity would shift Lindblad resonances outward compared to
the situation where the vortex would migrate in a non self-gravitating
disc.  This would weaken the vortex's outer wave, strengthen the inner
wave, and explain the slowing down of the migration (with some
analogy, again, with planet-disc interactions, see \citealp{BM08b}).

The location of dust particles of various sizes is marked by filled
circles in the left and middle panels of Figure~\ref{fig:g13_rf2}, and
their instantaneous Stokes number is shown in the panels. For model g2
again, dust grains quickly converge to specific locations in the disc
where a force balance is reached. However, particle concentration
operates differently with gas self-gravity: the particles shift angle
relative to the vortex centre increases monotonically with Stokes
number, ranging from $\approx 0$ degrees for St $\sim 0.02$ to
$\approx$ 90 degrees for St $\sim 16$. As shown again in the middle
panel of Figure~\ref{fig:g13_rf2} for model g2n, shift angles without
self-gravity increase with St up to St $\sim 1$, and decrease to zero
beyond.  The comparison between models g2 and g2n is further
highlighted in the right panel of Figure~\ref{fig:g13_rf2}, which
displays the shift angle and Stokes number at 500 orbits with and
without self-gravity.  We stress the rapid increase of the shift angle
with self-gravity when St $\gtrsim 1$. This transition occurs for
particles of a few centimetres in size when using the set of physical
units given as example in Section~\ref{sec:codeunits}. The location
where particles concentrate in model g2 is found to be independent of
their initial distribution (e.g., inside or outside the vortex's
radial location), at least for the particle sizes shown in
Figure~\ref{fig:g13_rf2}. Larger particles than those shown in the
figure can display a different behaviour, which we describe below.

To get more insight into the behaviour of the largest particles with
gas self-gravity, we show the evolution of particles 4 times larger
than the largest particles displayed in Figure~\ref{fig:g13_rf2}.
Their physical radius is 2 metres using the set of physical units
given in Section~\ref{sec:codeunits}, and their initial distribution
is again uniform between $r=1.2$ and 1.4. The time evolution of their
location relative to the vortex is illustrated in the three leftmost
panels in Figure~\ref{fig:ph}. As particles drift in toward the
vortex, they make a horseshoe U-turn ahead of the vortex centre
(leftmost panel).  Later, particles go on to a second U-turn (outward
this time, second panel on the left). Because of gas drag, the second
U-turn is closer to the vortex's radial location, but further away in
the azimuthal direction. Particles eventually drift toward their
equilibrium location about 90-100 degrees ahead of the vortex (third
panel). Gas streamlines, which are overplotted by white curves in the
panels, recall that gas does not make horseshoe U-turns relative to
the vortex. So do not the smallest particles, which concentrate
towards the vortex centre. Gas streamlines also indicate that the
equilibrium location of the largest particles is very close to the tip
of the elliptical streamline that lies furthest from the vortex. The
rightmost panel in Figure~\ref{fig:ph} shows the result of the same
simulation without self-gravity, and it is clear that the large
particles drift toward the vortex centre upon adopting elliptical
trajectories relative to the vortex.

The horseshoe U-turns performed by large particles is most likely the
consequence of the particles feeling the self-gravitating acceleration
of the gas and interacting with the vortex partly as if it was a
massive body. Like if it was a planet. Using the analogy with
disc-planet interactions, we interpret our results based on the
comparison between three timescales: (i) the timescale for dust
particles to drift across the half-width of the vortex's horseshoe
region ($\tau_{\rm drift}$), (ii) the timescale for particles to make
a U-turn relative to the vortex ($\tau_{\rm U-turn}$), and (iii) the
timescale for particles to complete a full horseshoe orbit relative to
the vortex (libration timescale, $\tau_{\rm lib} >
  \tau_{\rm U-turn}$). When $\tau_{\rm drift} \leq \tau_{\rm
  U-turn}$, gas drag prevents dust particles from making horseshoe
U-turns, and dust drifts towards the vortex's centre much like when
self-gravity is discarded. When $\tau_{\rm lib} > \tau_{\rm drift} 
\gtrsim \tau_{\rm U-turn}$, dust particles can
embark on, but not complete, horseshoe
trajectories. Gas drag still acts to damp the relative velocity
between dust and gas, and particles thus progressively concentrate at
or near the vortex's radial location (as can be seen in the two middle
panels in Figure~\ref{fig:ph}). When $\tau_{\rm drift} \gg \tau_{\rm
  lib}$, that is when the drag force is much smaller than the
effective gravitational force from the vortex, particles should
undergo horseshoe orbits very similar to those a nearly inviscid gas
would adopt in the presence of a planet. This is shown in the limit of
fully decoupled particles ($\tau_{\rm drift} \rightarrow \infty$) in
Section~\ref{sec:pinf} of the Appendix.

The particle Stokes number for which $\tau_{\rm drift} \gtrsim
\tau_{\rm U-turn}$, and for which dust can embark on horseshoe U-turns
relative to the vortex, can be estimated as follows.  We have
$\tau_{\rm drift} = x_{\rm s} / |v_{\rm r,dust}|$ with $x_{\rm s}$ the
half-width of the vortex's horseshoe region and $v_{\rm r,dust}$ the
particle radial velocity. In a 1D isothermal gas disc, and neglecting
the contribution of gas self-gravity to the azimuthal velocities of
the gas and dust, $v_{\rm r,dust}$ is given by
\citep[e.g.,][]{TakeuchiLin02}
\begin{equation}
v_{\rm r,dust} = \frac{1}{1+{\rm St}^2} \left( 
v_{\rm r,gas} + h^2 v_{\rm K} {\rm St} \frac{\partial\log\Sigma}{\partial\log r}
\right),
\label{eq:vrdust}
\end{equation}
where $v_{\rm r,gas}$ is the radial velocity of the gas and $v_{\rm
  K}$ is the Keplerian velocity. All quantities in
Eq.~(\ref{eq:vrdust}) should be evaluated at the separatrices of the
vortex's horseshoe region (that is, at $r = r_{\rm v} \pm x_{\rm s}$,
with $r_{\rm v}$ the vortex's radial location). Progress can be made
by considering the 2D structure of the vortex. For model g2, our
numerical simulations show that $v_{\rm r,gas}$ is fairly well
approximated by
\begin{equation}
v_{\rm r,gas} = -\frac{5h^2}{2} v_{\rm K} \frac{\partial\log\Sigma}{\partial\varphi}, 
\end{equation}
which is actually 5 times the first-order correction for the gas
radial velocity to maintain a geostrophic flow about the vortex
(balance between Coriolis and pressure forces; see, e.g., equation 16
in \citealp{MC15}). The difference accounts for the zero-order radial
velocity of the gas. We then find that $v_{\rm r,gas}(r = r_{\rm v}
\pm x_{\rm s}) \approx \pm h^2 v_{\rm K}$, with $v_{\rm r,gas}(r =
r_{\rm v} - x_{\rm s})$ evaluated behind the vortex in the azimuthal
direction (where outward U-turns occur), and similarly $v_{\rm
  r,gas}(r = r_{\rm v} + x_{\rm s})$ is evaluated ahead of the vortex
in the azimuthal direction (where inward U-turns occur). The radial
density profile of the gas being weakly altered by the vortex's
formation, the last term in the bracket in Eq.~(\ref{eq:vrdust}) can
be calculated with the initial surface density profile, which gives
$\partial\log\Sigma / \partial\log r (r = r_{\rm v} \pm x_{\rm s}) =
\mp r_{\rm v} x_{\rm s} / 4H^2$. Eq.~(\ref{eq:vrdust}) can finally be
recast as
\begin{equation}
v_{\rm r,dust}(r = r_{\rm v} \pm x_{\rm s}) = \frac{v_{\rm K}}{1+{\rm St}^2} \left( 
\pm h^2
\mp \frac{{\rm St}}{4} \frac{x_{\rm s}}{r_{\rm v}} 
\right).
\label{eq:vrdust2}
\end{equation}
We first note that, for dust horseshoe U-turns to be possible, we need
$v_{\rm r,dust}(r = r_{\rm v} - x_{\rm s}) > 0$ (outward U-turn) and
$v_{\rm r,dust}(r = r_{\rm v} + x_{\rm s}) < 0$ (inward U-turn), which
implies ${\rm St} > 4h^2 r_{\rm v}/x_{\rm s}$.  For the vortex in
model g2, $x_{\rm s} / r_{\rm v} \approx 0.04$ and above inequality
becomes St $\gtrsim 1$. From our numerical simulations, we find
$\tau_{\rm U-turn} \sim (2-3)\times T_{\rm orb}$
which, as expected, is a small fraction of $\tau_{\rm lib}$
($\tau_{\rm lib} = 8\pi r_{\rm v} / 3\Omega x_{\rm s} \approx 35
T_{\rm orb}$). Using Eq.~(\ref{eq:vrdust2}), we finally find that
$\tau_{\rm drift} \gtrsim \tau_{\rm U-turn}$ for St $\gtrsim 2-4$. It
is in good agreement with the range of Stokes numbers from which gas
self-gravity considerably increases the particle shift angle in model
g2 (see right panel of Figure~\ref{fig:g13_rf2}).

For completeness, we mention that the same self-gravitating simulation
as shown in the three leftmost panels in Figure~\ref{fig:ph}, but
where particles are placed initially between $r=0.5$ and 0.7, result
in the particles concentrating both ahead (at $\sim$$100$ degrees) and
behind (at $\sim$$-100$ degrees) the vortex centre. This shows that
the concentration of the largest particles, either far ahead of far
behind the vortex, depends on the relative flow between the dust and
the vortex, that is on the initial location of the particles (inside
or outside the vortex's radial location), as well as the direction of
vortex migration (not shown here). We point out again the analogy with
disc-planet interactions, where gas drifting relative to a low-mass
planet -- because the planet migrates and/or the gas drifts viscously
-- can be trapped ahead or behind the planet \citep{masset02, SJP14,
  Pierens15}. We also stress the analogy with the
  trapping of solid particles on tadpole orbits around a planet in the
  presence of gas drag \citep[e.g.,][]{Peale93, Lyra09}. The
  equilibrium location of solid particles ahead or behind the planet
  depends on the planet's mass and on the particle size, and,
  interestingly, the maximum shift angle of solid particles ahead of a
  planet is 108 degrees \citep{Peale93}, which is very close to the
  maximum shift angle that we find for a self-gravitating vortex.
  
As a brief summary of our results for model g2, we find that gas
self-gravity controls the dynamics of large particles, whereas gas
drag determines the dynamics of small particles.  Self-gravity allows
the largest particles to concentrate far ahead of the vortex (or far
behind it) upon adopting horseshoe U-turns. Gas drag prevents small
particles from taking the U-turns. The transition between what could
be called a gas drag-dominated regime and a self-gravity-dominated
regime occurs quite neatly around St $\approx 1$ for model
g2. However, the Stokes number at this transition likely depends on
the disc mass. For example, in a disc with lower gas surface density
near the vortex's radial location, the magnitude of the drag force
will be approximately the same for a given Stokes number
(approximately since the vortex's shape and strength may vary with the
disc's surface density), but the gravitational force from the vortex
will be smaller since the vortex is less massive. Results of
simulations at lower gas surface densities will be presented at the
beginning of Section~\ref{sec:synthetic}, which indicate that the
transition between the gas drag and self-gravity dominated regimes
occurs around a critical particle size.

\subsubsection{Results of model g5}
\label{sec:g5}
\begin{figure*}
\centering
 \includegraphics[width=0.47\hsize]{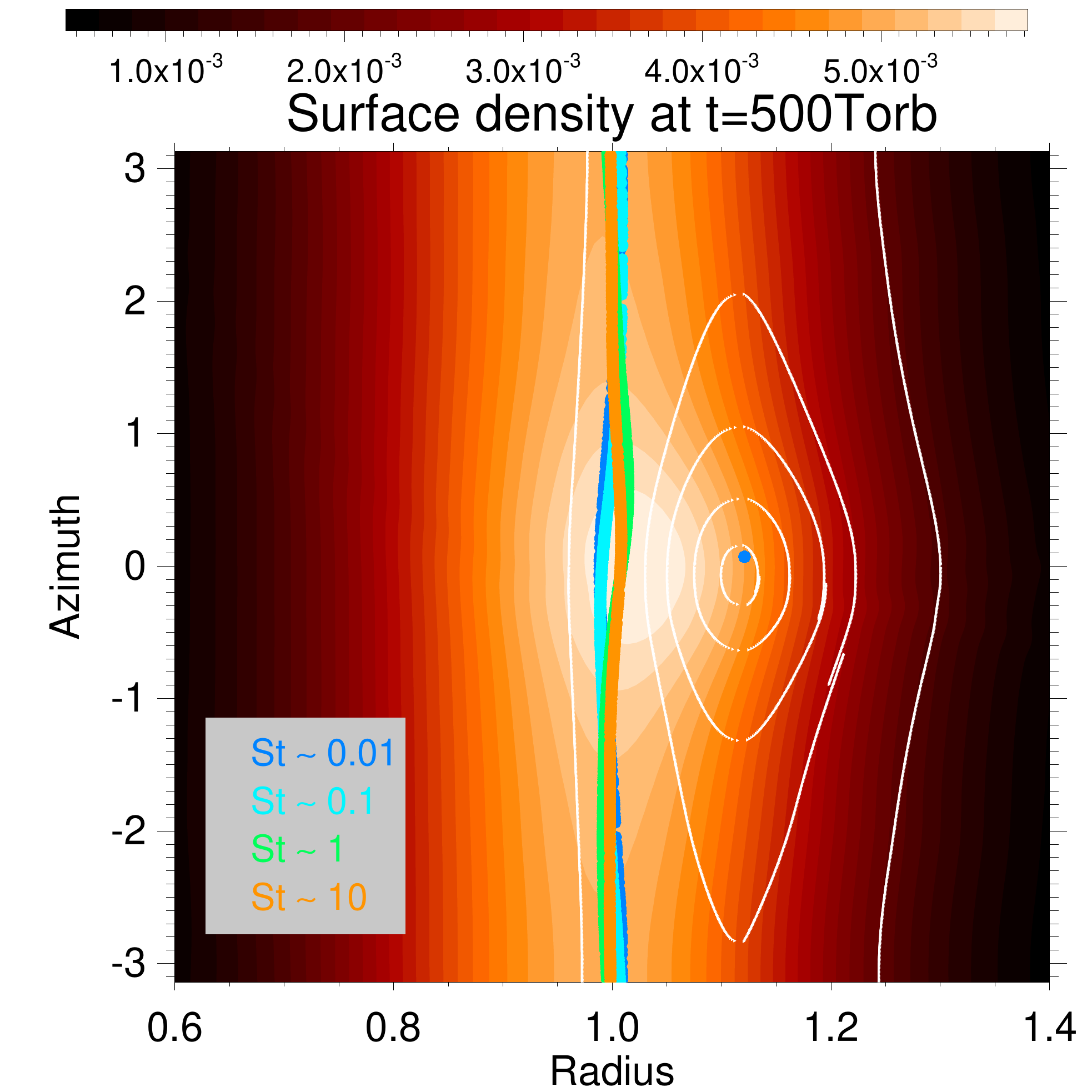}
  \includegraphics[width=0.47\hsize]{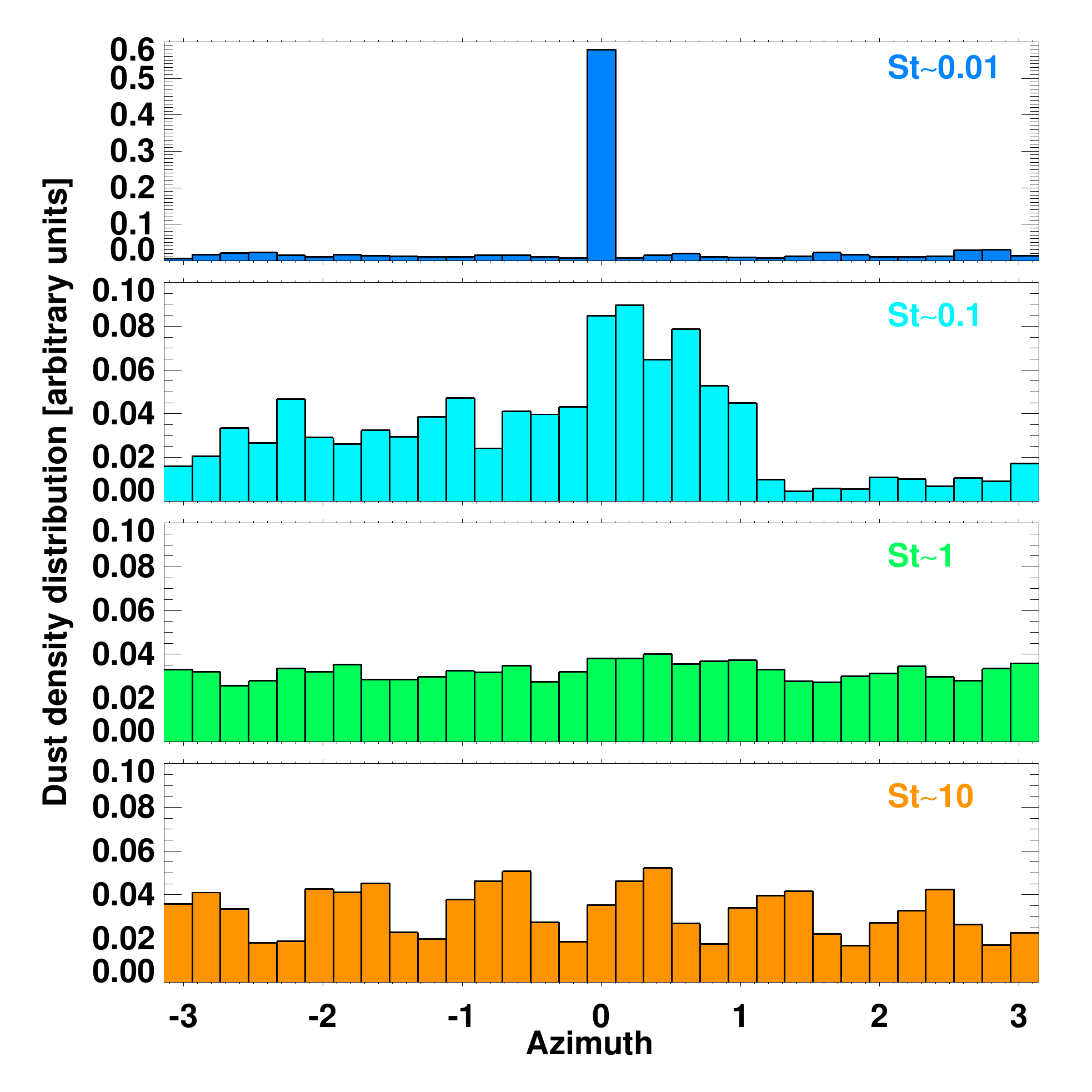}
  \caption{\label{fig:g1_rf2}Results of model g5 at 500 orbits. Left:
    contours of the gas surface density with gas streamlines relative
    to the vortex (white curves) and particles location (filled
    circles). Right: instantaneous dust density distribution with
    azimuth (see text). The colours used for the particles are the
    same in both panels, and the space-averaged Stokes number of the
    particles is indicated in the panels.}
\end{figure*}
The results of model g5 are displayed in Figure~\ref{fig:g1_rf2} at
500 orbits. We see that both the gas structure and the dust
distribution are markedly different from those of model g2, while the
initial gas densities differ by only a factor of 2.5 between both
models (compare with the left panel of Figure~\ref{fig:g13_rf2}). In
model g5, the vortex is more elongated along the azimuthal direction
and the density contrast is about 20\%. For comparison, in model g2
the density contrast along the vortex is by a factor of
$\sim$$2$. Furthermore, in model g5 the gas density maximum in the
vortex is still very close to $r = 1$, indicating that the vortex has
not migrated over the duration of the simulation. This is supported by
a detailed inspection at the gas density which shows marginal
excitation of density waves by the vortex compared to model g2. Also,
we notice that the radial location where elliptical streamlines are
centred in the frame rotating at the vortex's pattern frequency, which
defines the vortex's corotation radius, is shifted outward from the
vortex's radial location (defined as the radial location where the gas
surface density peaks).  This shows that the vortex's pattern
frequency is slower than the Keplerian frequency (Paper I). This slow
pattern frequency is probably the reason why the vortex emits no or
very little spiral density waves in the disc. The slow pattern
frequency also directly impacts the particles distribution. Most of
the small particles get trapped onto the elliptical streamlines in the
rotating frame and progressively drift toward the centre of the set of
elliptical streamlines (which actually corresponds to a minimum in the
gas potential vorticity or vortensity). This is what we obtain for the
St $\sim$ 0.01 particles (see the single blue dot at $r \approx$ 1.12
in the left panel of Figure~\ref{fig:g1_rf2}).  Note, however, that
some of the St $\sim$ 0.01 particles form a ring-like structure around
the vortex's radial location at $r \approx$ 1.

Larger particles decouple from the gas and are not tied to the
elliptical streamlines: they converge toward the pressure maximum
located at $r \approx$ 1. They also form ring-like structures, though
the left panel in Figure~\ref{fig:g1_rf2} does not hint at the
particles azimuthal distribution. The latter is shown as an histogram
in the right panel of Figure~\ref{fig:g1_rf2}, with large bins used to
smooth high spatial frequency fluctuations (the y-values in the
histogram are simply the fraction of the total number of particles in
each bin). Azimuths are relative to the vortex centre.  More than half
of the smallest (St $\sim$ 0.01) particles are at $\varphi \sim$ 0,
with the vast majority of them at the centre of the elliptical
streamlines. Larger particles show quite distinct azimuthal
distributions.  St $\sim$ 0.1 particles have a lopsided distribution
that peaks slightly ahead of the vortex. The fact it peaks ahead of
the vortex is just a coincidence: since St $\sim$ 0.1 particles and
the vortex move at different speeds, the azimuth at which the
particles' azimuthal distribution peaks varies with time. Finally,
while St $\sim$ 1 particles have a nearly flat azimuthal distribution,
St $\sim$ 10 particles feature a periodic pattern of azimuthal
wavenumber $m=6$. The reason for this particular pattern is unknown,
but resembles that of the St $\sim$ 0.01 particles located around the
vortex's radial location.  We find the same pattern when adopting a
slightly different initial distribution (uniform in $r \in
[0.8-1.2]$). We also comment that the vortex's corotation radius
becomes less shifted as grid resolution is increased. It is located at
$r \approx 1.05$ for a grid resolution increased to $600\times 1000$,
compared to $r \approx 1.12$ for our nominal resolution of $300\times
600$. More detailed tests of convergence in resolution with
self-gravity are presented for model g10 in Section~\ref{sec:g10}.
\begin{figure*}
\centering
\includegraphics[width=0.245\hsize]{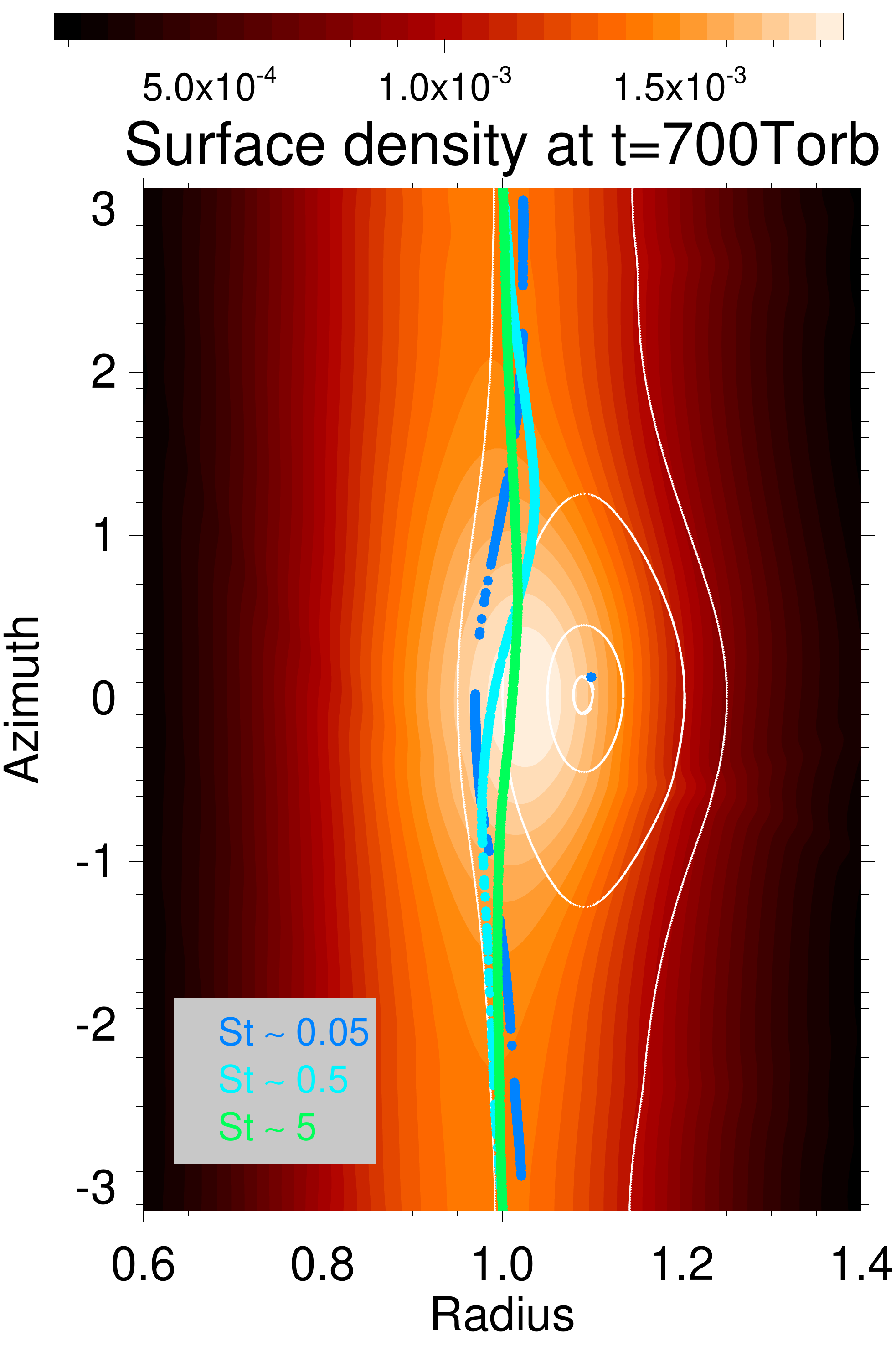}
\includegraphics[width=0.245\hsize]{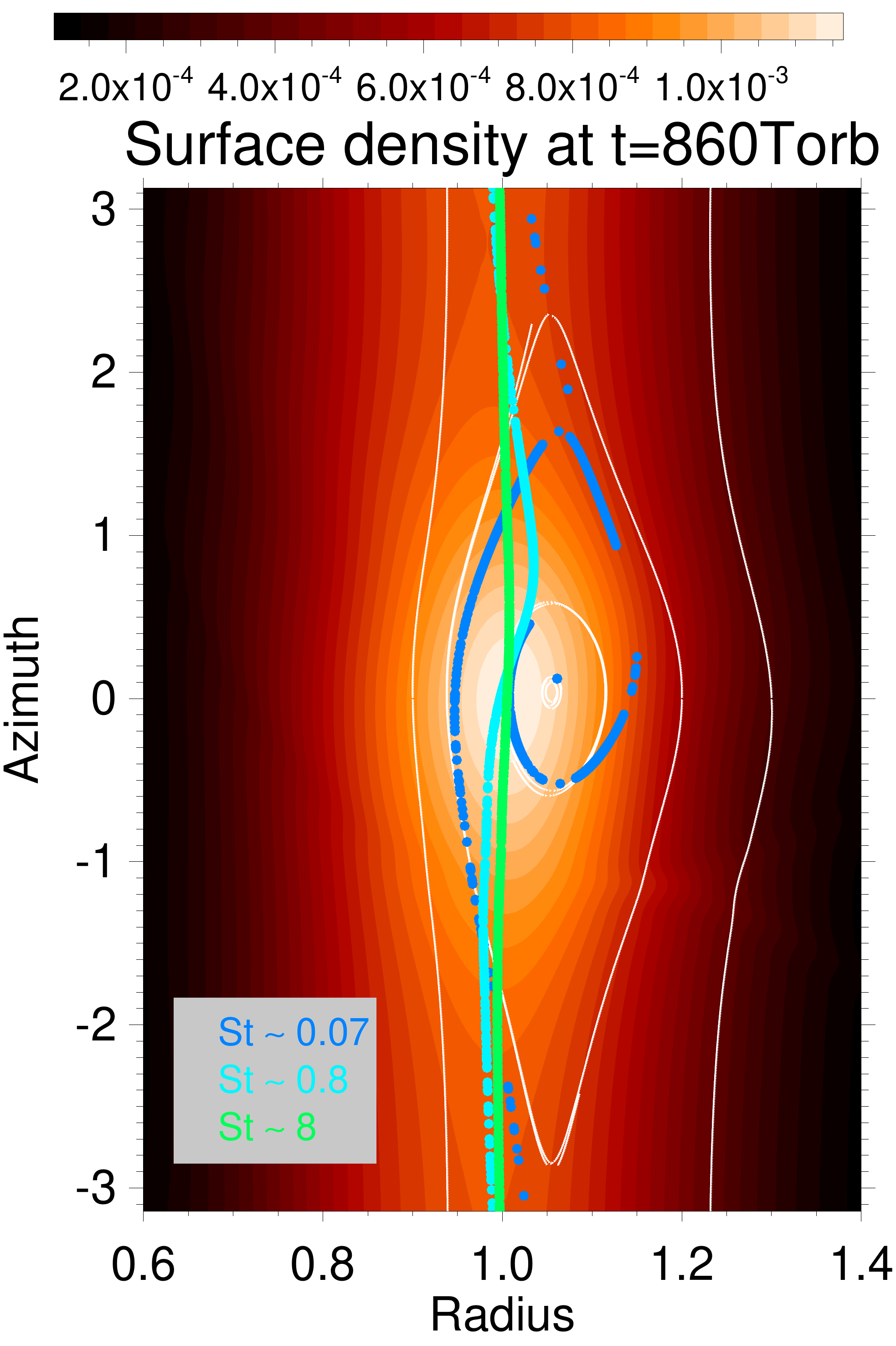}
\includegraphics[width=0.245\hsize]{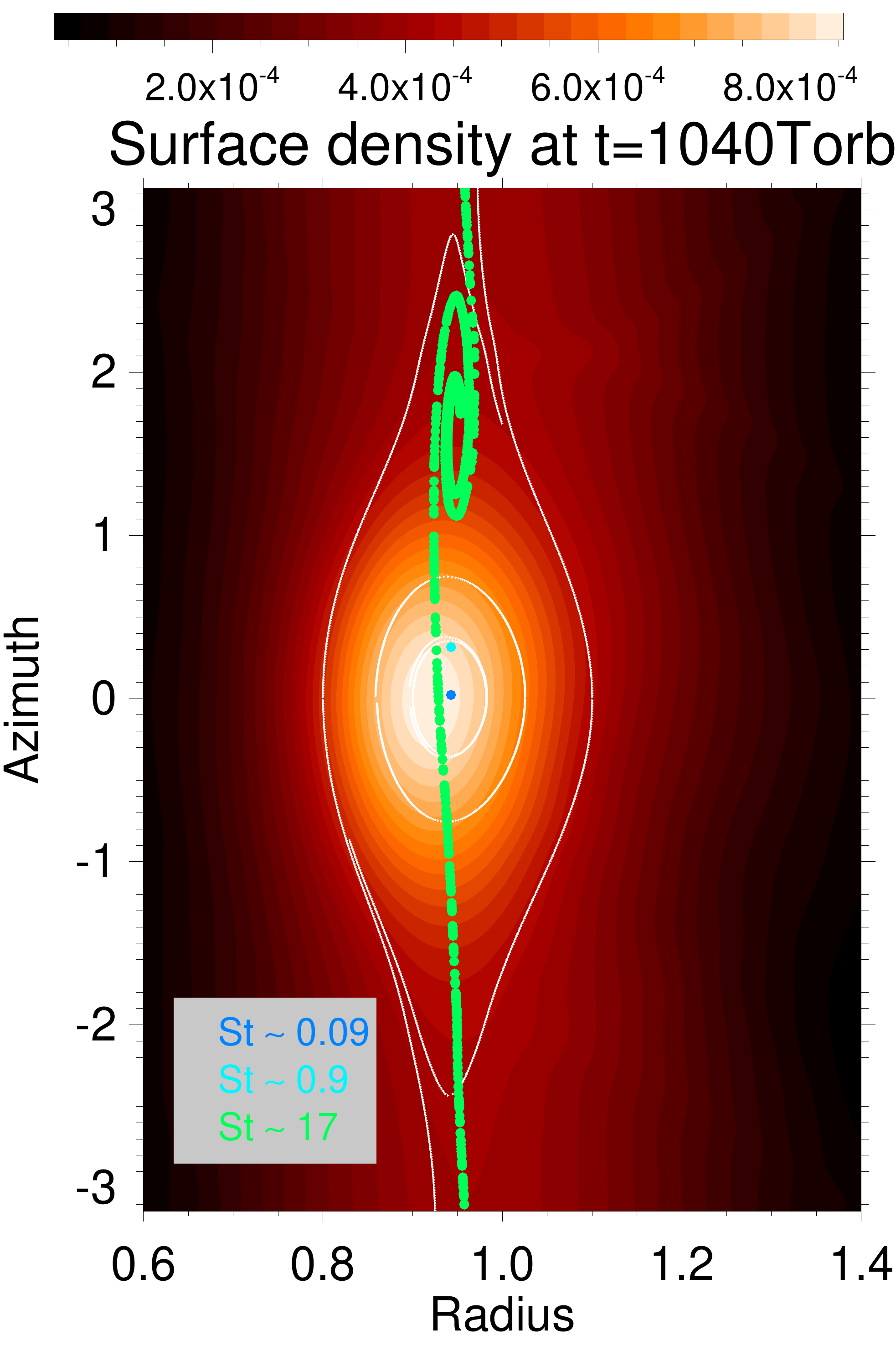}
\includegraphics[width=0.245\hsize]{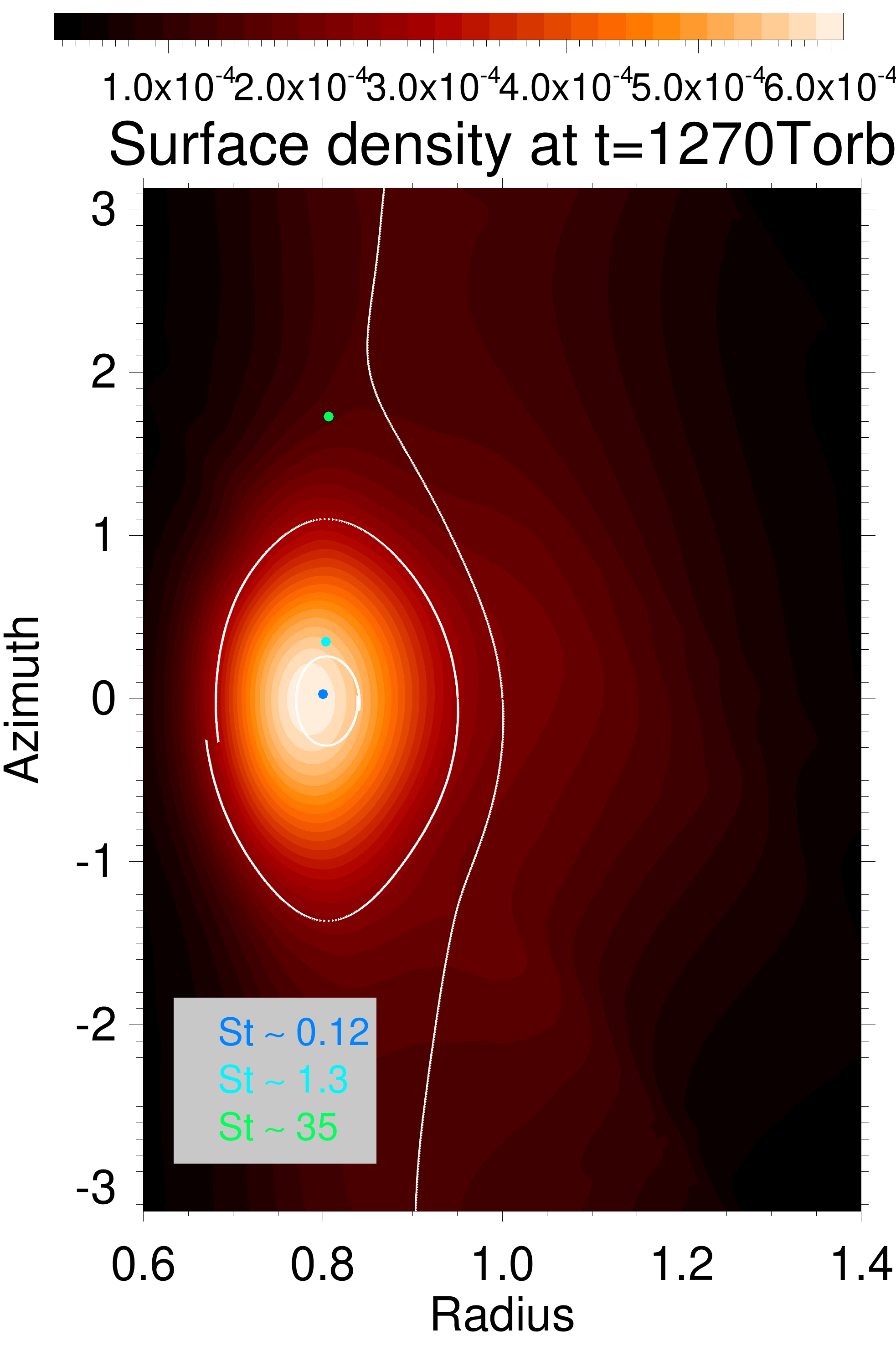}
\caption{\label{fig:g1r}Results of the restart simulation with model
  g5 where a slow decrease in the radial profile of the gas surface
  density is imposed (see text). As in previous figures, all panels
  show contours of the gas surface density with gas streamlines in the
  frame rotating at the vortex's pattern frequency (white curves) and
  the location of particles of different sizes (filled circles). The
  averaged Stokes number of the particles is shown in the lower-right
  corner in the panels.}
\end{figure*}

\begin{figure}
\centering
\resizebox{0.49\hsize}{!}
  {
 \includegraphics{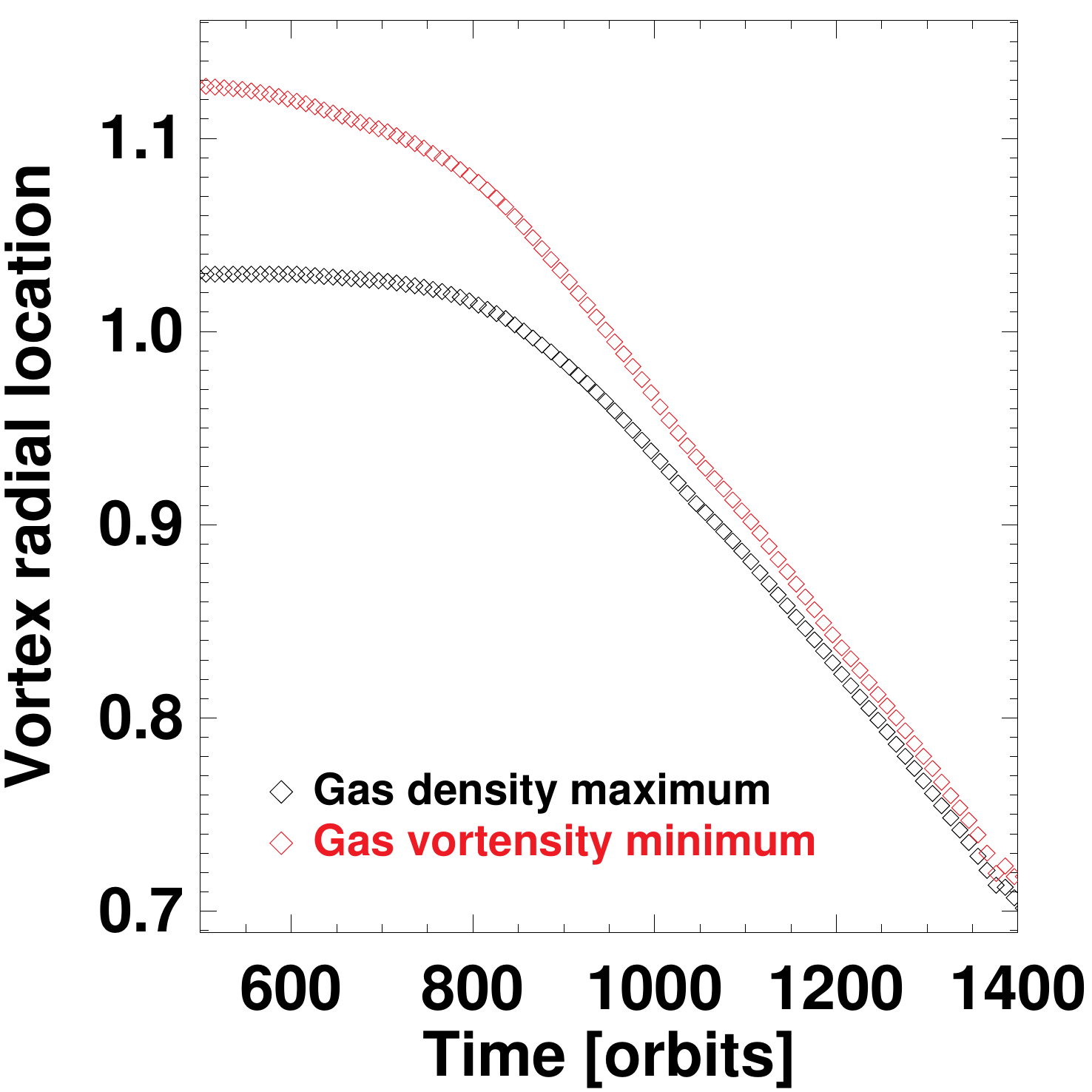}
  }
  \resizebox{0.49\hsize}{!}
  {
 \includegraphics{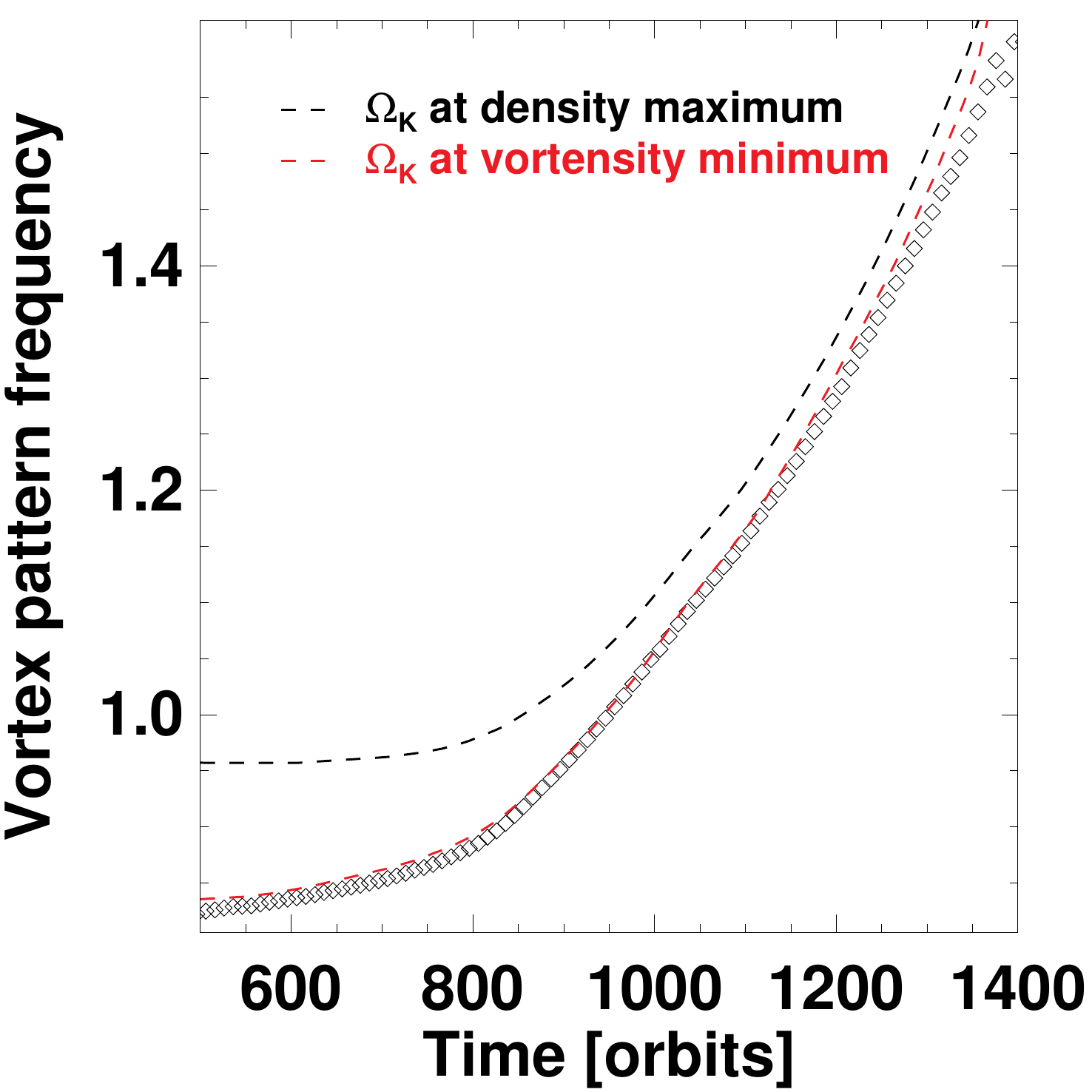}
  }
  \caption{\label{fig:vps}Results of the restart simulation with model
    g5 where a slow decrease in the radial profile of gas surface
    density is imposed. Left: time evolution of the radial location
    where the gas surface density peaks in the vortex (black symbols),
    and where the gas potential vorticity is minimum, which is the
    centre of the elliptical streamlines in the rotating frame (red
    symbols). Both locations coincide when the local disc mass becomes
    small enough. Right: vortex's pattern frequency (black symbols)
    and Keplerian frequency at both previous locations (dashed
    curves).}
\end{figure}
\begin{figure*}
\centering
 \includegraphics[width=0.27\hsize]{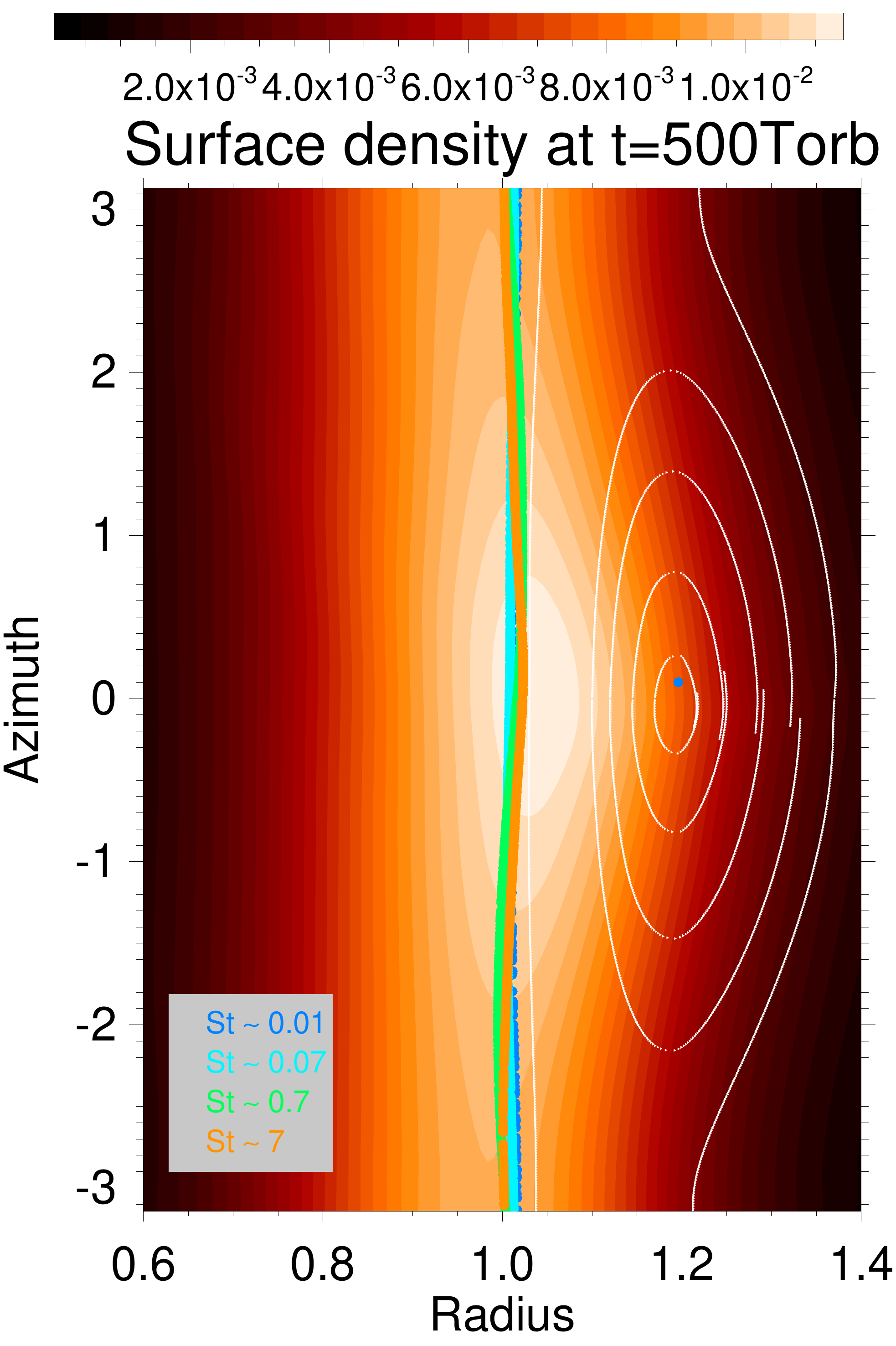}
 \includegraphics[width=0.27\hsize]{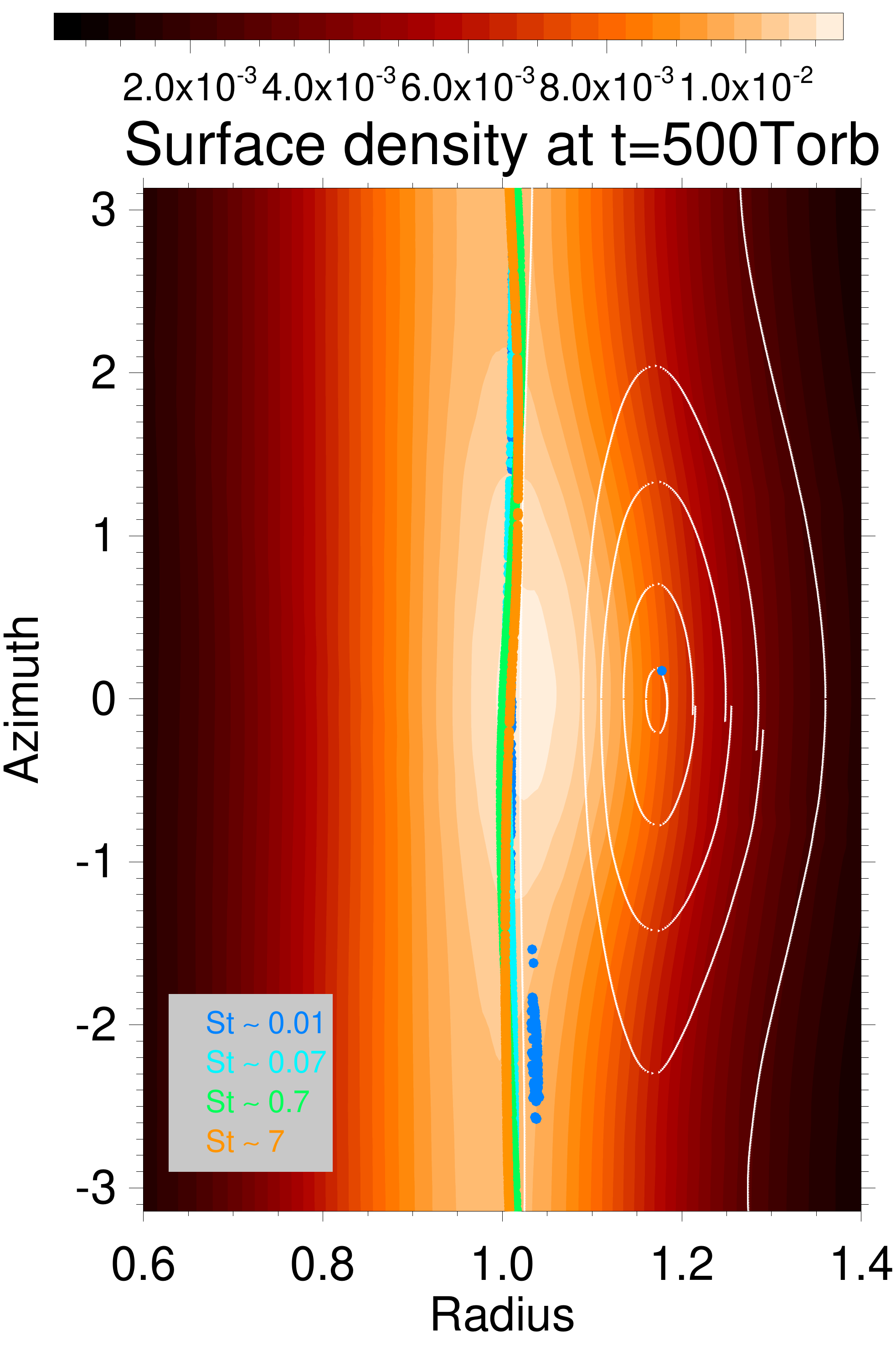}
 \includegraphics[width=0.42\hsize]{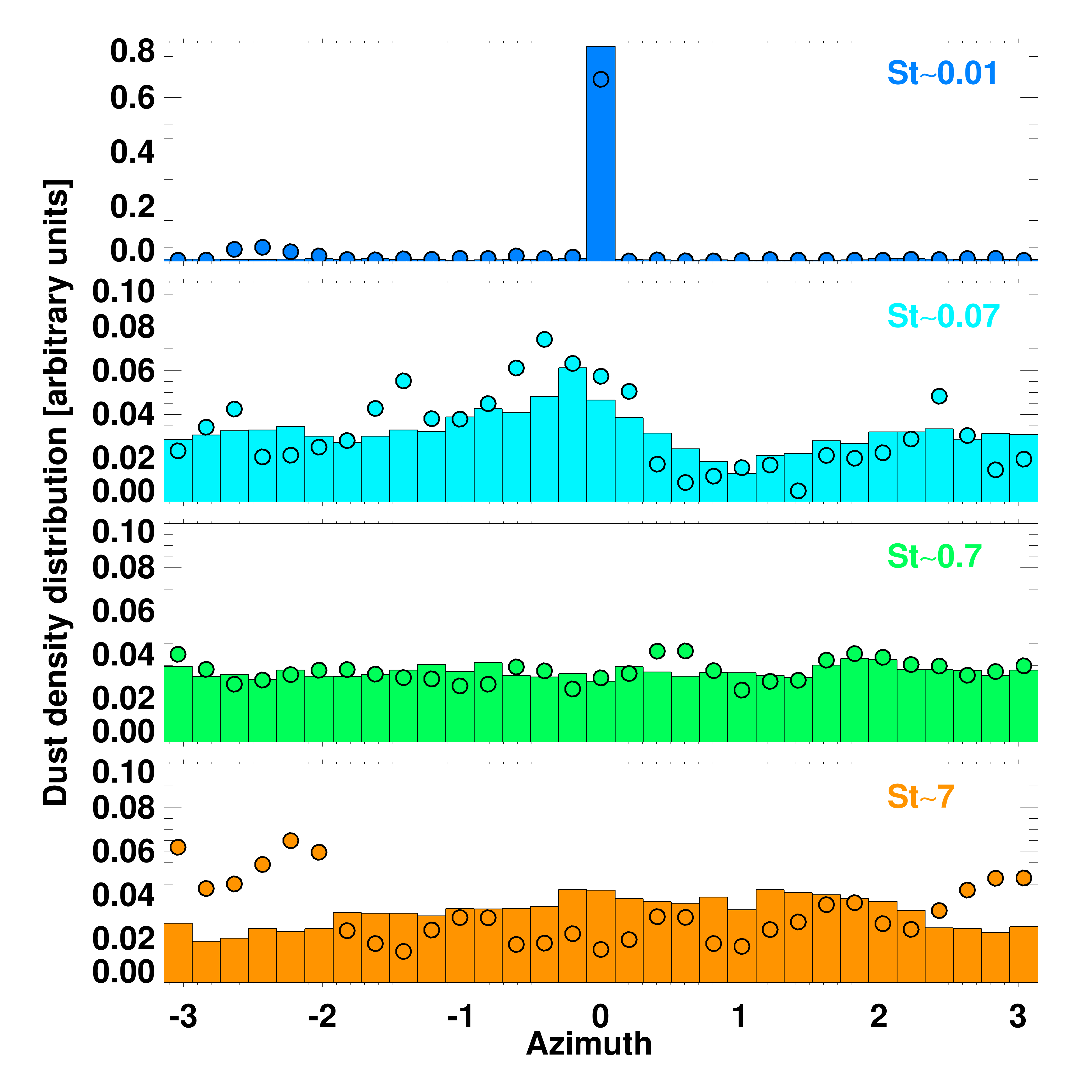}
 \caption{\label{fig:g10_rf2} Results of model g10 at 500 orbits with
   our fiducial grid resolution (300$\times$600, left panel) and at
   higher resolution (600$\times$1000, middle panel). Left and middle
   panels: contours of the gas surface density with gas streamlines
   relative to the vortex (white curves) and particles location
   (filled circles).  Right: instantaneous dust density distribution
   versus azimuth, with histogram bars for the run at nominal
   resolution and filled circles for the run at high resolution. The
   colours used for the particles are the same in the three panels,
   and the space-averaged Stokes number of the particles is indicated
   in the panels.  }
\end{figure*}
To get more insight into the different gas and dust behaviours between
models g2 and g5, we have restarted the simulation with model g5 at
350 orbits by imposing a slow decrease in the radial profile of the
gas surface density. By doing this we expect to see a smooth
transition in the concentration of the large dust grains from
ring-like structures (as in model g5) to point-like structures (as in
model g2). Specifically, as in \cite{BP13} we adopt a simple
exponential decay of the density profile by solving $\partial_t \Sigma
= -(\overline{\Sigma} - \Sigma_{\rm target})/\tau$ alongside the
hydrodynamical equations, with $\overline{\Sigma}$ the
azimuthally-averaged density profile at restart time, $\Sigma_{\rm
  target} = 10^{-3}\overline{\Sigma}$ an arbitrarily small density
profile that we take to be not zero for numerical convenience, and
$\tau$ is chosen to be 300 orbits. In practice, the axisymmetric
profile of gas density is decreased by a factor $\sim$10 in about 700
orbits after the restart. The results of this simulation are displayed
at four different times in Figure~\ref{fig:g1r}, where the Stokes
numbers shown in the panel are averaged in space (they increase with
time as the gas surface density decreases). The first panel from the
left is at 700 orbits (350 orbits after the restart).  Interestingly,
the gas density profile now takes smaller values than in model g2, yet
both the gas and the dust distributions are very similar to those of
model g5 prior to the restart. We note a few differences, though.
First, the centre of the gas elliptical streamlines in the rotating
frame is slightly closer to the vortex's centre (its radial location
has decreased from $r \approx 1.12$ to $r \approx 1.08$). Second and
probably related point, the smallest particles that are located around
the vortex's radial location now form a disrupted ring (see the
dark-blue dots in the panel). In the second panel (at 860 orbits), the
centre of the elliptical streamlines has become close enough to the
vortex centre to significantly perturb the smallest particles, which
become trapped onto elliptical streamlines. In the third panel (1040
orbits), the centre of the elliptical streamlines now coincides with
the vortex centre, which indicates that the vortex's pattern frequency
is now very close to the Keplerian frequency.  In addition, the vortex
is now less elongated along the azimuthal direction (the density
contrast along the vortex increases) and it has started to migrate
inward (the vortex's radial location has decreased from $r \approx 1$
to $\approx 0.92$ from 860 to 1040 orbits). Particles with the two
smallest sizes now have point-like distributions (dark-blue and cyan
dots in the panel), while the largest particles are getting trapped
much ahead of the vortex. In the fourth panel (1270 orbits) particles
have point-like distributions which are very close to those shown in
Figure~\ref{fig:g13_rf2} for model g2.

Before closing this subsection, we come back to the evolution of the
vortex's pattern frequency with gas self-gravity. We have seen that,
with increasing strength of self-gravity, the radial location where
the gas potential vorticity is minimum (the vortex's corotation
radius) shifts outward from the radial location where the gas density
peaks (which we have defined throughout this study as the vortex's
radial location). This is further illustrated and quantified in the
left panel of Figure~\ref{fig:vps} for the restart simulation with
model g5 where the gas density decreases over time. We have calculated
the vortex's pattern frequency by measuring the angle through which
the density maximum rotates over time (it is the same angle as that
through which the minimum potential vorticity rotates over time). The
time evolution of the vortex's pattern frequency is displayed in the
right panel of Figure~\ref{fig:vps}. Dashed curves show the Keplerian
frequency at the radial location where the gas density is maximum (in
black) and where the potential vorticity (vortensity) is maximum (in
red). The vortex's pattern frequency nearly equals the Keplerian
frequency at the radial location of the vortensity minimum, as
expected, and catches up with the Keplerian frequency at the density
maximum as the strength of self-gravity decreases.

\subsubsection{Results of model g10}
\label{sec:g10}
The results of model g10 are displayed in Figure~\ref{fig:g10_rf2} at
500 orbits. Overall, the gas and dust distributions are similar to
those of model g5 at the same time (compare with
Figure~\ref{fig:g1_rf2}). The smallest particles also tend to
concentrate at the centre of the elliptical streamlines located at the
vortex's corotation radius. The latter is shifted further outside the
vortex's radial location compared to model g5 (the vortex's corotation
radius is at $r \approx 1.2$ for model g10 and at $r \approx 1.12$ for
model g5). This goes along the trend illustrated at the end of the
previous subsection that the larger the disc mass (that is, the
stronger self-gravity), the slower is the vortex's pattern frequency
compared to the Keplerian frequency. Larger particles concentrate
along rings with different azimuthal structures depending on the
particle size (and thus Stokes number). Figure~\ref{fig:g10_rf2} also
displays the results of model g10 with a grid resolution increased
from $300\times600$ to $600\times1000$ (middle panel, filled circles
in the right panel). Results at both resolutions are in fairly good
agreement.  The vortex's corotation radius decreases from $r \approx
1.2$ at a resolution of $300\times600$ to $r \approx 1.18$ at a
resolution of $600\times1000$. Convergence in resolution is much
better than for model g5, which suggests that the transition to the
regime where self-gravity renders the vortex's pattern frequency
sub-Keplerian is very sensitive to grid resolution.
\begin{figure}
\centering
\includegraphics[width=0.83\hsize]{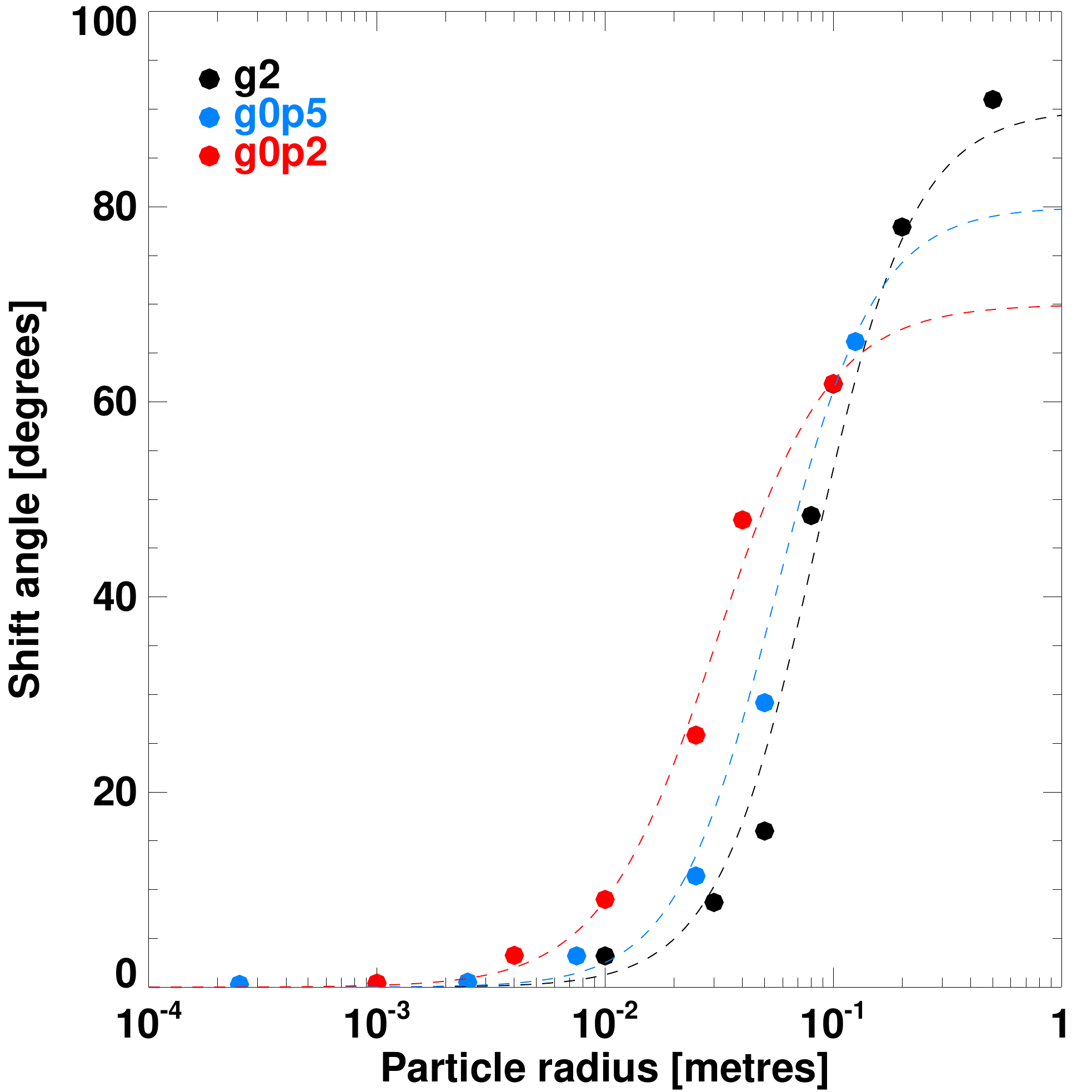}
\caption{\label{fig:depth} Shift angle relative to the vortex centre
  for various particle sizes in models g2, g0p5 and g0p2. Dashed
  curves are simple fitting functions (see text).  }
\end{figure}
\begin{figure*}
\centering
\resizebox{0.99\hsize}{!}
{
\includegraphics[width=0.24\hsize]{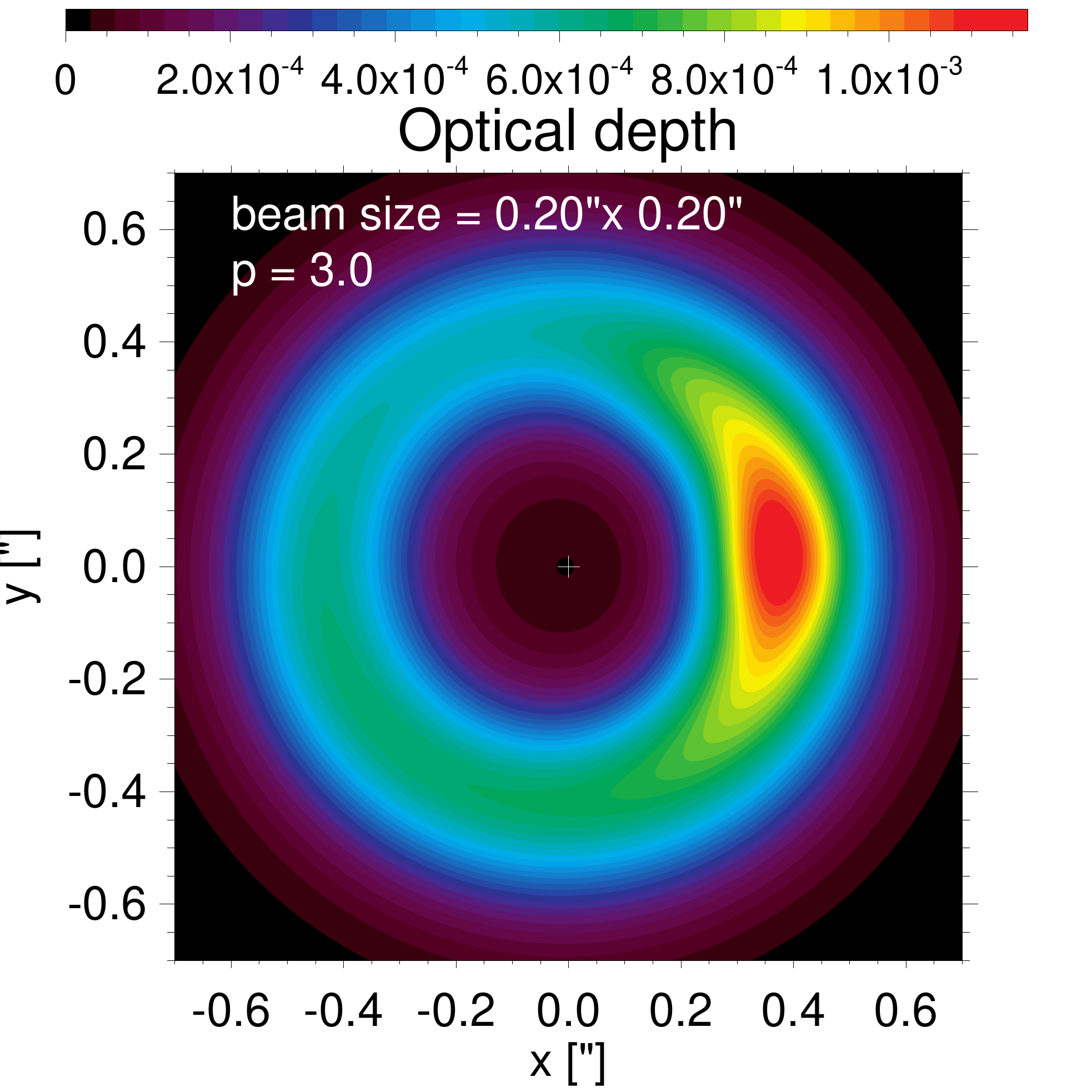}
\includegraphics[width=0.24\hsize]{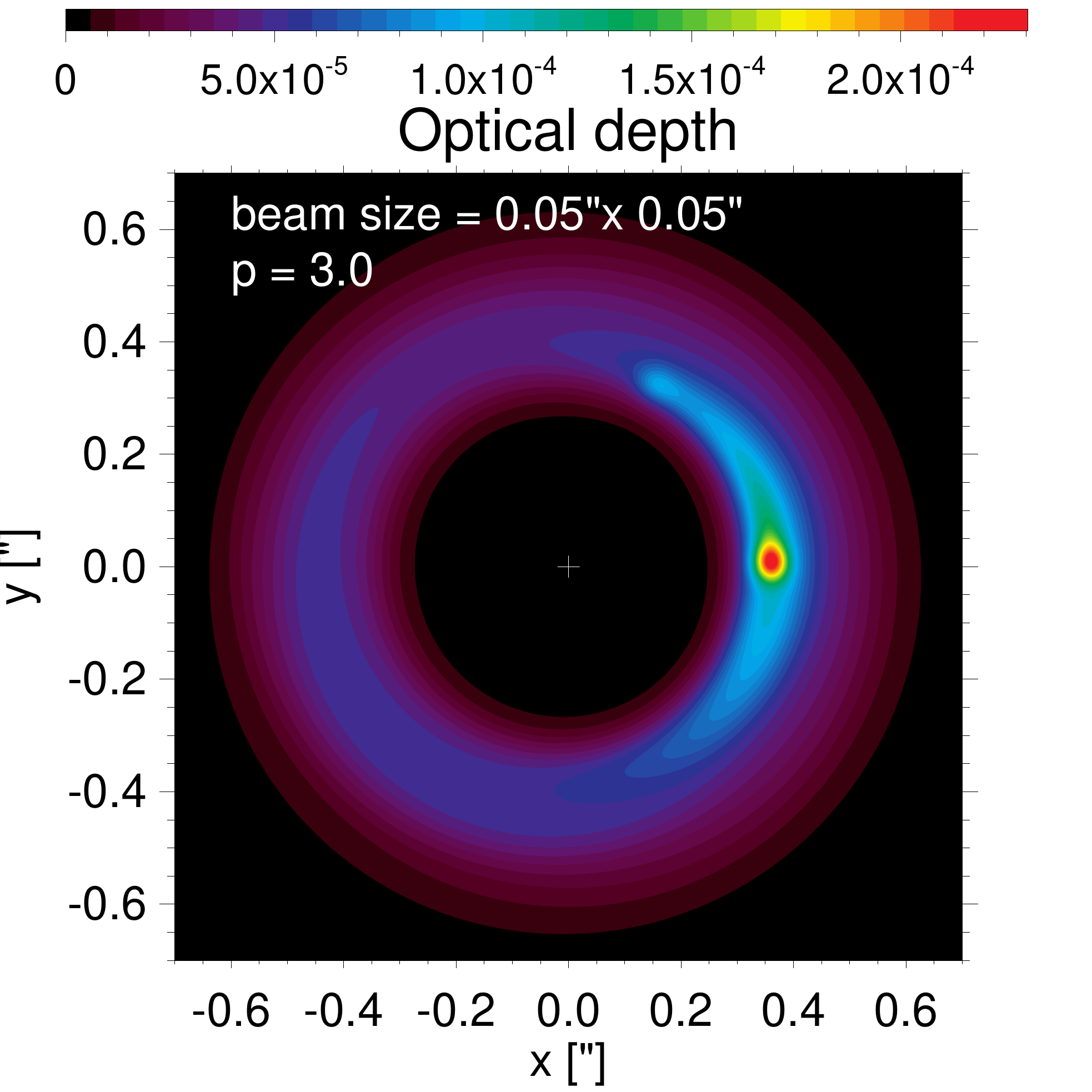}
\includegraphics[width=0.24\hsize]{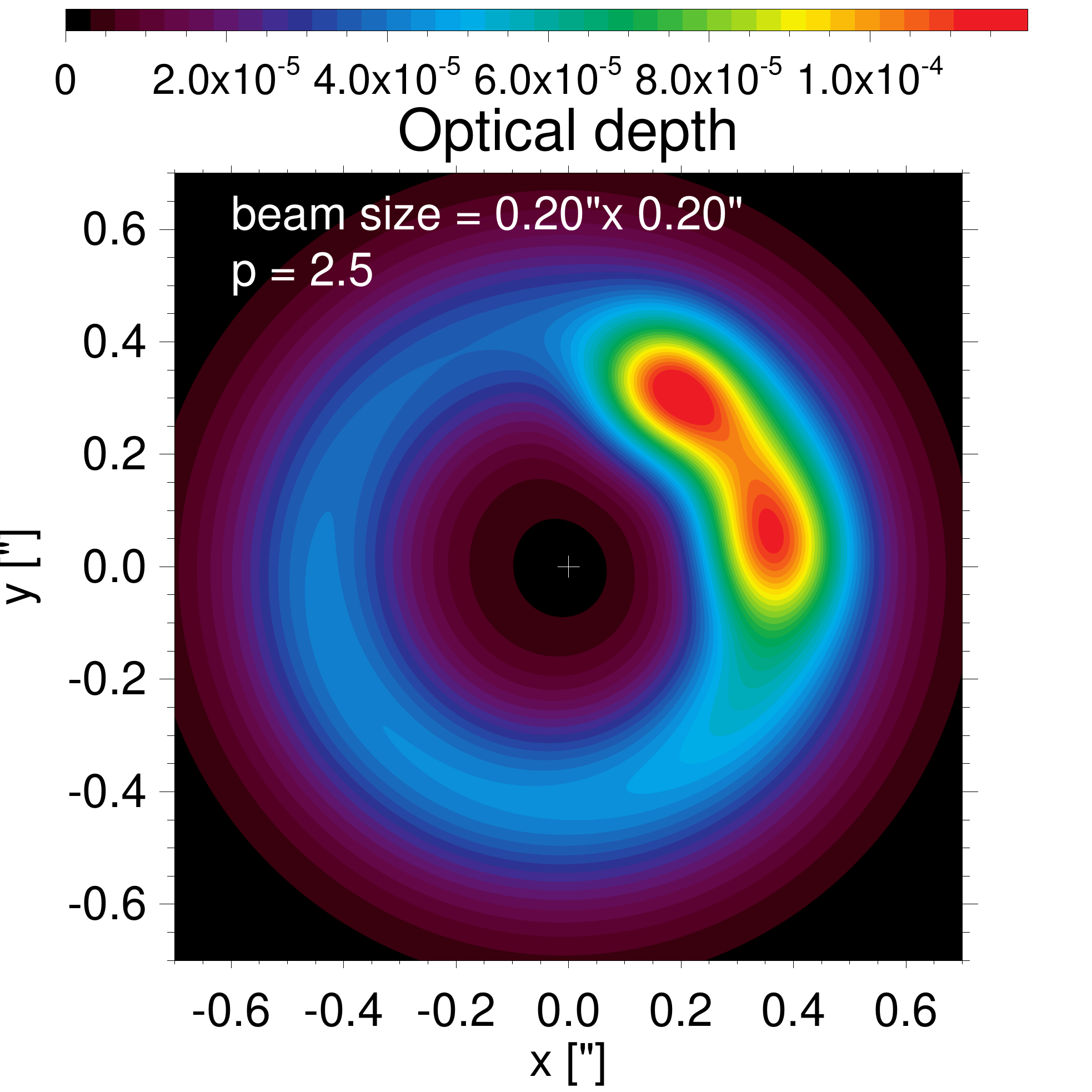}
\includegraphics[width=0.24\hsize]{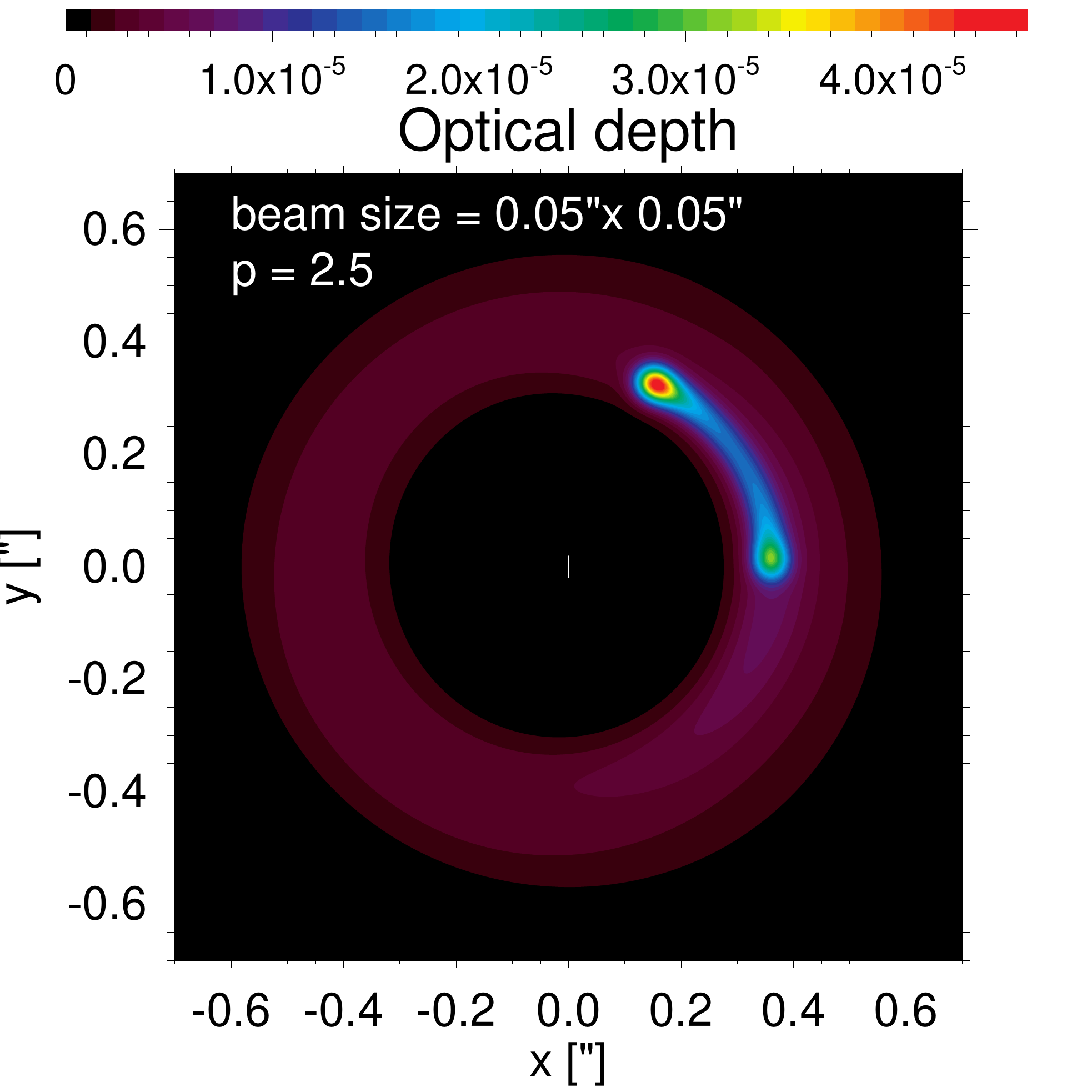}
}
\resizebox{0.99\hsize}{!}
{
\includegraphics[width=0.24\hsize]{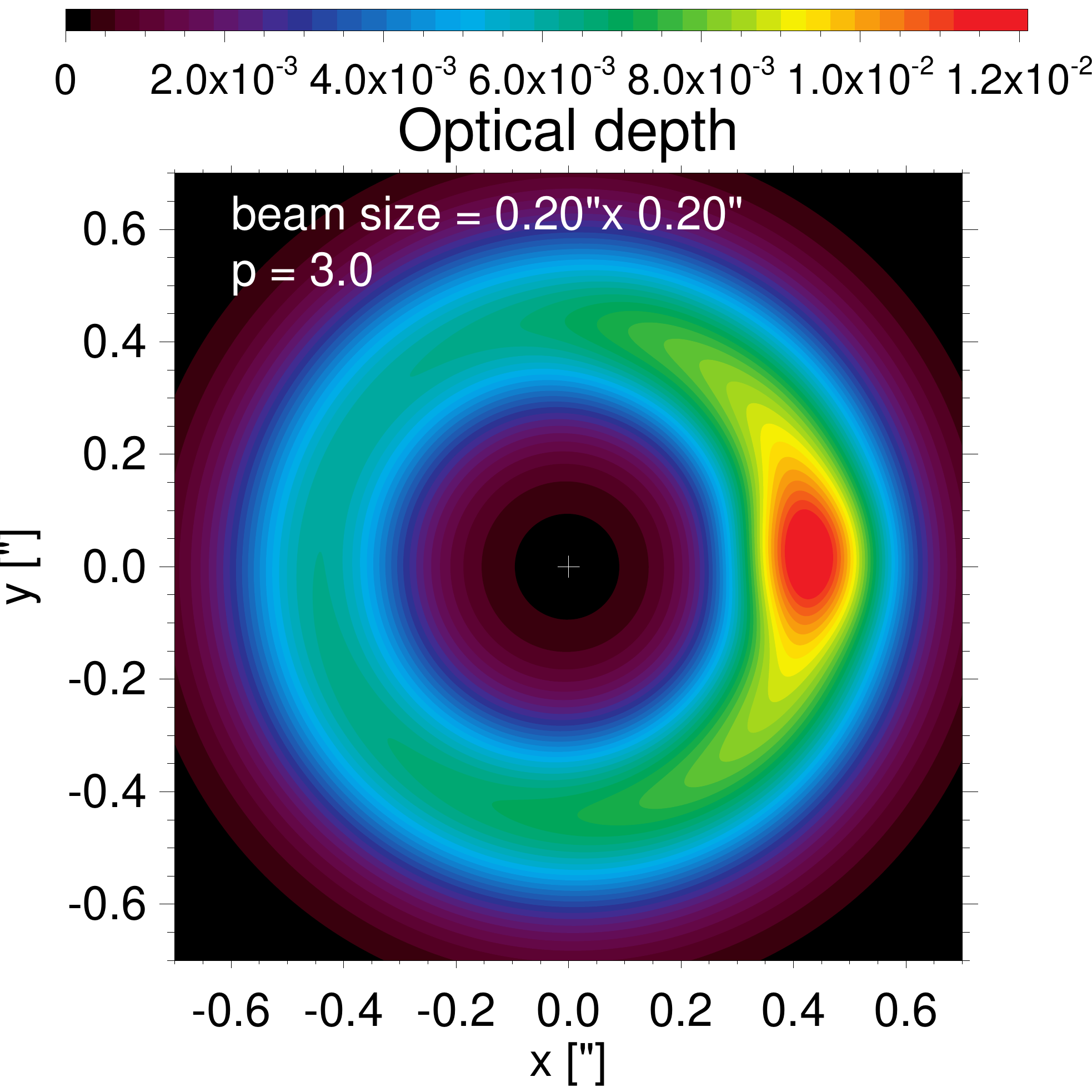}
\includegraphics[width=0.24\hsize]{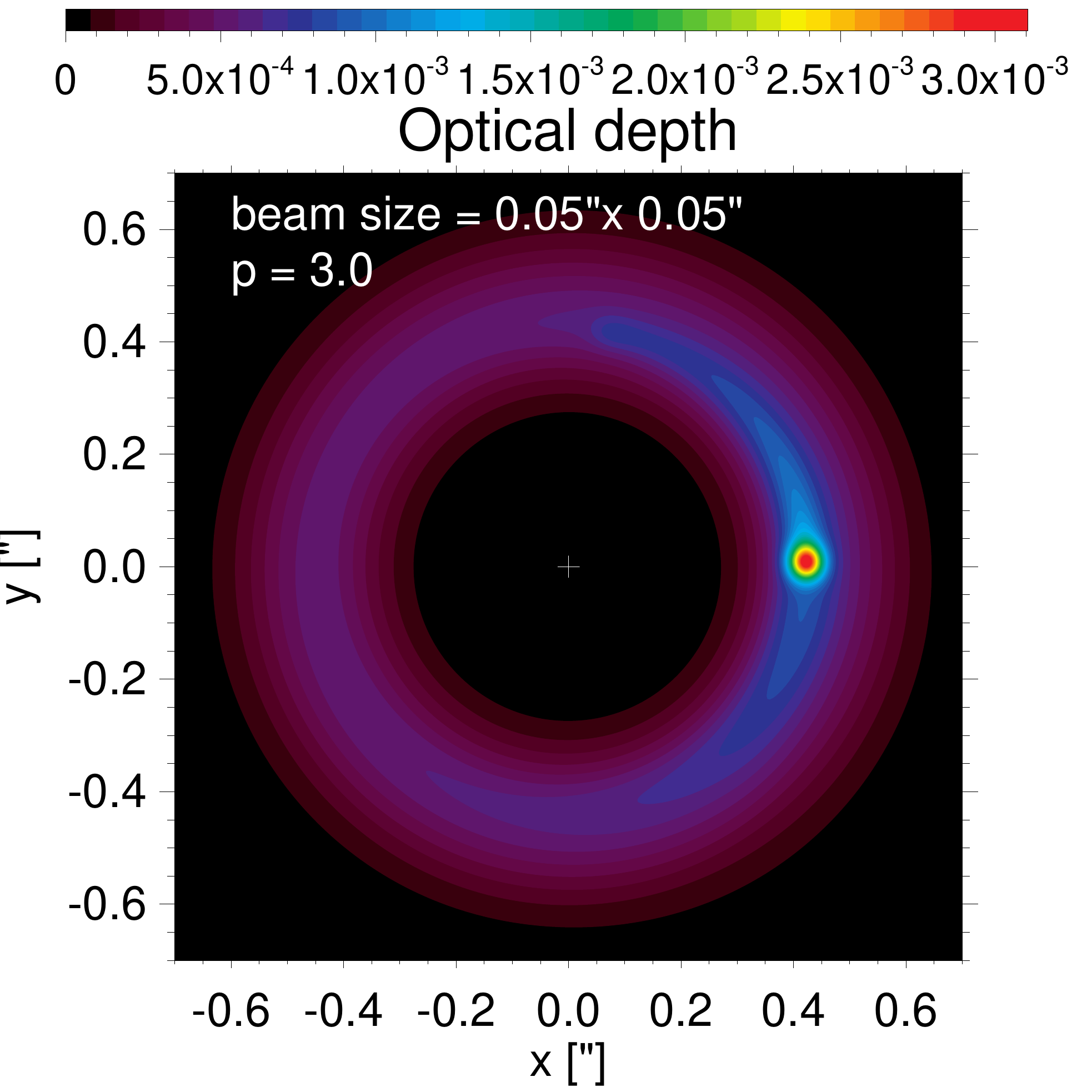}
\includegraphics[width=0.24\hsize]{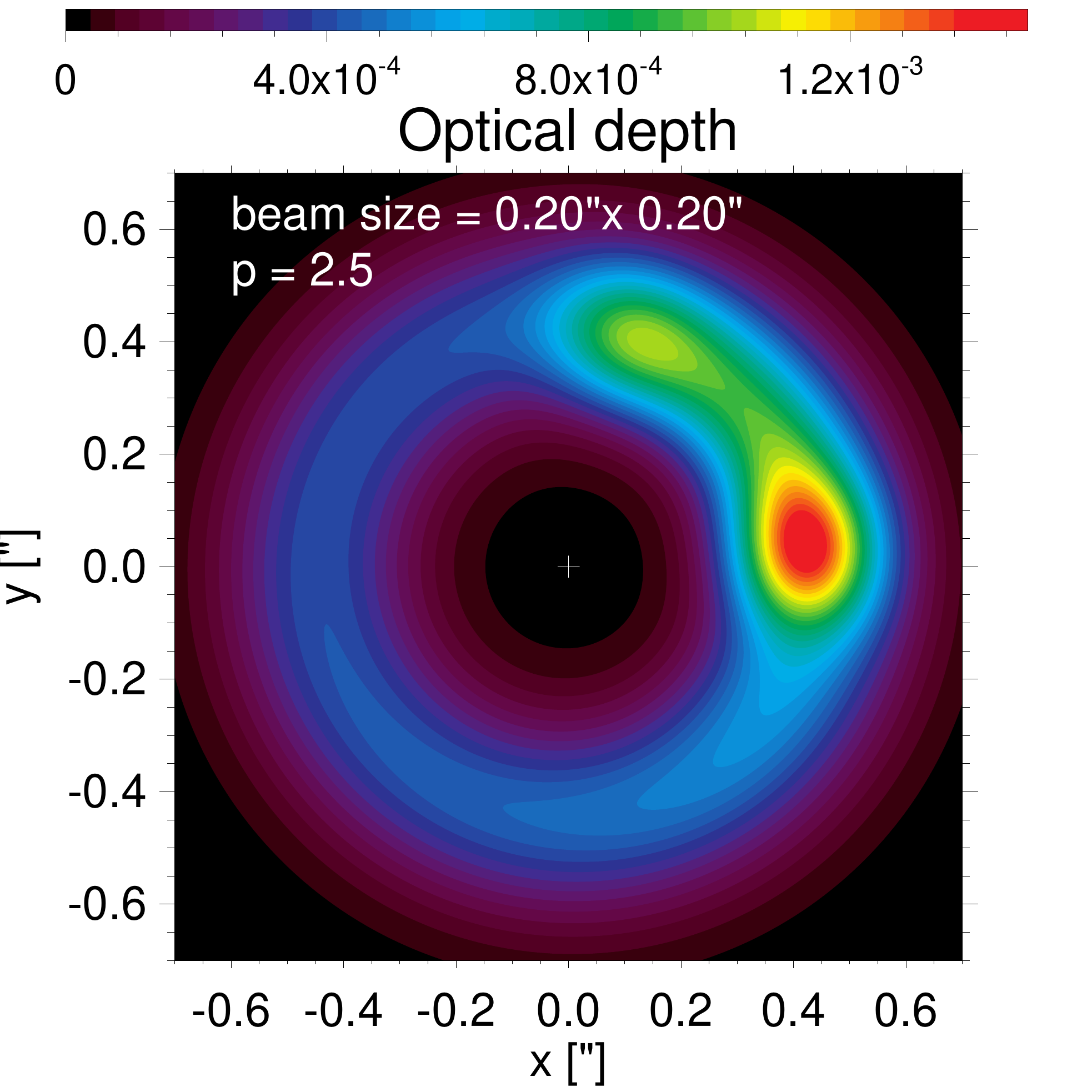}
\includegraphics[width=0.24\hsize]{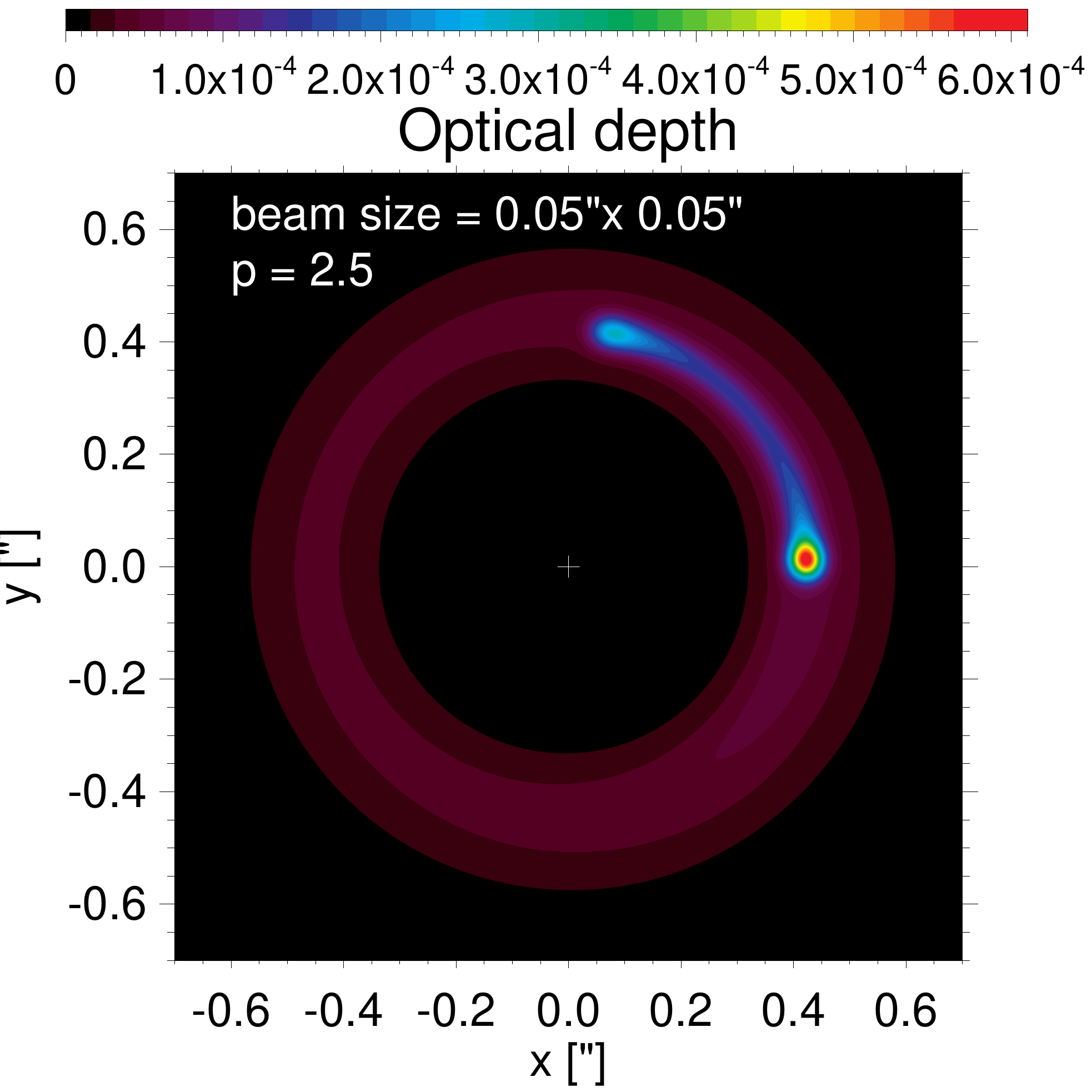}
}
\caption{\label{fig:synthetic} Synthetic images at 37.5 GHz (8 mm) of
  the disc's optical depth obtained from our simulations with models
  g0p2 (upper panels) and g2 (lower panels). The value of $p$ (minus
  the power-law index of the grains size distribution) equals 3 in the
  two columns on the left-hand side, and 2.5 on the right-hand
  side. Convolution is done with a Gaussian beam size of FWHM set to
  0.2" or 0.05". The plus sign at the origin marks the star's
  location. }
\end{figure*}

\section{Synthetic dust continuum observations}
\label{sec:synthetic}

In this section, we use the dust distribution in our simulations of
self-gravitating lopsided discs to produce synthetic images of dust
continuum observations. We focus on the models where the gas surface
density at the vortex location is lower or equal than in model g2, for
which, using the physical units of Section~\ref{sec:codeunits}, grains
up to few centimetres in size have nearly stationary point-like
distributions around the vortex's radial location. (In contrast, in
models with larger gas densities like models g5 and g10, grains larger
than about 1 cm have non-axisymmetric ring-like distributions with
non-trivial dependence on grain size, and are non-stationary in a
frame rotating at the vortex's frequency, which makes it more delicate
to predict the corresponding dust continuum emission). Particle shift
angles relative to the vortex centre are displayed in
Figure~\ref{fig:depth} for models g0p2, g0p5 and g2 at 500 orbits
after the beginning of the simulations (250 orbits after introduction
of the particles). For each model, particles larger than the largest
size shown in the figure describe ring-like structures centred around
the vortex's radial location.

To translate the dust distribution in our simulations into dust
surface density, we assume that dust grains have a size distribution
$n(s) \propto s^{-p}$ from 30 $\mu$m to 20 cm. The value of $p$ is
very uncertain, especially in the context of dust trapping in a
vortex, and we allow it to vary from 3 to 2.5. We divide the above
size range into small size bins with a logarithmic spacing. We assume
that grains smaller than 0.5 mm follow the gas distribution, and that
larger grains have point-like distributions shifted ahead of the
vortex. Particles are assigned shift angles ($\delta\varphi$) through
fits of the results of simulations shown in Figure~\ref{fig:depth}.
For model g2, we find good agreement with $\delta\varphi(s) =
90^{\circ} \times 0.5 \{1 + \tanh(2.6 \log_{10}(14s))\}$ with $s$ in
metres (see the black dashed curve in Figure~\ref{fig:depth}).  For
model g0p5, we use $\delta\varphi(s) = 80^{\circ}\times 0.5 \{1 +
\tanh(2.3 \log_{10}(18s))\}$ (blue dashed curve in the figure), and
for model g0p2, $\delta\varphi(s) = 70^{\circ} \times 0.5 \{1 +
\tanh(2 \log_{10}(33s))\}$ (red dashed curve). The dust's surface
density in each size bin is then derived assuming a total dust-to-gas
mass ratio of 1\%, and converting our results of simulations into
physical units using the set of units in Section~\ref{sec:codeunits}.

We use Mie theory to determine the dust opacity in each size bin at a
wavelength of 8 mm (37.5 GHz). See \cite{ZhuStone14} for more details
(their section 4.3). The dust opacities per unit dust mass at 8 mm are
$1.2\times10^{-3}$, $1.7\times10^{-3}$, $1.4\times10^{-2}$, 0.79,
0.67, 0.15, $5.2\times10^{-2}$ and $1.0\times10^{-2}$ cm$^2$ g$^{-1}$
for grains of radius 30 $\mu$m, 100 $\mu$m, 300 $\mu$m, 1 mm, 3 mm, 1
cm, 3 cm and 20 cm. We finally multiply the dust's surface density in
each size bin with the interpolated opacity and add these products
together to get the disc's optical depth. Synthetic images are
produced by convolving the disc's optical depth with a Gaussian kernel
of FWHM equal to 0.05" or 0.2". The disc is assumed to be located at
140 parsecs and to be seen face-on.

Synthetic images are displayed in Figure~\ref{fig:synthetic} for
models g0p2 (upper panels) and g2 (lower panels). Values of $p$ (minus
the power-law index of the dust size distribution), and of the FWHM of
the Gaussian kernel by which the total optical depth is convolved, are
indicated in the top-left corner in the panels. These images show
that, for steep size distributions ($p \geq 3$), the dust continuum
emission at 8 mm displays a lopsided distribution with a peak emission
at the vortex centre (where the gas surface density is maximum). For
shallower size distributions ($p \lesssim 2.5-3$), however, the dust
continuum emission at 8 mm features a double-peaked distribution, with
one peak at the vortex centre, and a second peak shifted ahead of the
vortex centre in the azimuthal direction, which corresponds to the
emission of grains of a few centimetres in size.  The azimuthal shift
of the second peak depends on the azimuthally-averaged surface density
of the gas at the vortex's radial location, $\langle\Sigma_{\rm
  V}\rangle$ (Figure~\ref{fig:depth}). It is about 70 degrees in model
g0p2, where $\langle\Sigma_{\rm V}\rangle \approx 0.5$ g cm$^{-2}$ at
$\approx 50$ AU, and goes up to nearly 90 degrees in model g2, where
$\langle\Sigma_{\rm V}\rangle \approx 5$ g cm$^{-2}$ at $\approx 60$
AU.

Such double peak continuum emissions, if observed at mm/cm wavelengths
like with ALMA or the VLA, could reveal the size segregation due to
gas self-gravity between small and large dust grains trapped in a
crescent-shaped vortex. There is so far no clear evidence for double
peak emissions similar to what our dusty horseshoe model predicts. The
double peak emission observed with ALMA in the HD 142527 disc is most
probably caused by shadows cast by the tilted inner disc over the
lopsided outer disc \citep{Casassus15,Marino15shadows}. VLA (0.9 cm)
and ALMA Band 7 (0.8 mm) continuum emissions of the MWC 758 disc show
two intensity peaks located to the north-west and to the south of the
star \citep{Marino15MWC758}. These authors show that two dust-trapping
vortices could account for the double peak continuum emissions, and
the fact the VLA emission is more compact than the ALMA emission. In
their model, the vortex to the south would be slightly closer to the
star (at $\sim 90$ AU) than the one to the north-west (at $\sim 120$
AU). It would be interesting to check whether our dusty horseshoe
model, which comprises a single vortex, could also reproduce the
double peak distribution in the MWC 758 disc. The large-scale spirals
observed in near-IR \citep{Benisty15} could greatly complicate the
dust concentration properties, whatever the number of dust-trapping
vortices to reproduce the double peak continuum emission at mm/cm
wavelengths. We also stress that in the synthetic sub-mm images of
\cite{Regaly12}, a single elongated vortex in the gas disc can produce
either one or two peaks of emission due to beam dilution effects
(especially when angular resolution is moderate and the beam is highly
elongated; Zs. Reg{\'a}ly, priv. comm.).

\section{Concluding remarks}
\label{sec:conclusion}
We examine in this paper the concentration properties of large dust
grains in protoplanetary transition discs characterised by a lopsided
crescent-shaped distribution in the gas surface density. 2D
hydrodynamical simulations have been carried out including both gas
and dust, with dust modelled as test particles undergoing gas drag
(Epstein regime). The lopsided distribution in the gas is obtained by
adopting a ring-like gas density profile that becomes unstable against
the Rossby-wave instability.  The instability saturates into a
large-scale vortex that efficiently traps the grains. This trapping
mechanism is often invoked to interpret the crescent-shaped continuum
emission imaged by sub-millimetre interferometers in some transition
discs \citep[e.g.,][]{Regaly12}.

In accompanying Paper I (Zhu \& Baruteau, submitted) we explore the
role of gas self-gravity and of the disc mass in shaping the vortex
structure.  Their impact on the trapping of dust grains is the scope
of the present paper. We first present results of models without gas
self-gravity, for which we find that the indirect force resulting from
the displacement of the star by the vortex plays a prominent role in
the gas and dust dynamics, in agreement with \cite{MC15}.  While
grains with Stokes number (stopping time to orbital time ratio, St)
much smaller and much larger than unity are trapped near the vortex
centre, particles with St $\sim 1$ are shifted ahead of the vortex in
the azimuthal direction. This is in good qualitative agreement with
the results of test particle integrations by \cite{MC15}.  We find
particle shift angles relative to the vortex centre that go up to
20-25 degrees for St $\sim 2-4$ particles, depending on the disc mass,
which strongly affects the shape and orbital evolution of the vortex.
Our maximum shift angles are a factor $\sim 2$ smaller than in
\cite{MC15}.

We then present results with gas self-gravity, and find that it
largely affects dust trapping in the vortex. At low to moderate gas
surface densities at the vortex's radial location (local Toomre
parameter $\gtrsim 10$), we find gas self-gravity to be most important
for large grains. Grains with St $\lesssim 1$ (typically smaller than
1 cm) have similar shift angles with and without self-gravity, as gas
drag plays a major role in determining the dynamics of small
particles.  The dynamics of larger particles (St $\gtrsim 1$) becomes
increasingly dominated by gas self-gravity, and we find that shift
angles increase very rapidly with increasing grain size
(Figure~\ref{fig:depth}). We find shift angles of about 50 degrees for
grains of 3-5 cm in size, and up to 90 degrees for $\sim$ 10 cm
grains.  We show that these large shift angles result from the large
particles undergoing horseshoe U-turns relative to the vortex before
drifting to their equilibrium location via gas drag. The horseshoe
U-turns arise because of the self-gravitating acceleration of the gas
exerted on the particles, which therefore interact with the vortex as
if it was a massive body. The limit case of fully decoupled particles
actually shares a number of analogies with the evolution of inviscid
gas near a low-mass planet. Synthetic images produced from our
simulations show that the size segregation caused by self-gravity
between the smallest and largest grains at the vortex's radial
location could cause a double peak in the dust's continuum emission,
depending on the dust's size distribution. The same size segregation,
which does not mix the smallest and largest grains at the vortex
centre, could have interesting consequences on core growth inside the
vortex, and on dust feedback onto the gas.

At large gas surface densities at the vortex's radial location (local
Toomre parameter $\lesssim 10$), self-gravity renders the vortex's
pattern frequency slower than the Keplerian frequency.  This shifts
the vortex's corotation radius outside the radial location where the
lopsided gas density distribution is maximum. Consequently, (most of)
the smallest particles concentrate in point-like distributions that
are shifted slightly outside the gas density maximum. Conversely,
larger particles (St $\gtrsim 0.1$) describe ring-like structures,
some of them lopsided, which are centred about the radial location
where the gas density peaks (Figures~\ref{fig:g1_rf2}
and~\ref{fig:g10_rf2}).

Our study highlights that the concentration of large dust grains is
sensitive to the shape of the vortex, which depends on the gas model.
The main result of Paper I is that self-gravity plays a prominent role
in structuring the vortex in massive discs. The main result of the
present paper is that self-gravity plays a prominent role in setting
particle concentration within the vortex even at low disc masses.

There are several ways that the physical model used in this work can
be improved. Inclusion of an energy equation for the gas, of the
disc's vertical stratification (via 3D simulations), and of possibly
important feedback effects of dust concentration onto the gas are some
of them. Another outstanding question is that of the particles
concentration in a vortex formed at the outer edge of a gap-opening
planet, which could be significantly impacted by the planet wakes.
More generally, a pressing, yet unsolved issue is that of the
excitation and long-term maintenance of a large-scale vortex through
the RWI. In this work, we have assumed a narrow ring-like density
profile for the (nearly inviscid) gas plus an ad hoc $m=1$ seed
perturbation; however, there is no robust way to generate
either. Non-axisymmetric seed perturbations could be
induced by disc turbulence, embedded planet companions or nearby
stars. Several mechanisms mentioned in the third-to-last paragraph in
the introduction could provide the necessary axisymmetric density
(vortensity) extremum, like for instance an effective viscosity jump
at the transition between magnetically active and inactive
regions. But, again, more work is needed to assess whether these
mechanisms, and the vortices that they trigger, have a long lifetime.

\appendix

\section{Convergence with grid resolution}
\label{sec:conv}
\begin{figure}
\centering
\includegraphics[width=0.49\hsize]{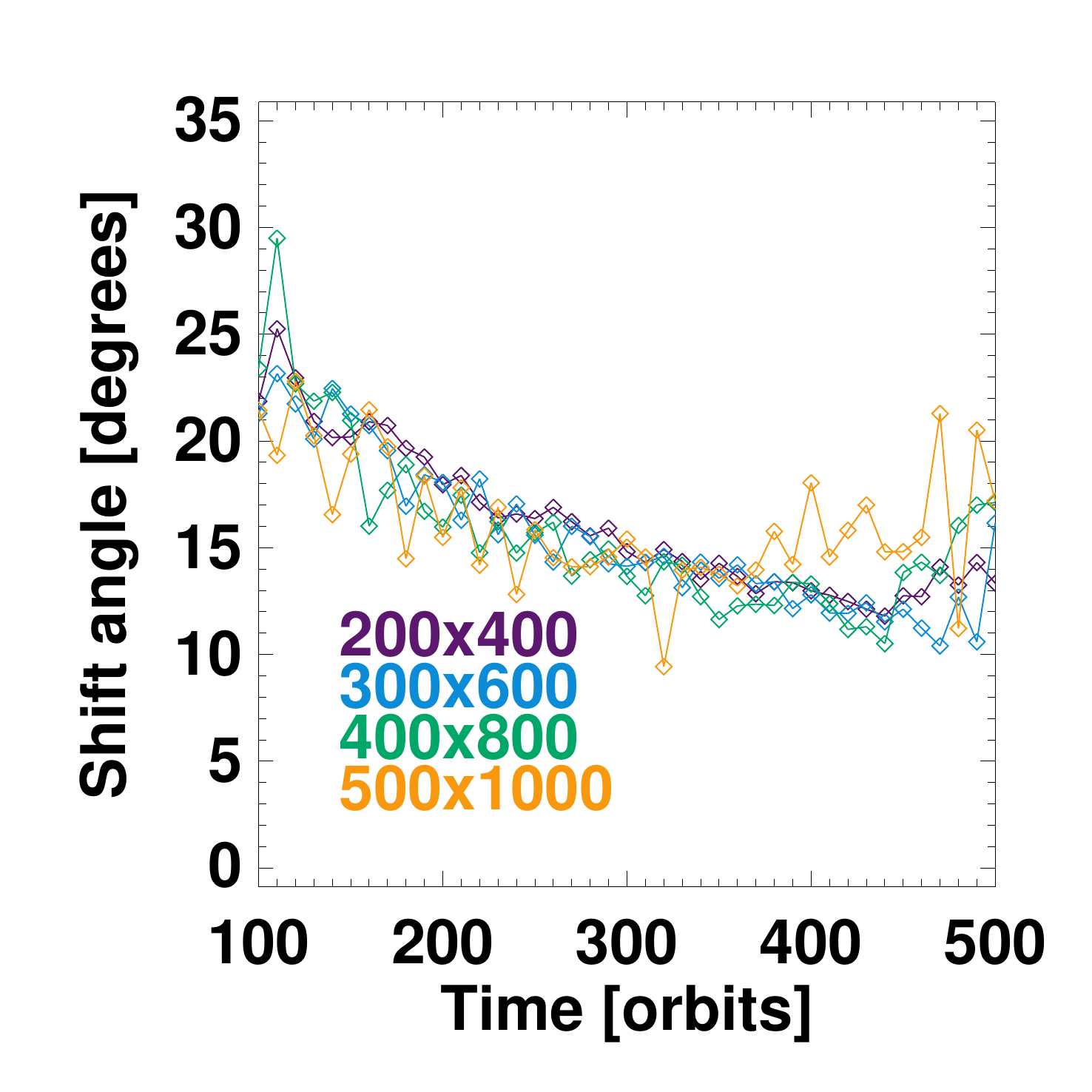}
\includegraphics[width=0.49\hsize]{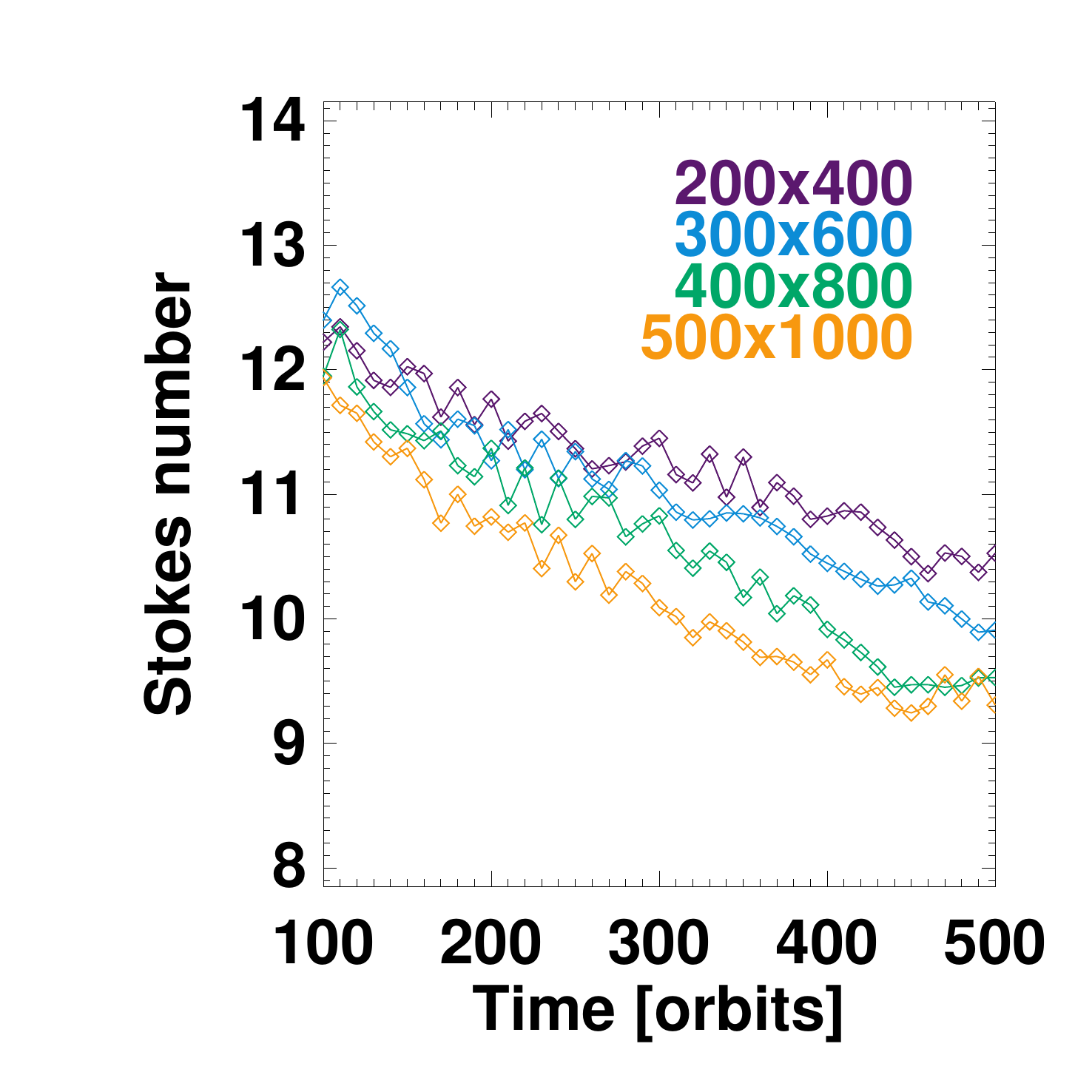}
\includegraphics[width=0.49\hsize]{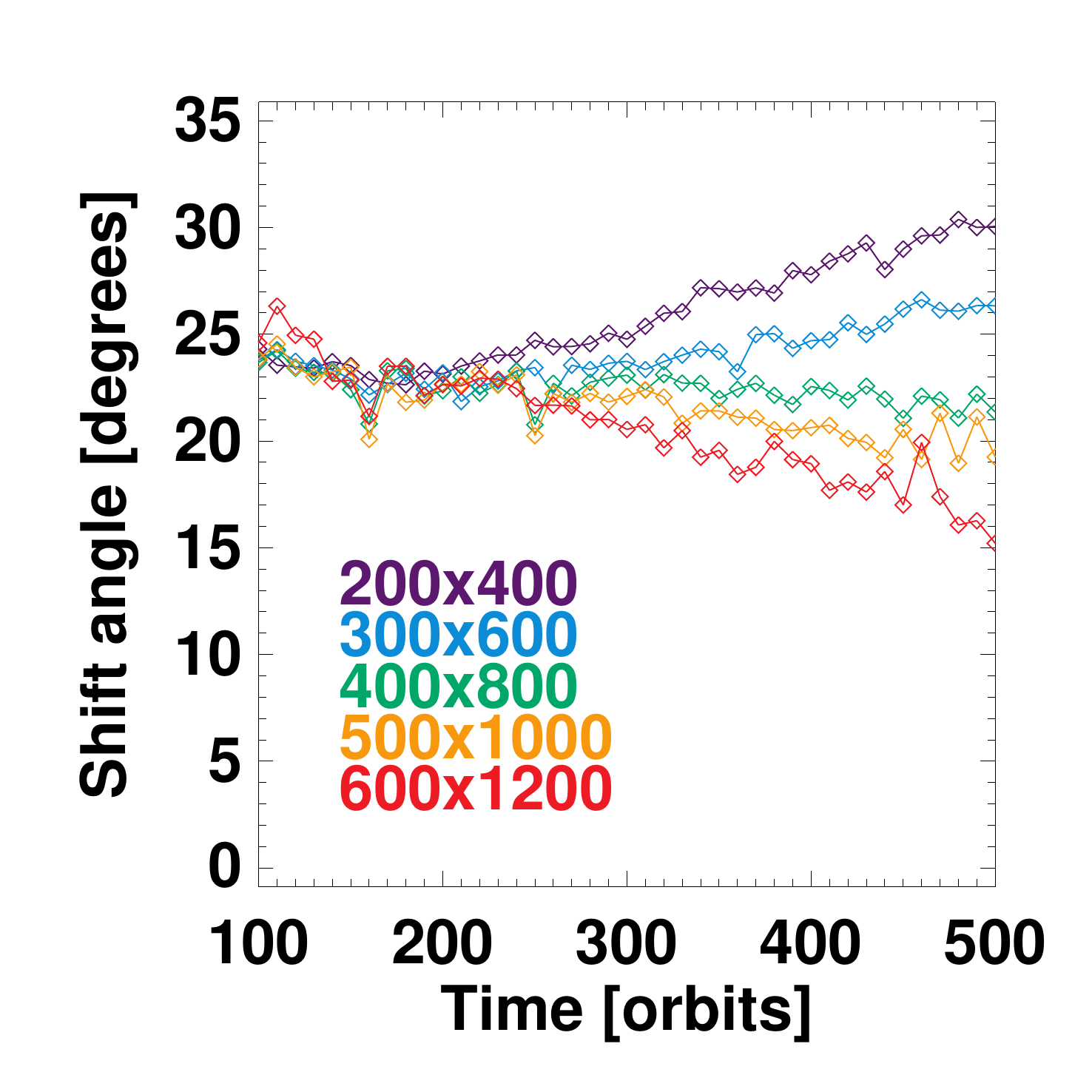}
\includegraphics[width=0.49\hsize]{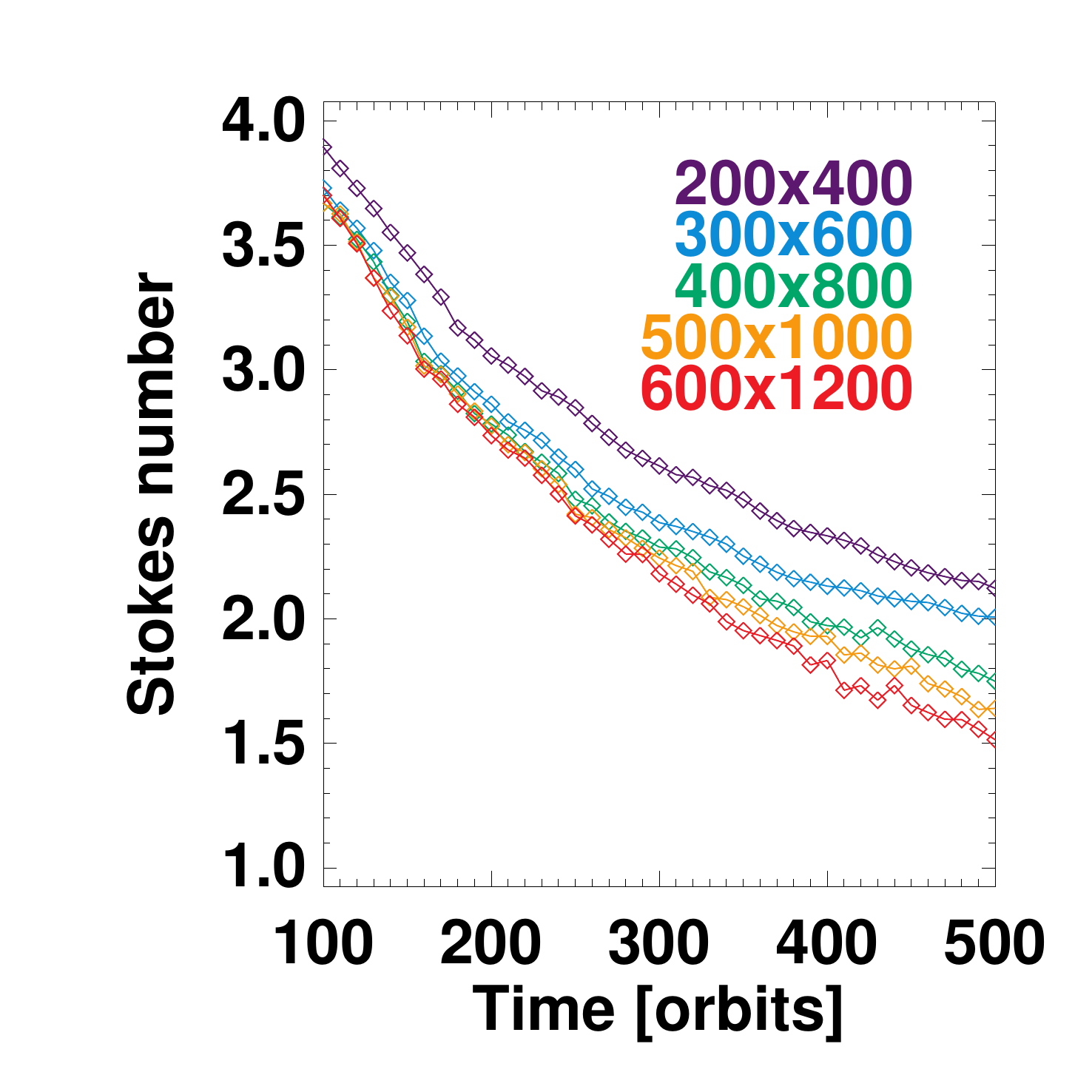}
\caption{\label{fig:conv}Time evolution of the particles shift angle
  and Stokes number in models g2n (upper panels) and g5n (lower
  panels) when varying the grid resolution. The grid resolution is
  shown in the panels as $N_{\rm r} \times N_{\rm s}$ with $N_{\rm r}$
  and $N_{\rm s}$ the number of grid cells in the radial and azimuthal
  directions, respectively.}
\end{figure}

We have carried out tests of convergence with increasing grid
resolution for the non self-gravitating models g2n and g5n. Each model
was simulated with the same particle size, and with the number of grid
cells in the radial times azimuthal directions increased from $200
\times 400$ to $500 \times 1000$ for model g2n, and up to $600 \times
1200$ for model g5n. Results are displayed in
Figure~\ref{fig:conv}. For model g2n (upper panels) convergence in
resolution is fairly good, with the particles shift angle and Stokes
number varying by less than 10\% between our nominal resolution
($300\times 600$) and maximum resolution. This fairly good convergence
is not surprising since our gas model aims at producing a large-scale
lopsided distribution in the gas surface density ($m=1$
mode). Convergence is not excellent either. It is due to the fact that
vortex inward migration occurs at a pace that slightly increases with
increasing resolution. Visual inspection at the gas density at
different resolutions points to a larger density contrast of the
vortex waves with increasing resolution, which relates to a larger
angular momentum flux extracted from the vortex and carried away by
the waves, and therefore faster inward migration (note again the
analogy with planetary migration). Increasing resolution therefore
tends to strengthen the vortex and increase the gas density inside the
vortex, which causes the particles Stokes number to slightly decrease,
explaining our results. This behaviour is also apparent in model g5n
up to about 300 orbits, after which discrepancies with varying
resolution are found to increase, due to the proximity of the vortex
to the inner edge of the computational domain and the resulting
partial reflection of waves at that location. Apart from when the
vortex gets too close to the grid's inner edge, convergence with grid
resolution is satisfactory and our results show that the moderate grid
resolution adopted throughout this work is sufficient for our
purposes. Tests of convergence in resolution with self-gravity are
presented for model g10 in Section~\ref{sec:g10}.

\section{Self-gravitating simulations with decoupled dust particles}
\label{sec:pinf}
\begin{figure}
\centering
\includegraphics[width=0.49\hsize]{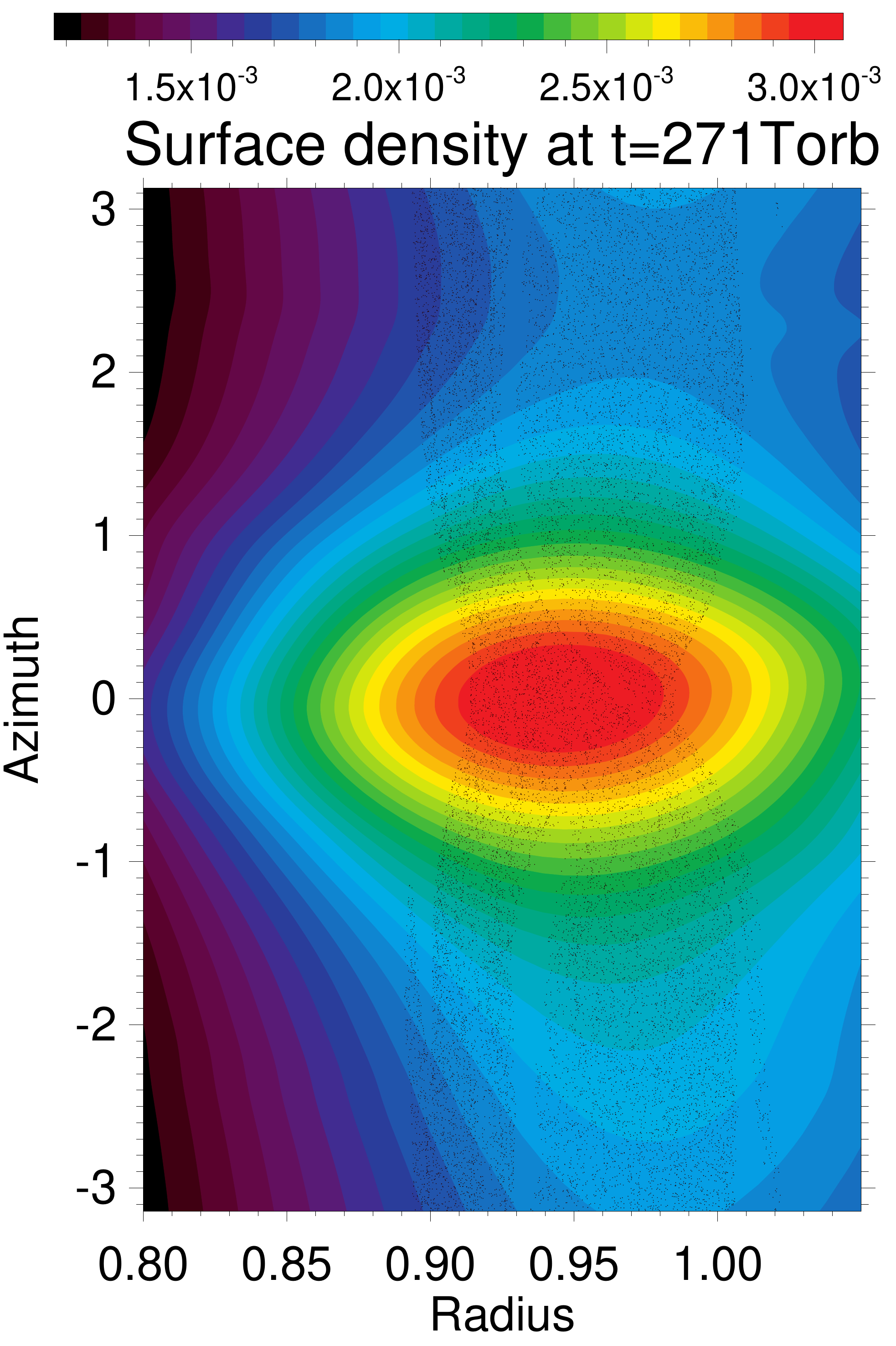}
\includegraphics[width=0.49\hsize]{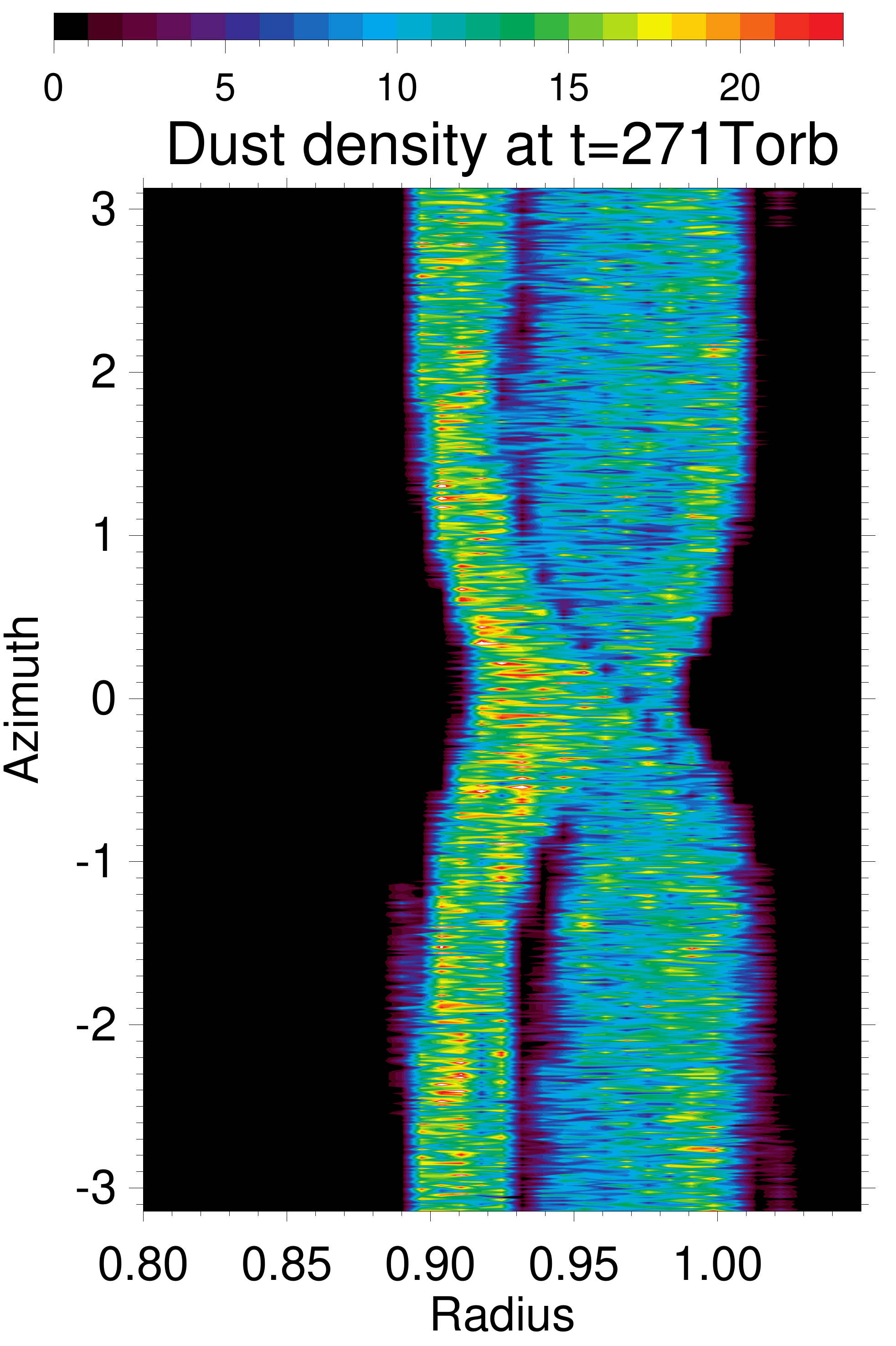}
\includegraphics[width=0.49\hsize]{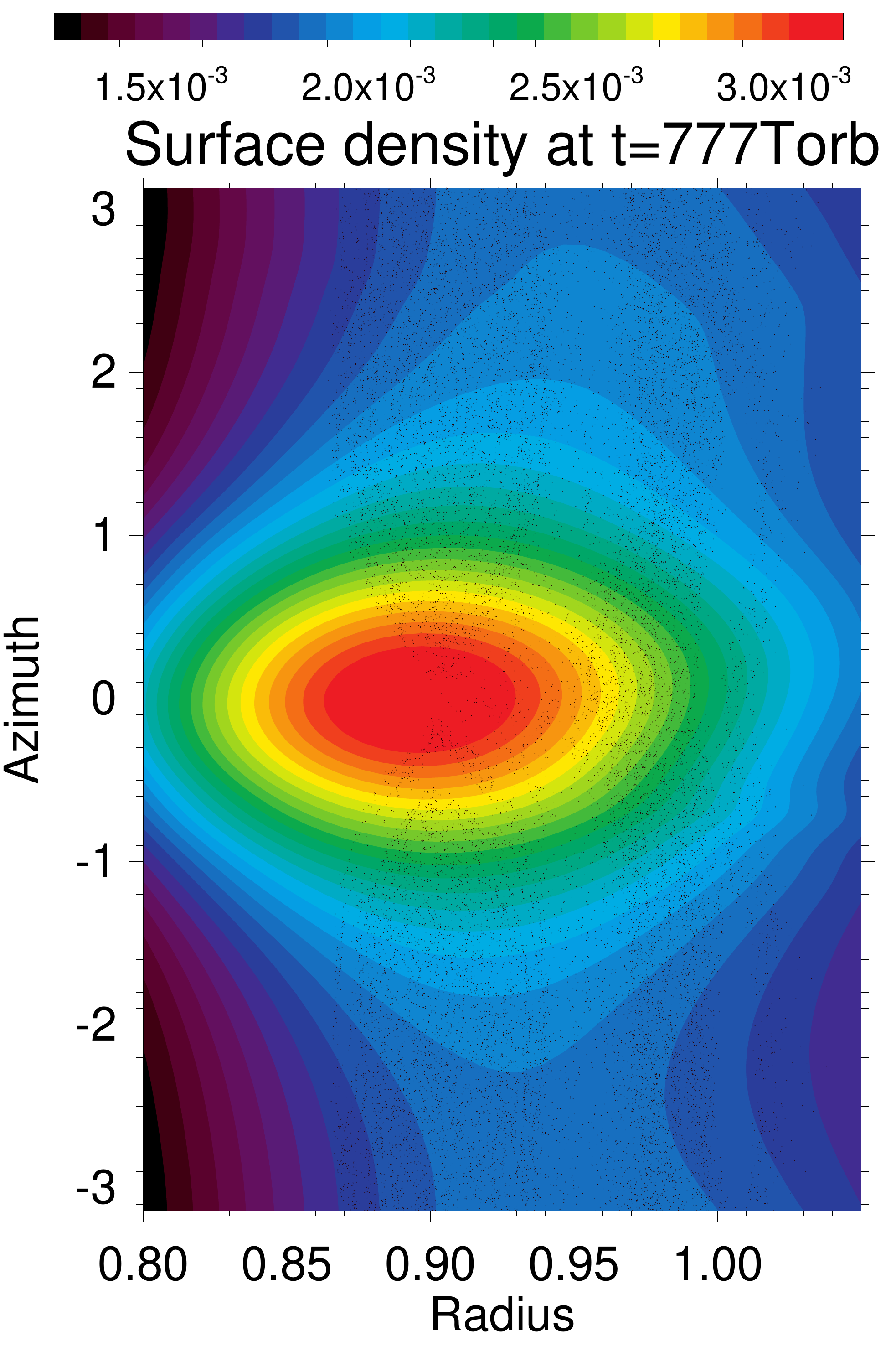}
\includegraphics[width=0.49\hsize]{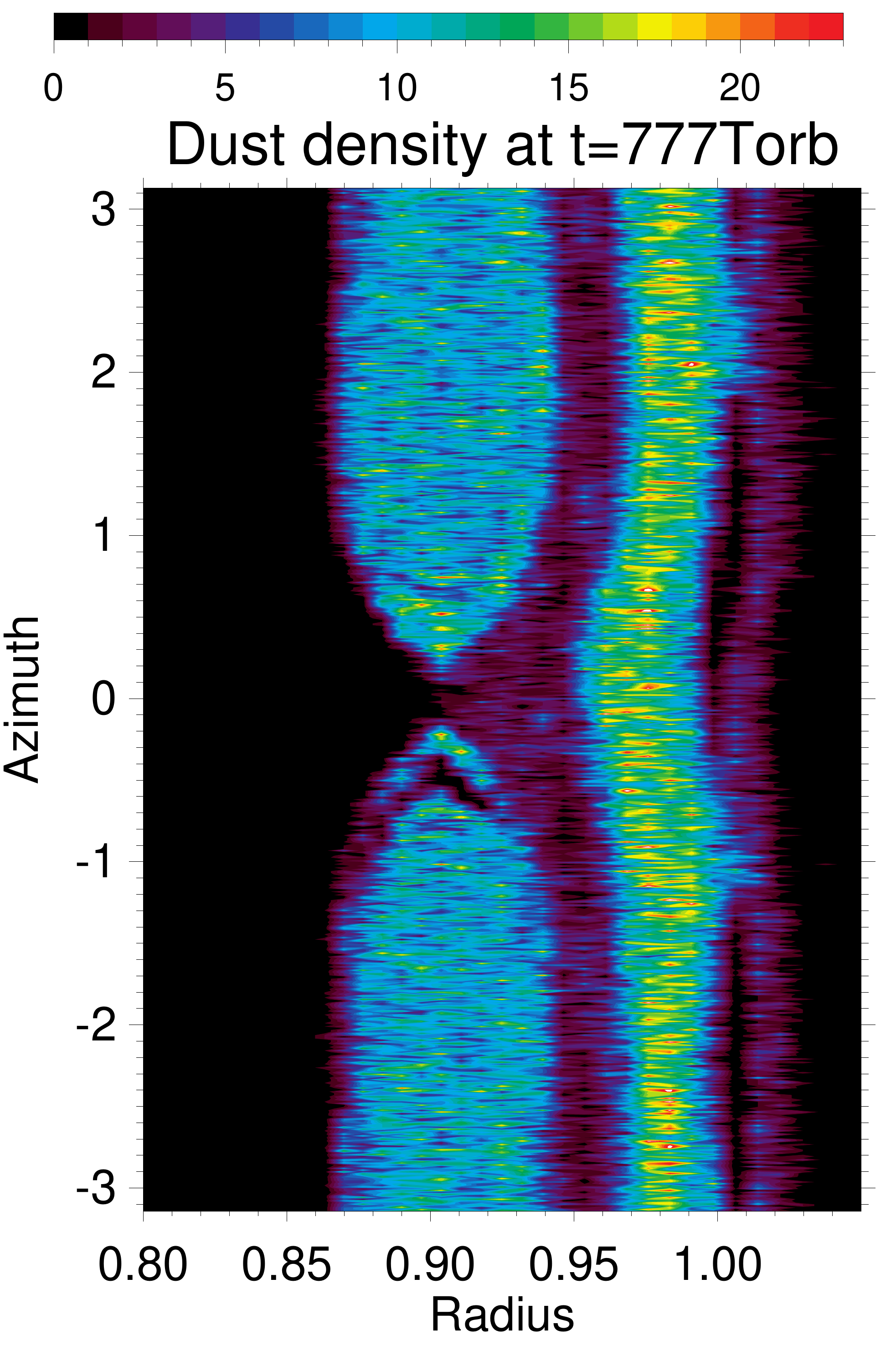}
\caption{\label{fig:pinf}Results of model g2 with dust particles fully
  decoupled from the gas. Contours of the gas surface density with the
  particles location overplotted by black dots are shown in the left
  panels. The dust density distribution is displayed in the right
  panels (see text). Results are shown at 271 and 777 orbits in the
  upper and lower panels, respectively.}
\end{figure}
We have shown in Section~\ref{sec:w} that with gas self-gravity and
moderate disc mass (model g2), the larger the particles, the further
they concentrate ahead of the vortex. We have shown this behaviour to
be related to horseshoe U-turns performed by the large particles
relative to the vortex before they drift to their equilibrium location
through gas drag.  In this section, we report the results of a
simulation with model g2 and fully decoupled dust particles. At 250
orbits, we restarted the simulation by introducing 100 000 particles
with a sharply decreasing radial distribution in
$r\in[0.9-1]$. Results are displayed in Figure~\ref{fig:pinf} at 21
orbits after the restart (upper panels) and 527 orbits after (lower
panels). While the left panels help locate the particles relative to
the vortex, the right panels display the dust's density distribution,
which we simply calculate as the number of particles per grid
cell. The upper-right panel is strongly suggestive that the particles
number density is advected inside what already appears as the vortex's
horseshoe region. In comparison, in the lower panels the vortex has
slightly migrated inward, with two notable consequences: (i) the dust
density is nearly uniform in the distinct horseshoe region of the
vortex, and (ii) the region of highest dust density, which is
initially inside the vortex's orbit, ends up circulating relative to
the vortex beyond its horseshoe region. These results are analogous to
the advection of gas potential vorticity and/or entropy in the
presence of a low-mass planet \citep[e.g.,][]{BaruteauPP6}.

In the lower panels of Figure~\ref{fig:pinf}, the full width of the
vortex's horseshoe region is $\approx 0.07$. Although the vortex is
not a point mass, it is instructive to compare its horseshoe width to
that expected for a planet with the same mass. We
  estimate the mass of the vortex ($m_{\rm v}$) as the mass between $r
  = 0.84$ and $r=0.95$, $\varphi = \pm \pi/4$, and having uniform
  surface density $\approx 3\times 10^{-3}$ in code units (see
  lower-left panel in Figure~\ref{fig:pinf}).  We obtain $m_{\rm v}
  \approx 4.6\times10^{-4}M_{\star}$. In the circular restricted
  three-body problem, the maximum full width of the horseshoe region
  of a planet with mass $m_{\rm v}$ and orbital distance $r_{\rm v}$
  is $2\sqrt{12}\times r_{\rm v}(m_{\rm v}/3M_{\star})^{1/3}$
  \citep{MurrayDermottBook}, which amounts here to $\approx 0.33$.
  Using three-body integrations, we find that the maximum full width
  for test particles to undergo periodic horseshoe trajectories with a
  small epicyclic motion is $\approx 0.10$, which is about twice the
  planet's Hill radius ($2r_{\rm v} ( m_{\rm v}/
  3M_{\star})^{1/3}$). Finally, the maximum full width of the planet's
  horseshoe region in the gas with same aspect ratio ($h = 0.1$) and
  same viscosity as in the simulations presented throughout this study
  is $\approx 0.14$. Not surprisingly, the width of the vortex's
  horseshoe region is smaller than all above estimates for a planet
  with the same mass as the vortex's. We note that, since $m_{\rm v} /
  M_{\star} \lesssim h^3$, the planet's horseshoe width in the gas
  scales as $\sqrt{m_{\rm v}}$ \citep[e.g.,][]{mak2006}, which
  indicates that the vortex's horseshoe width in the simulation shown
  in Figure~\ref{fig:pinf} is very similar to that of the gas around a
  planet of mass $m_{\rm v}/4 \approx 10^{-4}M_{\star}$.

\section*{Acknowledgments}
Z.Z. acknowledges support by NASA through Hubble Fellowship grant
HST-HF-51333.01-A awarded by the Space Telescope Science Institute,
which is operated by the Association of Universities for Research in
Astronomy, Inc., for NASA, under contract NAS 5-26555. We thank
Olivier Bern{\'e}, Simon Casassus and Zsolt Reg{\'a}ly for helpful
discussions, and Eugene Chiang, the reviewer, for a prompt and
constructive report.

\bibliographystyle{mnras}

\bsp
\label{lastpage}
\end{document}